\documentclass[aip,amsmath,amssymb, reprint]{revtex4-1}%
\usepackage[pdftex]{graphicx}
\usepackage[english]{babel}

\usepackage[T1]{fontenc}
\usepackage[colorlinks,linkcolor=blue,citecolor=blue,urlcolor=blue]{hyperref}
\hypersetup{pdftitle={Turbulent transport in 2D collisionless guide field reconnection},pdfauthor={P. A. Mu\~noz}}

\begin{document}

\title[]{Turbulent transport in 2D collisionless guide field reconnection}

\author{P. A. Mu\~noz}
\email{munozp@mps.mpg.de}
\affiliation{Max-Planck-Institut f\"ur Sonnensystemforschung, D-37077 G\"ottingen, Germany}
\affiliation{Max-Planck/Princeton Center for Plasma Physics, D-37077 G\"ottingen, Germany}

\author{J. B\"uchner}
\affiliation{Max-Planck-Institut f\"ur Sonnensystemforschung, D-37077 G\"ottingen, Germany}
\affiliation{Max-Planck/Princeton Center for Plasma Physics, D-37077 G\"ottingen, Germany}

\author{P. Kilian}
\affiliation{Centre for Space Research, North-West University, 2520 Potchefstroom, South Africa}

\date{\today}

\begin{abstract}

	Transport in hot and dilute, i.e., collisionless, astrophysical and space plasmas is called ``anomalous''. This transport is due to the interaction between the particles and the self-generated turbulence by their collective interactions. The anomalous transport has very different and not well known properties compared to the transport due to binary collisions, dominant in colder and denser plasmas.
	Because of its relevance for astrophysical and space plasmas, we explore the excitation of turbulence in current sheets prone to component- or guide-field reconnection, a process not well understood yet. This configuration is typical for stellar coronae, and it is created in the laboratory for which a 2.5D geometry applies. In our analysis, in addition to the immediate vicinity of the X-line, we also include regions outside and near the separatrices.
	We analyze the anomalous transport properties by using 2.5D Particle-in-Cell code simulations. We split off the mean slow variation (in contrast to the fast turbulent fluctuations) of the macroscopic observables and determine the main transport terms of the generalized Ohm's law. We verify our findings by comparing with the independently determined slowing-down rate of the macroscopic currents (due to a net momentum transfer from particles to waves) and with the transport terms obtained by the first order correlations of the turbulent fluctuations.
	We find that the turbulence is most intense in the ``low density'' separatrix region of guide-field reconnection. It is excited by streaming instabilities, is mainly electrostatic and ``patchy'' in space, and so is the associated anomalous transport. Parts of the energy exchange between turbulence and particles are reversible and quasi-periodic.
	The remaining irreversible anomalous resistivity  can be parametrized by an effective collision rate ranging from the local ion-cyclotron to the lower-hybrid frequency.
	The contributions to the parallel and the perpendicular (to the magnetic field) components of the slowly varying DC-electric fields, balanced by the turbulence, are similar.
	This anomalous electric field is, however, smaller than the  contributions of the off-diagonal pressure and electron inertia terms of the Ohm's law.
	This result can now be verified by in-situ measurements of the turbulence, in and around the magnetic reconnection regions of the Earth's magnetosphere by the multi-spacecraft mission MMS and in laboratory experiments like MRX and VINETA-II.\\
	\textit{This article may be downloaded for personal use only. Any other use requires prior permission of the author and the American Institute of Physics.\\
	The following article appeared in  P.A. Mu\~noz, J. B\"uchner and P. Kilian, Physics of Plasmas \textbf{24}, 022104 (2017),  and may be found at
	\href{http://dx.doi.org/10.1063/1.4975086}{http://dx.doi.org/10.1063/1.4975086}
	}
\end{abstract}

\maketitle

\section{Introduction}\label{sec:intro}

In the hot and dilute astrophysical plasmas, transport is mainly due to the interaction of the charged particles with their collectively self-generated electromagnetic turbulence. The latter replaces the binary particle collisions responsible for the transport in denser and colder plasmas.
Different from those collision-dominated plasmas, the properties of the collisionless self-generated turbulence depend, however, on the parameters and configuration of the plasma.
``Anomalous'' (or turbulent) transport depends, therefore, also on the plasma parameters and configuration.
Here, we focus on the role of anomalous (turbulent) transport in collisionless guide-field magnetic reconnection.  In space and astrophysical plasmas, reconnection ubiquitously converts magnetic energy into particle acceleration, bulk flows, and heat.
In many environments, e.g., stellar atmospheres, magnetic reconnection develops in external, current-aligned, magnetic fields.
Because of this, we investigate the influence of the strength of the external guide-magnetic-field on the generation of the turbulence in collisionless current sheet (CS) regions prone to magnetic reconnection.
Note that guide field reconnection is now investigated in situ by the MMS space mission\cite{Burch2016} and by laboratory experiments.\cite{Yamada1997,Bohlin2014}.

So far, the properties of collisionless guide-field reconnection were analyzed theoretically and by numerical simulations using a number of different simplifying plasma models (see, e.g., Refs.~\onlinecite{Buchner2007a,Zweibel2009, Yamada2010,Treumann2013b,Gonzalez2016a} and references therein).
A number of micro-instabilities were identified, which can provide, in principle, anomalous transport for collisionless reconnection. Their dependence on the guide field strength, however, has not been completely understood yet.
This is particularly true with respect to regions outside the immediate vicinity of the X-line, near the separatrices, and in the exhaust (outflow) region of reconnection. Also, it appeared to be necessary to include the consequences of the non-linear evolution of the unstable plasma waves beyond the limits of the quasi-linear theory.

To understand these issues, we carried out 2.5D Particle-in-Cell (PiC) code simulations varying the strength of the external (guide-) field. To verify our approach, we compare our results with those obtained in the limiting case of antiparallel reconnection. We self-consistently obtain the non-linear evolution of micro-instabilities to turbulence and investigate its influence on the balancing of macroscopic electric fields in collisionless guide-field reconnection (see Sec.~\ref{sec:ohm}). To further verify our findings, we independently  calculate also the correlations of electromagnetic field and the turbulent plasma fluctuations (see Sec.~\ref{sec:resistivity_fluctuations}) as well as the slowing-down rate of the mean current due to a net momentum transfer from particles to waves (see Sec.~\ref{sec:momentum_exchange}).

Since the 1970s, different mathematical approaches and formalisms have been developed to describe the anomalous transport in collisionless  plasmas. They mainly assumed weak turbulence so that quasi-linear approximations could be applied (see, e.g., Refs.~\onlinecite{Davidson1975,Davidson1977,Davidson1977a, Papadopoulos1977,Galeev1979,Sagdeev1979}, and references therein).
According to these results, collective-collisionless-plasma-transport effects can be described by an effective anomalous resistivity $\eta_{\rm anom}$ relating linearly (or tensorially) electric field and current density as $\vec{E}=\eta_{\rm anom}\;\vec{j}$, providing a term for a resistive MHD-Ohm's law.
Such an anomalous resistivity term  (which includes $\eta_{\rm anom}$) of the MHD-Ohm's law  is meant to describe the consequences of the interaction of the charged particles with the self-generated turbulence, replacing the binary Coulomb collisions which dominate in cold and dense plasmas.\cite{Spitzer1965,Braginskii1965}
Early simulations aiming to obtain self-consistently $\eta_{\rm anom}$  used simplified setups, designed to explicitly consider specific instabilities and to verify predictions of the quasi-linear theory (see, e.g., Refs.~\onlinecite{Boris1970a,Biskamp1971a,Winske1978,Sato1981}). The investigations of more realistic scenarios, including the consideration of the nonlinear development in the course of magnetic reconnection, had to wait until more computational power became available.
Note that micro-turbulence can break the frozen-in condition of ideal MHD in the momentum equation and heat the plasma as collisional Joule heating not only at the X-lines, but also near the separatrices and in the outflow region of reconnection.

CLUSTER\cite{Panov2006,Panov2006a,Lui2007} and more recently the MMS space mission\cite{Torbert2016} provided positive observational evidence for the role of micro-turbulence and anomalous resistivity in reconnection regions.  There is, however, a controversy that some authors claim that the contribution of anomalous resistivity to the reconnection electric field is negligible, interpreting measurements of the Polar,\cite{Bale2002} THEMIS,\cite{Eastwood2009} and also CLUSTER\cite{Mozer2011,Mozer2011a} missions.
Evidence for anomalous resistivity was, however, found in magnetic reconnection experiments such as MRX (see Refs.~\onlinecite{Zweibel2009,Yamada2010} and references therein). Note that these investigations revealed that models of anomalous resistivity do not explain sufficiently well the experimental results.\cite{Dorfman2014} Considering 2.5D configurations we reach similar conclusions.

In this paper, we will apply a  mean-field approach to derive the terms of a generalized Ohm's law.
Let us first  briefly review the mathematical framework for the calculation of the terms of such an equation. In our mean-field approach, we split off the fast fluctuations of the evolution of the slowly varying mean quantities.
In fact, macroscopically observable electromagnetic fields, plasma parameters and variables, are mean quantities slowly varying compared to fast fluctuations, which include microscopic variables.
Hence, let us represent any macroscopically relevant physical quantity $\vec{A}$  by a slowly varying and spatially averaged (over the micro-scales of fast variation) mean value $\langle \vec{A}\rangle$ and a local and instantaneous deviation from it $\delta \vec{A}$
\begin{align}\label{eq:fluctuations}
	\vec{A} =  \langle \vec{A}\rangle  + \delta \vec{A}. 
\end{align}
In Eq.~\eqref{eq:fluctuations}, the brackets  $\langle \rangle $ stand for an appropriate time and spatial averaging, which reveal mean quantities.
The spatial and time scales of the averaging should be chosen in a way that they do not exceed the scales of approximate homogeneity and stationarity of the variation of $\vec{A}$. On the other hand, they should be large and long enough,  to average over the local and fast variations, including microscopic fluctuations, of the turbulence (see Sec.~\ref{sec:convergence_fluctuations}).

For a collisionless plasma, the Vlasov equation describes the evolution of the single particle distribution function $f_{\alpha}(\vec{x},\vec{v},\vec{t})$ for every specie $\alpha = e, i, \dots$ in the self-consistently determined electromagnetic fields $\vec{E}$ and $\vec{B}$.
The Vlasov equation includes the detailed variations including that of the fast turbulent fluctuations. Its right-hand-side (r.h.s.) vanishes if the particle collisions are not explicitly considered. From the Vlasov equation, the following Boltzmann equation can be derived after writing each term in the form given by Eq.~\eqref{eq:fluctuations}:
\begin{align}\label{vlasov_anomalous}
	\nonumber & \left[\frac{\partial}{\partial t}+\vec{v}\cdot\frac{\partial}{\partial \vec{x}}+\frac{q_{\alpha}}{m_{\alpha}}\left(\langle\vec{E}\rangle + \vec{v}\times\langle \vec{B} \rangle \right)\cdot\frac{\partial}{\partial \vec{v}}\right]\langle f_{\alpha}\rangle \\
	\nonumber & =  \left(\dfrac{\partial \langle f_{\alpha}\rangle}{\partial t}\right)_{\rm anom},                                                                                                                                                                            \\
	& =-\frac{q_{\alpha}}{m_{\alpha}}\left\langle \left( \delta \vec{E} + \vec{v} \times  \delta \vec{B}\right)\cdot \frac{\partial \delta f_{\alpha}}{\partial \vec{v}} \right\rangle.
\end{align}
The left hand side (l.h.s.) of Eq.~\eqref{vlasov_anomalous} contains the mean field quantities $\langle f_{\alpha} \rangle$, $\langle \vec{E}\rangle$,  $\langle \vec{B}\rangle$. The r.h.s. of this Boltzmann equation resembles a collision term, describing the higher order correlations between the split-off fluctuations of the electromagnetic fields ($\delta \vec{E}$,  $\delta \vec{B}$) and the distribution function $\delta f_{\alpha}$ (e.g., density $\delta n_{\alpha}$, bulk flow velocity $\delta \vec{V}_{\alpha}$, pressure tensor $\delta P_{\alpha,ij}$, etc.) and its derivatives $\partial \delta f_{\alpha}/\partial v$.

Taking the first order velocity momenta of Eq.~\eqref{vlasov_anomalous} and considering only first-order fluctuations (see the articles Ref.~\onlinecite{Silin2005,Yoon2006}, the reviews Refs.~\onlinecite{Buchner2006b,Buchner2007}, or the textbook Ref.~\onlinecite{Treumann2001a} and references therein), one obtains,  for an electron-ion plasma (with subscripts ``e'' and ``i'', respectively), a generalized two-fluid Ohm's law for the mean field quantities
\begin{align}\label{eq:ohm_2fluid_meanfield}
	\langle E_i\rangle +
	\varepsilon_{ijk} \langle V_{e,j}\rangle  \langle B_k \rangle
	&=
	- \frac{1}{e \langle n_e\rangle }\frac{\partial \langle P_{e,ij}\rangle }{\partial x_j}
	- \frac{m_e}{e}\frac{d\langle V_{e,i}\rangle }{dt} \nonumber \\
	&+ \eta_{ij,{\rm anom}}j_j.
\end{align}
In Eq.~\eqref{eq:ohm_2fluid_meanfield}, $j$ and $k$ are summation indices, $\varepsilon_{ijk} \langle V_{e,j}\rangle  \langle B_k\rangle=(\vec{V}_e\times\vec{B})_i$ is the $i-$component of the macroscopic convective electric field, $1/(e \langle n_e\rangle) \partial \langle P_{e,ij}\rangle/\partial x_j$ is the electron pressure-tensor term, while $(m_e/e)d\langle V_{e,i}\rangle /dt=(m_e/e)\langle V_j\partial V_{e,i}/\partial x_j + \partial  V_{e,i}/\partial t\rangle $ is usually called the ``(electron) inertial term''.
In particular, the ``anomalous resistivity'' term $ \eta_{ij,{\rm anom}}j_j$ can be written as\cite{Silin2005,Yoon2006}
\begin{align}\label{ohm_anomalous}
	E_{\text{\rm anom}, i} & :=-\eta_{ij,{\rm anom}}j_j,                                                                                                                                               \\
	                       & = - \frac{1}{n_e}\int dv^3\left( \langle \delta E_i \delta f_e\rangle +  \langle \varepsilon_{ijk} v_{j} \delta f_e  \delta B_k\rangle   \right), \label{ohm_anomalous2} 
	\\
	                       & = -\frac{1}{\langle n_e \rangle}\left(\langle \delta n_e \delta E_{i} \rangle
	+  \langle \varepsilon_{ijk} \delta (n_e V_{j})  \delta  B_k  \rangle \label{ohm_anomalous3}
	\right).
\end{align}
Note that Eq.~\eqref{ohm_anomalous} corresponds to the assumption that the electric field is linearly (or tensorially) related to the current density. Meanwhile, Eqs.~\eqref{ohm_anomalous2} and ~\eqref{ohm_anomalous3} are directly derived from the averaged Vlasov equation.
In this work, we calculate the r.h.s. term of Eq.~\eqref{ohm_anomalous3} for the numerically simulated quantities by appropriately averaging (see above) over the correlated fluctuations.
While these quantities   were previously determined  mainly by quasi-linear calculations based on estimates of a (weak) turbulence saturation in idealized scenarios (see the review Ref.~\onlinecite{Galeev1984, Treumann2001}, the article Ref.~\onlinecite{Yoon2006}, or the textbooks Ref.~\onlinecite[(Sec.~12.1)]{Treumann2001a} or Ref.~\onlinecite[(Sec.~7.1.4)]{Biskamp2000}, and references therein), we obtain them directly and self-consistently.

\section{Simulation setup}\label{sec:setup}

The setup of our simulations is described in Ref.~\onlinecite{Munoz2016a}. We use the PiC-code ACRONYM.\cite{Kilian2012} Since we aim at the investigation of quasi-2.5-dimensional reconnection, we neglect variations along the $z$ direction. Two CSs are initialized forming a double Harris-sheet equilibrium.\cite{Harris1962} The external guide-field strength $b_g$ along the  $z$ direction (perpendicular to the $x$--$y$ reconnection plane), is varied in a range from $b_g=0$ (for the antiparallel reconnection limit) and $b_g=8$ to simulate scenarios like stellar coronae and laboratory experiments.
Here, we express the guide field strength normalized to the amplitude of the antiparallel component (in the  $y$ direction) of the Harris-CS $B_{\infty y}$ (i.e., $b_g=B_z/B_{\infty y}$).

The CS parameters were halfwidth $L/d_i=0.5$, mass ratio $m_i/m_e=100$, frequency ratio $\omega_{pe}/\Omega_{ce}=4.16$,  temperature ratio of $T_i/T_e=1.0$ and a background plasma density $n_b/n_0=0.2$. These parameters give an electron thermal speed of $v_{th,e}/c=\sqrt{k_B T_e/m_e}/c=0.12$, where $c$ is the speed of light.
Other physical parameters  are defined as follows: $n_0=n_e=n_i$ is the electron/ion plasma density of the current-carrying population at the center of the CS, $d_{i/e}=c/\omega_{pi/pe}$ is the ion/electron skin depth, $\omega_{pi/pe}$ the ion/electron plasma frequency calculated with the density $n_0$, and $\Omega_{ce}$ is the electron plasma frequency in the Harris CS magnetic field amplitude $B_{\infty y}$.

The numerical parameters of our simulations were: $250$ and $50$ particles per cell for the current-carrying and background population, respectively (for both electron and ion population); a simulation box size of $L_x\times L_y =(20.94\,d_i\times 12.56\,d_i)$ in the $x$--$y$ plane,  the grid spans $2500\times 1500$ grid points, periodic boundary conditions  apply in the $x$ and $y$ directions, and the spatial (grid) resolution was $\Delta x=0.7\lambda_{De}$, where $\lambda_{De}$ is the electron Debye length in the center of the CS (calculated with $n_0$). The timestep is chosen to satisfy the Courant-Friedrichs-Lewy (CFL) condition for light wave propagation at a level of $c\Delta t/\Delta x=0.5$.

Reconnection is initialized by a small-amplitude long-wavelength perturbation to quickly reach a fully developed stage with X-lines at $x=\pm L_x/4$ and $y=L_y/2$ for every CS.  Note that we illustrate our analysis in this paper using the results obtained for the CS centered around $x=-L_x/4$. We define a reference electric field as  $E_0 = B_{\infty y}V_A/c$ used for the normalizations, with $V_A$ the Alfv\'en speed calculated with the central density $n_0$ and the asymptotic in-plane magnetic field strength $B_{\infty y}$.

\section{Results}\label{sec:results}

\subsection{Balance of Ohm's law terms}\label{sec:ohm}

We calculate the mean  quantities appearing in the generalized Ohm's law (Eq.~\eqref{eq:ohm_2fluid_meanfield}) by means of a (running) time average with a time windows of $\Delta T=0.25\Omega_{ci}^{-1}=6.3\Omega_{LH}^{-1}$, where $\Omega_{LH}=\omega_{pi}/\sqrt{1+\omega_{pe}^2/\Omega_{ce}^2}$ is the lower hybrid  frequency calculated in the total magnetic field. This filtering damps out fluctuations with frequency significantly higher than the inverse of the time windows length. The windows length is chosen to cover all relevant plasma frequencies, while numerically caused fluctuations are removed. More consequences of this choice are discussed in Sec.~\ref{sec:convergence_fluctuations}.

\begin{figure*}[!ht]
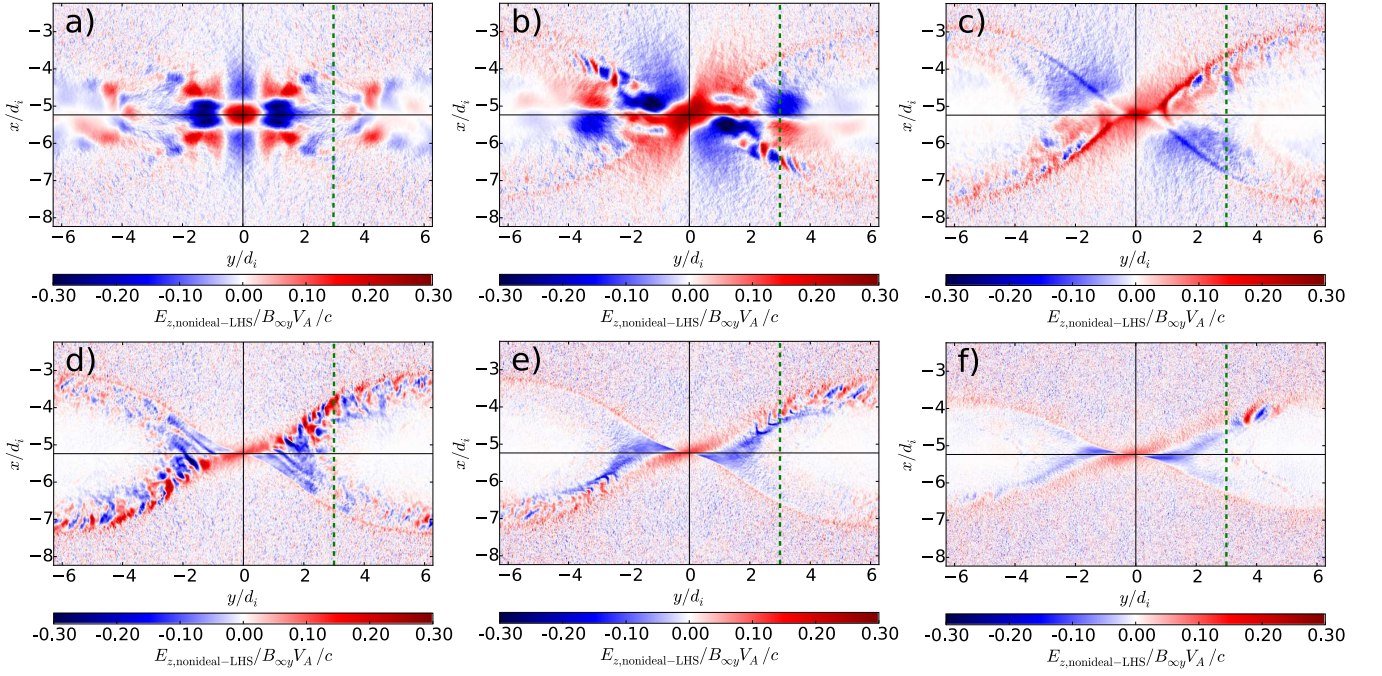

	\centering
	\includegraphics[width=1.0\linewidth]{{{./ohm_lhs_bgs}}}
	\caption{Color-coded contour plot of the of the $z-$component of the non ideal electric field $\langle\vec{E}\rangle + \langle\vec{V}_e\rangle\times\langle\vec{B}\rangle$ in the time averaged generalized Ohm's law (Eq.~\eqref{eq:ohm_2fluid_meanfield}) for different guide fields and times: a) $b_g=0$ at $t=12\Omega_{ci}^{-1}$, b) $b_g=0.26$ at $t=12\Omega_{ci}^{-1}$, c) $b_g=1.0$ at $t=14\Omega_{ci}^{-1}$ d) $b_g=3.0$ at $t=18\Omega_{ci}^{-1}$ e) $b_g=5.0$  at $t=20\Omega_{ci}^{-1}$ f) $b_g=8.0$ at $t=20\Omega_{ci}^{-1}$. The time average is over a window length of  $\Delta T=0.25\Omega_{ci}^{-1}=6.3\Omega_{LH}^{-1}$. \label{fig:ohm_lhs}}
\end{figure*}

First, in Fig.~\ref{fig:ohm_lhs}, we show the guide field dependence of the spatial distribution of the (out-of-plane) $z-$component (time averaged) DC electric fields,  $\langle\vec{E}\rangle + \langle\vec{V}_e\rangle\times\langle\vec{B}\rangle$, in the left hand side (l.h.s.) of the  (time averaged) generalized Ohm's law (Eq.~\eqref{eq:ohm_2fluid_meanfield}). This sum describes the mean non-ideal electric field. It can, therefore, be used as signature of the non-ideal processes  violating the frozen-in condition of the electron fluid to the plasma. Note that we use different times in each case to compensate for the delay in the reconnection saturation time for stronger guide-fields (see, e.g., Refs.~\onlinecite{Horiuchi1997,Pritchett2005,Ricci2004,Huba2005}). The plots are shown when the magnetic islands reach comparable sizes. In all these contours plots, we use, as a post-processing step, a spatial (low-pass) Gaussian filter with a small width $\Delta x=0.3d_e$ to get rid of the PiC shot noise with small wavelengths.

Fig.~\ref{fig:ohm_lhs}(a) shows that in antiparallel reconnection and for very small guide fields, the non-ideal terms are located mostly in the diffusion region near the X-line of reconnection. There are also structures visible in the exhaust region away from the X-line, whose magnitude is, however, smaller than the electric field at the X-line. These structures are due to an instability driven by temperature anisotropy and are explained in detail in the Appendix~\ref{sec:app}, while  we focus here on the strong guide-field cases.

Previous findings have shown that the non-ideal contributions due to the off-diagonal terms of the electron pressure tensor or the electron inertia terms can balance the reconnection electric field $E_z$ in the neighbourhood of the X-line (see a review in, e.g., Ref.~\onlinecite{Treumann2013b}).
However, for guide fields of the order of $b_g=1$ (Fig.~\ref{fig:ohm_lhs}(c)), the deviations from non-ideal behavior shift to the separatrices, in particular, the low-density one, with a ``patchy'' spatial distribution.  Therefore, dissipation and violation of the frozen-in condition can also occur away from the traditional diffusion region near the X-line, as discussed in some previous works\cite{Le2013}.
Note that the low-density separatrix is the region around the antisymmetrically located upper right and lower left separatrix arms in our setup, characterized by a diminished density in comparison with the other separatrix, as typically seen in guide-field reconnection.\cite{Kleva1995,Rogers2003} For larger guide fields of the order of $b_g=3$ (Fig.~\ref{fig:ohm_lhs}(d)), the non-ideal contribution close to the X-line is much less important and spatially smaller than in the low density separatrix.  Cases with guide field $b_g\gtrsim 5$, such as $b_g=8$  (Fig.~\ref{fig:ohm_lhs}(f)), show a diminished non-ideal activity even in the separatrices. The reason is because the (streaming) instability producing the non-ideal behavior is active only in the strong guide field regime $1.5\lesssim b_g\lesssim 6$. The upper limit seems to be a result of the reduced 2.5D geometry used in the simulation, as discussed in our previous work.\cite{Munoz2016a}

\begin{figure*}[!ht]
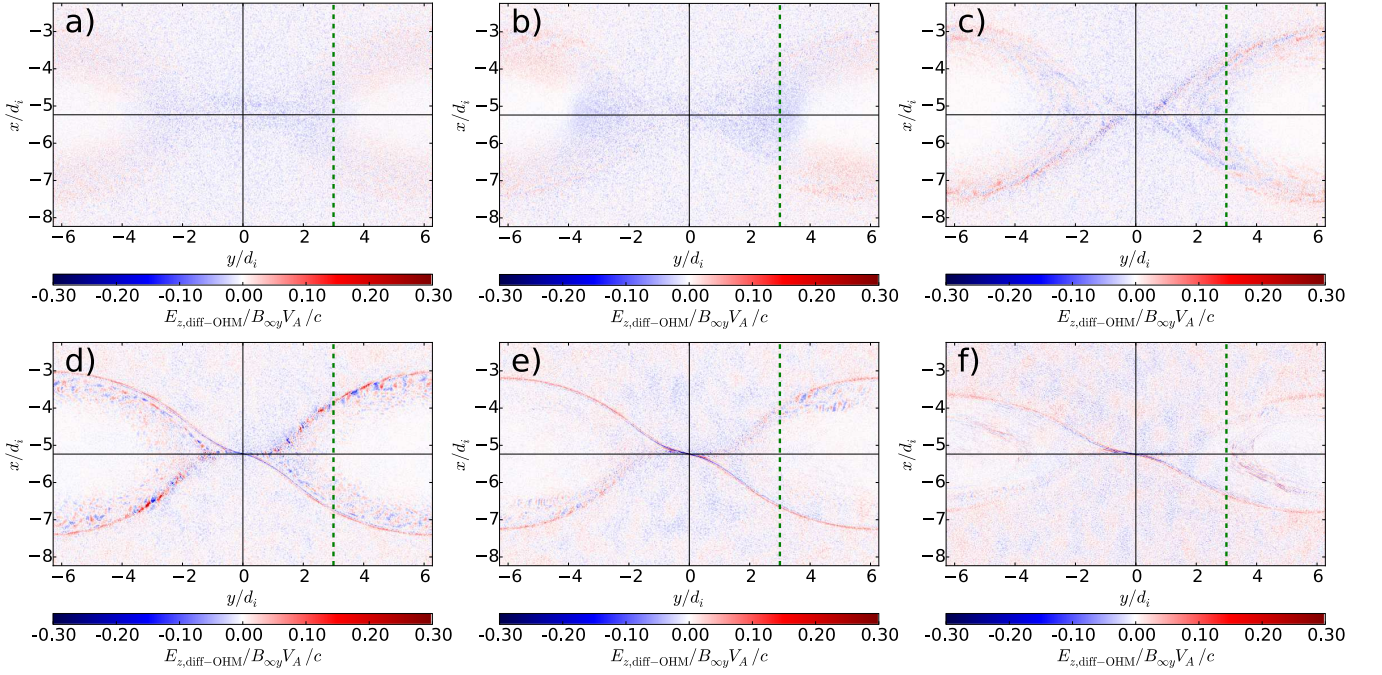

	\centering
	\includegraphics[width=1.0\linewidth]{{{./ohm_diff_bgs}}}
	\caption{Contour plot of the remaining term  (l.h.s. - r.h.s., excepting the anomalous term) of the $z-$component of the mean generalized Ohm's law (Eq.~\eqref{eq:ohm_2fluid_meanfield}) for different guide fields. a) $b_g=0$, b) $b_g=0.26$, c) $b_g=1.0$, d) $b_g=3.0$, e) $b_g=5.0$, f) $b_g=8.0$. The dashed line represents the $x$-cut used for the profile shown in Fig.~\ref{fig:ohm_diff_profile}. Same times and time average used as in Fig.~\ref{fig:ohm_lhs}. \label{fig:ohm_diff}}
\end{figure*}

Fig.~\ref{fig:ohm_diff} displays the difference between the l.h.s. and r.h.s. of the $z-$component of the mean generalized Ohm's law Eq.~\eqref{eq:ohm_2fluid_meanfield}, excepting the a priori unknown anomalous electric field term. Any deviation  can be attributed to non-ideal effects beyond that caused by the pressure or inertia term contributions to the electric field and are thus equal to $E_{z,{\rm anom}}=-\eta_{z,{\rm anom}}j_z$ in  Eq.~\eqref{eq:ohm_2fluid_meanfield}. For the antiparallel $b_g=0$ (Fig.~\ref{fig:ohm_diff}(a)) or small guide fields $b_g=0.26$ cases (Fig.~\ref{fig:ohm_diff}(b)), the deviations are small in comparison to the non-ideal electric field $E_z$ and can be attributed to the PiC noise. This is justified by comparing with the values of this difference at points in the inflow region, away from the CS. However, for stronger guide fields such as $b_g=3$  (Fig.~\ref{fig:ohm_diff}(d)), the deviations start to form localized and ``patchy'' structures near the separatrices. They have smaller spatial extent than the structures seen for the l.h.s. of the Ohm's law in Fig.~\ref{fig:ohm_lhs}(d). This is also in agreement with previous works showing the relative contributions of the instantaneous Ohm's law in the separatrices of guide-field reconnection.\cite{Pritchett2013a,Wendel2016}
All these observations prove that the non-ideal behavior seen in the latter can be explained mostly by the two-fluid effects of pressure and inertia contributions to the electric field (plots not shown here), being violated only very locally.

\begin{figure*}[!ht]
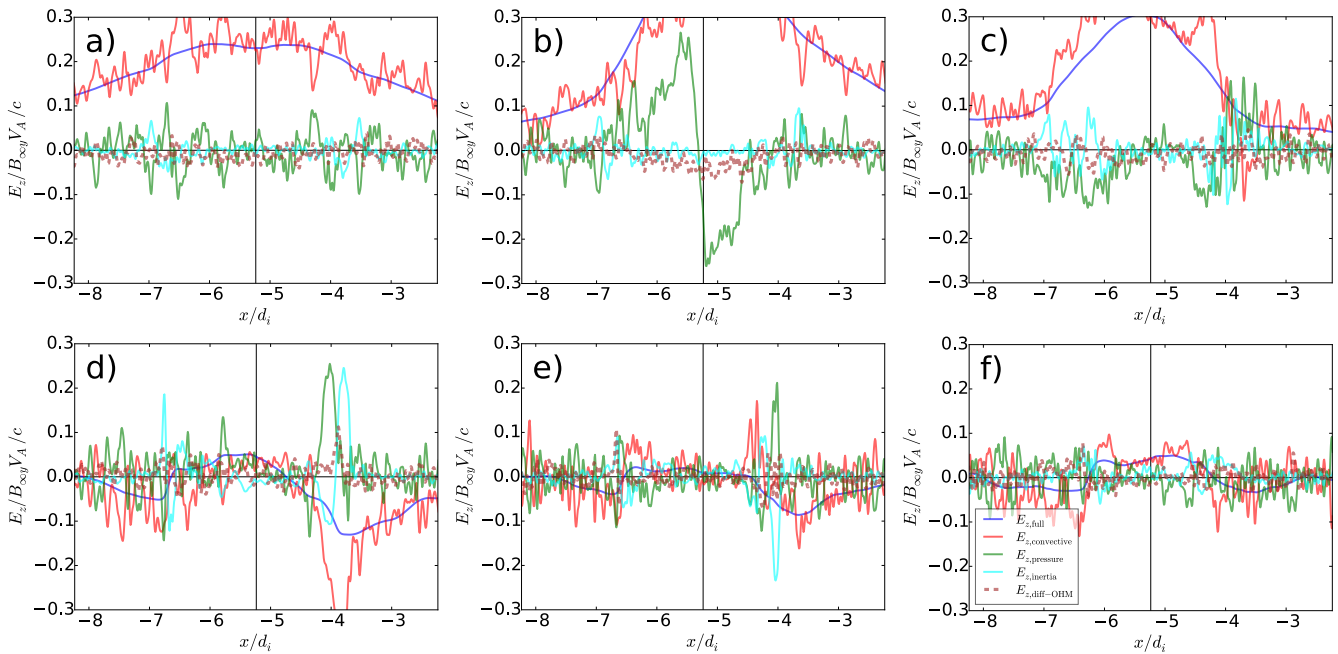

	\centering
	\includegraphics[width=1.0\linewidth]{{{./ohm_balance_bgs_new}}}
	\caption{$x$-cut of the $z-$component terms of the mean generalized Ohm's law for different guide fields. a) $b_g=0$, b) $b_g=0.26$, c) $b_g=1.0$, d) $b_g=3.0$, e) $b_g=5.0$, f) $b_g=8.0$. The $x$-cut shown was taken at $y=3.0d_i$ (shown in Fig.~\ref{fig:ohm_diff}). The black vertical line represents the CS midplane. The low-density separatrix is located close to $x\sim -4d_i$ and the high density separatrix close to $x\sim-7d_i$. \label{fig:ohm_diff_profile}}
\end{figure*}
In Fig.~\ref{fig:ohm_diff_profile}, we complement the information of Fig.~\ref{fig:ohm_diff} by analyzing the individual contributions to the balance of the mean Ohm's law terms for an $x$-cut at the separatrices. We have applied an additional Savitzky-Golay spatial  filter with a small width to smooth out those profiles, without affecting significantly the values of the local maxima/minima. In the past, many works have investigated the relative contributions of the generalized Ohm's law terms to the electric field in guide field reconnection near the X-line, in order to find the mechanism leading to the breaking of the frozen-in condition (see, e.g., Refs.~\onlinecite{Ricci2004,Pritchett2005,Pritchett2009,Che2011}). Note, however, that most of those works used the instantaneous instead of the time averaged version of the generalized Ohm's law used here, not allowing for additional terms coming from higher order fluctuations.

For guide fields $b_g\lesssim 1$ (Fig.~\ref{fig:ohm_diff_profile}(a)-(c)), most of the out-of-plane $E_z$ is balanced by the convective electric field, with small contributions from pressure or inertia terms. Note that even small guide fields ($b_g=0.26$, Fig.~\ref{fig:ohm_diff_profile}(b)) introduce an asymmetry with respect the CS mid-plane in all the electric field terms, with a significant contribution from the pressure term.
But it is only with guide fields stronger than unity (e.g., $b_g=3$, Fig.~\ref{fig:ohm_diff_profile}(d)) that the inertia term becomes relevant, taking similar peak values to both pressure and convective electric fields, well above the noise level. This is especially noticeably in the region close to the low density separatrix (at $x\sim -4.0 d_i$).  Associated with this fact are the largest deviations in the balance of the mean field generalized Ohm's law (on the order of $E_z/E_0\sim 0.1$). This implies that the anomalous term in Eq.~\eqref{eq:ohm_2fluid_meanfield} reaches the strongest values among all the cases analyzed here, with a corresponding largest anomalous resistivity $\eta_{z,{\rm anom}}$.

In this case of $b_g=3$, we verified that among the three contributions to the inertia term, i.e.,  $(m_e/e)\langle V_{e,x}\partial V_{e,z}/\partial x + V_{e,y}\partial V_{e,z}/\partial y +\partial  V_{e,z}/\partial t\rangle$, the most important is the (Eulerian) time derivative (third term). This indicates that the out-of-plane electron current $V_{e,z}$ is changing quickly, which is the signature of an efficient momentum transfer between electron and ions.

\subsection{Momentum exchange from the slowing-down rate of mean currents}\label{sec:momentum_exchange}

Indeed, by multiplying the mean field Vlasov equation~\eqref{vlasov_anomalous} by $m_{\alpha}v$ and integrating in the velocity space, it is possible to relate directly the slowing-down rate of the (mean) electron current density, due to a net momentum transfer from particles to waves, with an effective collision frequency $\nu_{j,\alpha}$ for each specie $\alpha$ along the direction $j$\cite{Davidson1975,Silin2005}
\begin{equation}\label{nu_anom_ddt}
	\nu_{j,\alpha} = -\frac{1}{\langle n_{\alpha}m_{\alpha}V_{\alpha,j}\rangle}\left(\frac{\partial \langle n_{\alpha}m_{\alpha}V_{\alpha,j}\rangle}{\partial t}\right).
\end{equation}
As a result, the effective resistivity along the $j$-direction is\cite{Moritaka2007}
\begin{equation}\label{eta_anom_ddt}
	\eta_{j,anom}= \frac{m_e\langle V_{e,j}\rangle \nu_{j,e} - m_i \langle V_{i,j}\rangle \nu_{j,i} }{\langle n_e\rangle e^2 (\langle V_{i,j}\rangle - \langle V_{e,j}\rangle )}.
\end{equation}
Note that, in general, a quantity $-{\partial p/\partial t}/p$, with $p$ a momentum, can be interpreted as the inverse of a slow-down time associated with, e.g., collisions, especially useful in the dicussion of collisions operators for the Boltzman equation (see, e.g., Sec.~13.2 in Ref.~\onlinecite{Bellan2006}). Expressions of the form of Eq.~\eqref{nu_anom_ddt} represent the rate of momentum transfer  between particles and, e.g., waves.
In general, this quantity can be either positive or negative.
It corresponds to an anomalous resistivity if there is a net loss of momentum of current-carrying particles so that the bulk motion of  current-carrying particles slows down (positive sign in  Eq.~\eqref{nu_anom_ddt}).
An effective anomalous resistivity implies a reduction of the current, the usual source of free energy for the instabilities generating waves and turbulence.
The ``slowing down rate of electron current" is, therefore, an appropiate expression to describe the process associated with the effective collisions (collision frequency defined in Eq.~\eqref{nu_anom_ddt}).

\begin{figure*}[!ht]
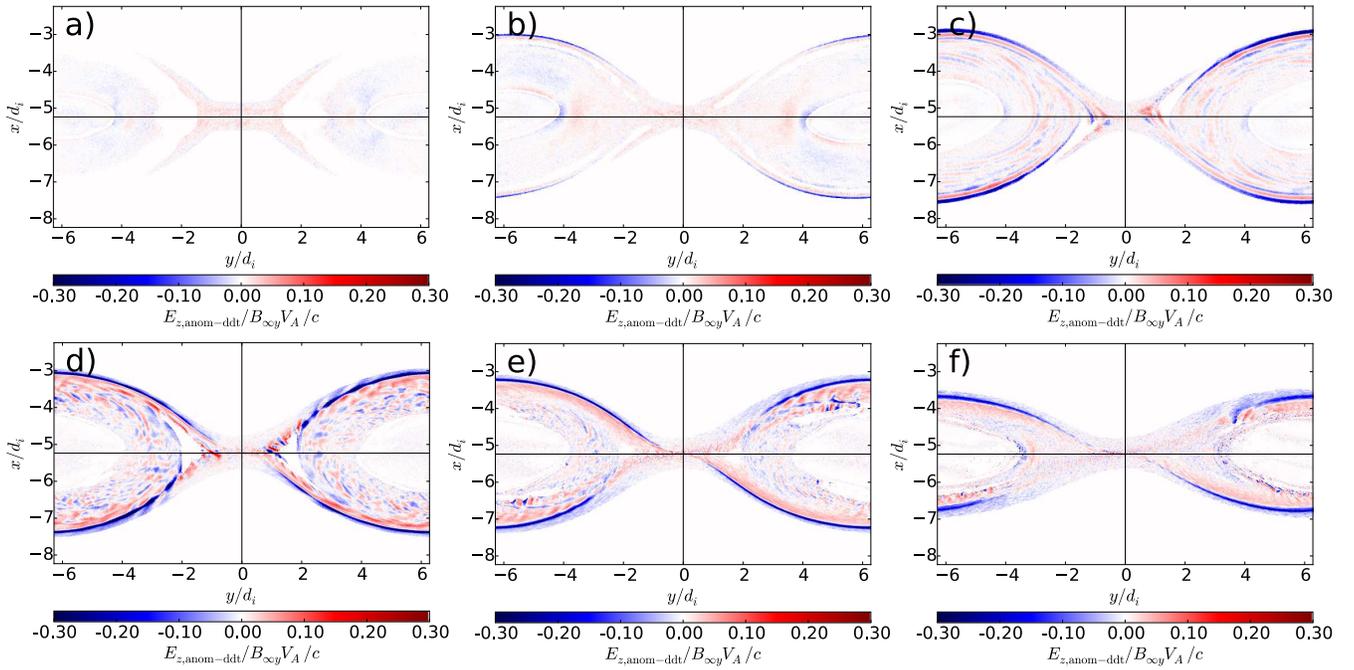

	\centering
	\includegraphics[width=0.99\textwidth]{{{./ez_anom_ddt_bgs}}}
	\caption{Color-coded distribution of the electric field corresponding to the slowing-down rate  of the electron current, due to a net momentum transfer from particles to waves, $E_{z,{\rm anom-ddt}}$ given by Eq.~\eqref{e_anom_ddt}. The different panels represent the guide field cases: a) $b_g=0$, b) $b_g=0.26$, c) $b_g=1.0$, d) $b_g=3.0$, e) $b_g=5.0$, and f) $b_g=8.0$. Same times and time average used as in Fig.~\ref{fig:ohm_lhs}. \label{fig:eanom_ddt_bgs}}
\end{figure*}

This approach has been used, e.g., by Refs.~\onlinecite{Watt2002,Petkaki2003,Hellinger2004,Petkaki2006,Buchner2005a,Buchner2006,Petkaki2008,Wu2010a} for the analysis of Vlasov code simulations of current-driven instabilities, in  Refs.~\onlinecite{Silin2005} for 3D Vlasov simulations of CS, and in Refs.~\onlinecite{Moritaka2007} for 2D PiC simulations of CS. As a comparison of the terminology used in those works, an expression equivalent to Eq.~\eqref{nu_anom_ddt} was used in Refs.~\onlinecite{Watt2002,Petkaki2003} called it ``rate of change of electron momentum'', in Ref.~\onlinecite{Silin2005} ``rate of momentum exchange'', in Ref.~\onlinecite{Buchner2006} ``anomalous momentum transfer rate'',
in Ref.~\onlinecite{Moritaka2007} ``damping rate of electron current density'' and in  Ref.~\onlinecite{Hellinger2004} ``slowing down of electrons''.
Because the contribution from the ion quantities is much smaller than the electron ones,\cite{Silin2005} the $j-$component of the anomalous electric field can be approximated as,
\begin{align}\label{e_anom_ddt}
	E_{j,{\rm anom-ddt}}
	&=\eta_{j,{\rm anom}}j_{j}\approx \frac{m_e \nu_{j,e}}{\langle n_e\rangle e^2 }  j_{j} \nonumber \\
	& = \frac{m_e}{\langle n_e\rangle e^2 }\left[-\frac{1}{\langle n_{e}V_{e,j}\rangle}\left(\frac{\partial \langle n_{e} V_{e,j}\rangle}{\partial t}\right) \right] j_{j}.
\end{align}
Then, $E_{j,{\rm anom-ddt}}$ is related to the time derivative of the electron current density and thus to the (Eulerian) partial time derivative term $\partial  \langle V_{e,j}\rangle /\partial t$ of the  inertia term of the mean generalized Ohm's law (Eq.~\eqref{eq:ohm_2fluid_meanfield}). The difference is the contribution  coming from the electron density $n_e$ and its time derivative, not appearing in the inertia term. The spatial distribution of  $E_{z,{\rm anom-ddt}}$  can be seen in  Fig.~\ref{fig:eanom_ddt_bgs}. We avoid the calculation in regions where the mean electron current density $\langle n_{e} V_{e,j}\rangle$ is very close to zero by assigning a zero value to the total quotient, in order to avoid unphysical diverging values of the electric field due to numerical noise. Figs.~\ref{fig:eanom_ddt_bgs}(a)-(b) show small contributions to the mean $\langle E_z\rangle$ in the separatrix region for small guide fields. For these cases, $E_{z,{\rm anom-ddt}}$ is mostly concentrated around the X-line. Larger guide fields ($b_g=1$, Fig.~\ref{fig:eanom_ddt_bgs}(c)) localize the contributions of this term to a very small area close to the X-line, with a slight increase at the separatrices. Again, only for $b_g\gtrsim 3$ (Fig.~\ref{fig:eanom_ddt_bgs}(d)), the contribution of this term mostly correlates with the locations at the separatrices where the balance of the mean field Ohm's law is violated (compare with Fig.~\ref{fig:ohm_diff}(d)). Note, however, that the spatial distribution of $E_{z,{\rm anom-ddt}}$ is more spread and with higher values than $E_{z,{\rm anom-diff}}$ given by the difference in the mean field Ohm's law. Therefore, it cannot account for all the deviations in the mean Ohm's law. But it proves the existence, instead, of an effective transfer of momentum between electron and ions with a ``patchy'' spatial structure (alternating sign). Stronger guide fields ($b_g=5-8$, Figs.~\ref{fig:eanom_ddt_bgs}(e)-(f)), reduce both values and ``patchiness'' of $E_{z,{\rm anom-ddt}}$. This correlates with the behavior shown by $E_{z,{\rm anom-diff}}$ in  Figs.~\ref{fig:ohm_diff}(e)-(f), due to the weakening of the instabilities producing the fluctuations.

In the remaining part of this paper, we are going to characterize in more detail the physics of the processes leading to the deviations in the mean field Ohm's law for the case $b_g=3$ at the separatrices.
\subsection{Perpendicular momentum exchange from the slowing-down rate of mean currents for $b_g=3$}\label{sec:momentum_exchange_bg3}

\begin{figure*}[!ht]
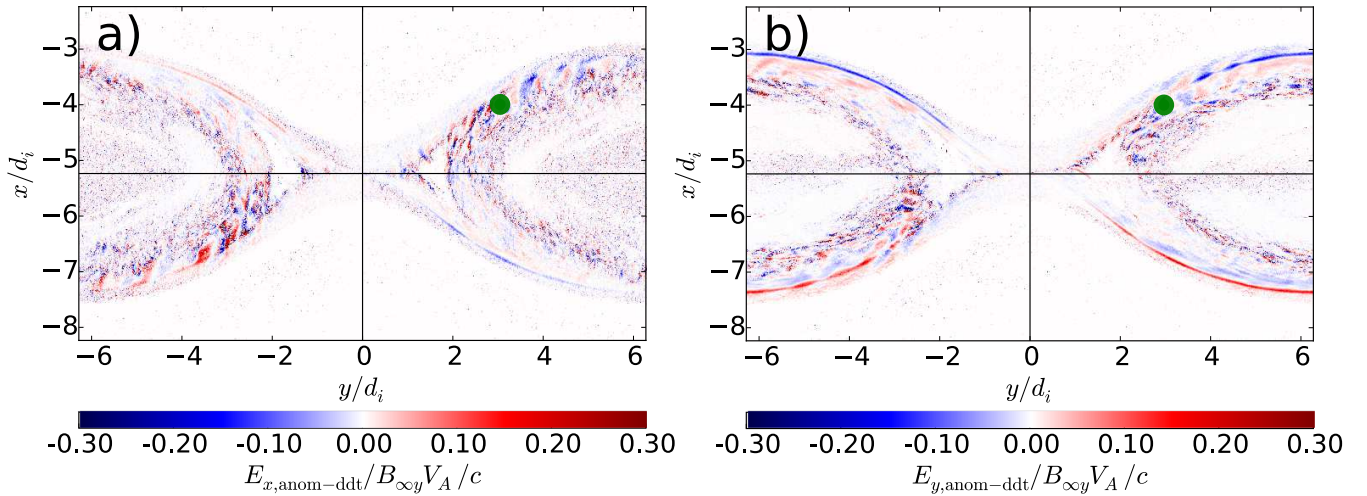
\centering
	\includegraphics[width=1.0\textwidth]{{{./exy_anom_ddt_bg3_wp}}}
	\caption{Color-coded distribution of the electric field corresponding to the slowing-down rate  of the electron current, due to a net momentum transfer from particles to waves, given by Eq.~\eqref{e_anom_ddt}. a) $E_{x,{\rm anom-ddt}}$ and b) $E_{y,{\rm anom-ddt}}$. Compare with the $z-$component in  Fig.~\ref{fig:eanom_ddt_bgs}(d). The large filled green circle indicates the location used for the time series in the next plots.  \label{fig:einertia_bg3_components}}
\end{figure*}

The enhanced wave-particle interactions in the separatrices for the case $b_g=3$ do not only lead to an  alternating effective resistivity in the $z-$direction between $j_z$ and $E_z$. Indeed, the relation between the $x$ and $y$ components of the current density and electric field also behaves in a very similar manner and with similar values as well. However, the deviations from the mean generalized Ohm's law do not provide a clear proof of this, possibly because the typical values of $E_x$ and $E_y$ are about 20 times larger than $E_z$. But the time derivative of the electron current density in these (in-plane) $x-$ and $y-$ directions provides an effective resistivity of the same order of magnitude as this for the $z-$component. This effective resistivity has a spatial distribution concentrated near the separatrices, while the noise is more dominant in the outflow region.  This can be seen in the contour plots of  $E_{x,{\rm anom-ddt}}$ and  $E_{y,{\rm anom-ddt}}$ shown in Fig.~\ref{fig:einertia_bg3_components} and comparing with  $E_{z,{\rm anom-ddt}}$ in Fig.~\ref{fig:eanom_ddt_bgs}(d). Note that both $x-$ and $y-$ components are noisier than the $z-$component because $\langle n_eV_{e,x}\rangle$ and $\langle n_eV_{e,y}\rangle$, appearing in the denominator of Eq.~\eqref{e_anom_ddt}, are closer to zero in larger regions inside the CS than  $\langle n_eV_{e,z}\rangle$. In summary, the similar behavior of all the components of the anomalous electric field implies that the momentum exchange between electrons and ions does not have a clear preferential direction as one would expect in a guide field situation but rather tends to be isotropic near the separatrices.

\begin{figure*}[!ht]
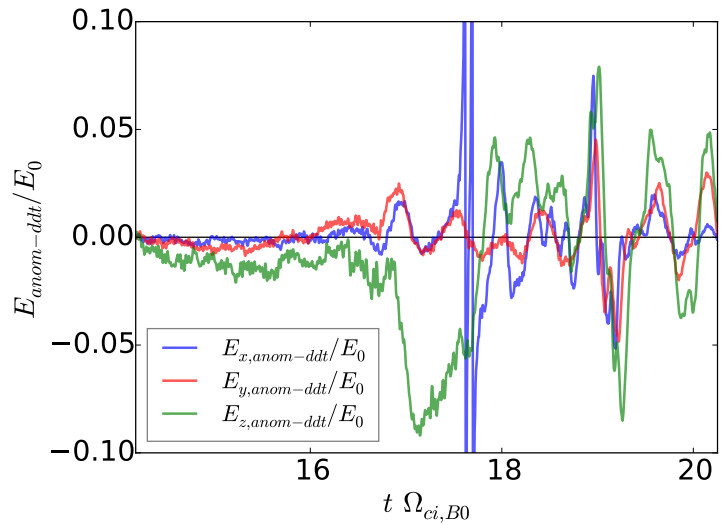
\centering
	\includegraphics[width=0.55\textwidth]{{{./e_anom_ddt_timeseries_new}}}
	\caption{Time series of the three spatial components of the electric field $ E_{j,{\rm anom-ddt}}=\eta_{j,{\rm anom-ddt}} j_j$ due to the slowing-down rate of the electron current density due to a net momentum transfer from particles to waves, for the case $b_g=3$. The tracked point in the low density separatrix is located at $x=-4d_i$, $y=3d_i$, indicated in Fig.~\ref{fig:einertia_bg3_components}.\label{fig:edamping_timeseries_bg3}}
\end{figure*}

The locations of preferential momentum transfer at the separatrices for the case $b_g=3$ do not  only alternate sign in space but also do in time. This can be easily seen by tracking the values of $\vec{E}_{{\rm anom-ddt}}$ at a point in the low density separatrix located at $x=-4d_i$, $y=3d_i$. The results for the three components of this anomalous electric field are shown in Fig.~\ref{fig:edamping_timeseries_bg3}. The values start to significantly increase after $t\gtrsim16\Omega_{ci}^{-1}$, with average peaks of about $5\times10^{-2}E_0$. This corresponds to an equivalent collision frequency $\nu_{j,e}$ (given by Eq.~\ref{nu_anom_ddt}) with values around $1/5$ of the lower hybrid frequency $\Omega_{LH}$ (plots not shown here). Note, however, that due to the quasi-periodic nature of these fluctuations, there is no net dissipation that can be associated with this anomalous electric field.

\begin{figure*}[!ht]
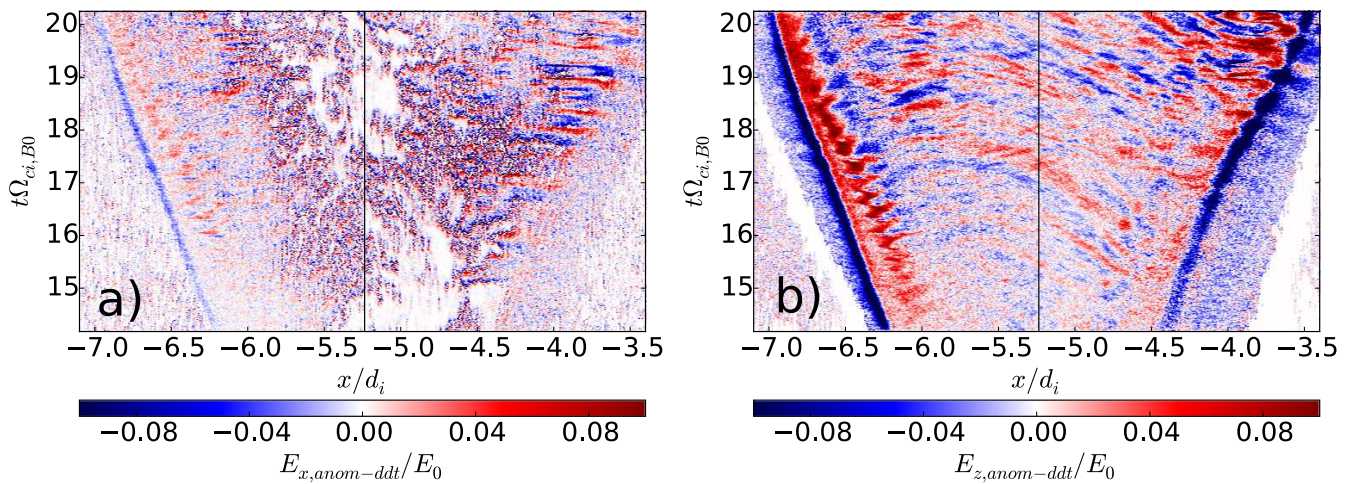
\centering
	\includegraphics[width=1.00\textwidth]{{{./e_anom_ddt_stack}}}
	\caption{Stack plot showing the time evolution of an $x$-cut of components of the anomalous electric field at $y=3d_i$, for the case $b_g=3$. a) in-plane $E_{x,{\rm anom-ddt}}=\eta_{x,{\rm anom-ddt}} j_x$. b) Out-of-plane $E_{z,{\rm anom-ddt}}=\eta_{z,{\rm anom-ddt}} j_z$ . Same time average used as in previous plots. We only show a zoomed region containing the CS with both separatrices. \label{fig:e_anom_ddt_stack}}
\end{figure*}

Fig.~\ref{fig:e_anom_ddt_stack} shows the time evolution of an $x$-cut of this anomalous electric field across the CS midplane during fully developed reconnection. The fluctuations of $\vec{E}_{{\rm anom-ddt}}$ move along the separatrices, outwards from the X-line, associated with the motion of the dominant electron beam producing the streaming instability that causes all these processes. The low density separatrix is located at $x=-4.5d_i$ for $t=14\Omega_{ci}^{-1}$ and reaches $x=-3.5d_i$ for $t=20\Omega_{ci}^{-1}$, due to the growth of the magnetic island. Note that a long enough time average following a point near the low density separatrix (with positive values of, e.g., $E_{z,{\rm anom-ddt}}$) would have a net positive value, implying a net effective dissipation.

In (our) previous work,\cite{Munoz2016a} we provided evidence that the instability leading to the development of electric field fluctuations at the separatrices (and also in the exhaust region) is a streaming instability due to counterstreaming electron beams generated only for a regime of guide fields  $1.5\lesssim b_g\lesssim 6$. Indeed, the entire region where this instability takes place shows significant deviations from gyrotropy. Ref.~\onlinecite{Munoz2016a} showed that the typical fluctuation frequency is broadband up to $\Omega_{LH}$, as expected from this kind of instability. Thus, fluctuations around the lower hybrid frequency can be characterized as an effective collision frequency of the same order of magnitude, even though they arise from very different physical processes. The fact that the effective collision frequency increases with the lower hybrid frequency was also shown observationally, experimentally, and numerically (see Ref.~\onlinecite{Silin2005} and references therein).

We also found that in the same locations where the peak values of the anomalous resistivity (either positive or negative) are maximal/minimal, the traditional dissipation $\vec{j}\cdot\vec{E}$ and the dissipation in the electron frame of reference of Ref.~\onlinecite{Zenitani2011}  $D_e=\vec{j}'\cdot\vec{E}$ are also maximal/minimal (plots not shown here). This further validates that there is an effective exchange of energy between the electromagnetic fields and the particles, a signature of local wave-particle interactions. Note, however, that there is no overall dissipation as it would be with only positive values of anomalous resistivity and so no net collisionless transport during the whole duration of the reconnection process.
\subsection{Anomalous resistivity due to fluctuations}\label{sec:resistivity_fluctuations}
Note that the two approaches mentioned before for the calculation of the anomalous resistivity ---the remaining term of the mean field generalized Ohm's law and the slowing-down rate of the electron current due to a net momentum transfer from particles to waves--- involve only mean quantities. Now, we show the information contained in the so far neglected part: the fluctuations  (given by the r.h.s. of Eq.~\eqref{ohm_anomalous3}) causing the anomalous resistivity in the separatrices for the case $b_g=3$.
Previous works have quantified mostly spatial fluctuations, in 2D\cite{Tanaka1981,Moritaka2007} and also 3D PiC simulations of CS.\cite{Drake2003, Che2011,Roytershteyn2012,Fujimoto2012,Pritchett2013a,Liu2013,Che2014} These works used a spatial average in the equilibrium current direction ($\hat{z}$), where turbulence due to micro-instabilities was assumed to be homogeneous. Instead, especially considering that a CS is inhomogeneous, we calculate these fluctuations as the RMS fluctuation values, defined as the standard deviation of each point in the corresponding time series, with the mean taken as the running average mentioned before.  The time average can be equivalent to the spatial one if the turbulence is both stationary and homogeneous (and both equivalent to an ensemble average if the processes are ergodic\cite{Tsinober2014}). Note that this time averaging approach is also practically universally used in spacecraft data analysis since it is often the only measurement available.

\begin{figure*}[!ht]
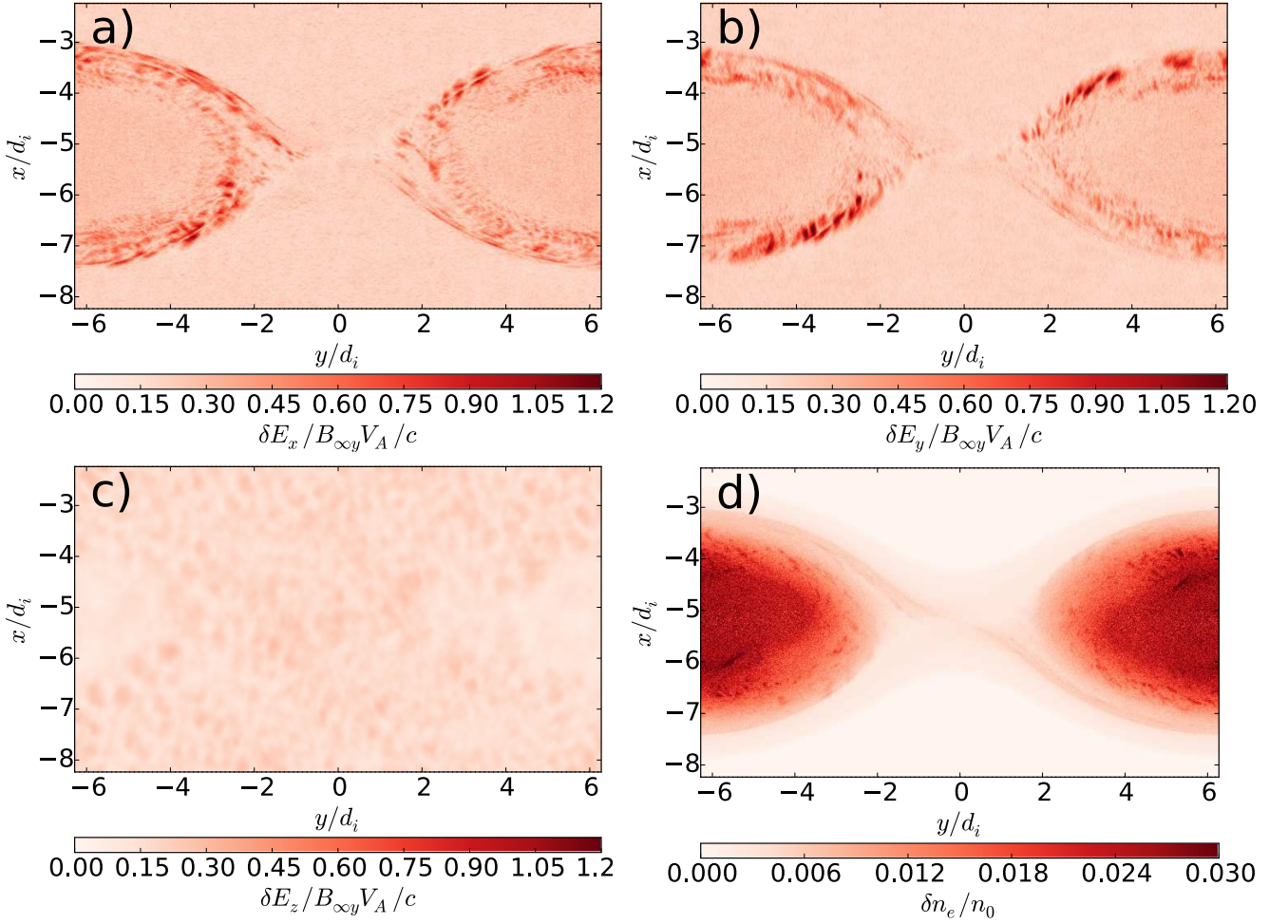
\centering
	\includegraphics[width=1.0\linewidth]{{{./fluctuations_rms_bg3}}}
	\caption{Color-coded contour plot of the RMS fluctuations for the components of $\vec{E}$ and $n_e$ for the case $b_g=3$ at $t=18\Omega_{ci}^{-1}$. We used the same time average as before, $\Delta T=0.25\Omega_{ci}^{-1}$. a) $\delta E_x$, b) $\delta E_y$, c) $\delta E_z$, d) $n_e$.\label{fig:location_fluctuations_bg3}}
\end{figure*}

The results of our time averaging procedure are shown in Fig.~\ref{fig:location_fluctuations_bg3}, using the same time average $\langle\rangle$ as before ($\Delta T=0.25\Omega_{ci}^{-1}=6.3\Omega_{LH}^{-1}$). The dominant in-plane electric field RMS fluctuations $\delta E_x$ and $\delta E_y$ (Fig.~\ref{fig:location_fluctuations_bg3}(a)-(b)) are located mostly in the separatrices. Density fluctuations $\delta n_e$ (Fig.~\ref{fig:location_fluctuations_bg3}(d)) are more significant only inside the exhaust region and less important near the separatrices. Therefore, correlated fluctuations $\langle\delta E_x\delta n_e\rangle$ and $\langle\delta E_y \delta n_e\rangle$ are expected to be important only in the separatrices. No significant level of in-plane electrostatic fluctuation is observed close to the X-line. On the other hand, there is no significant fluctuation of the out-of-plane electric field $\delta E_z$ at any location that can be clearly distinguished from the surrounding noise level. Note also that the typical RMS values of this component are much smaller than those of the in-plane components $x-y$ (by a factor of 2.5-3). Therefore, there is no significant contribution from the correlated term $\langle \delta E_z\delta n_e\rangle $ to the mean reconnection electric field $\langle E_z\rangle$.

\begin{figure*}[!ht]
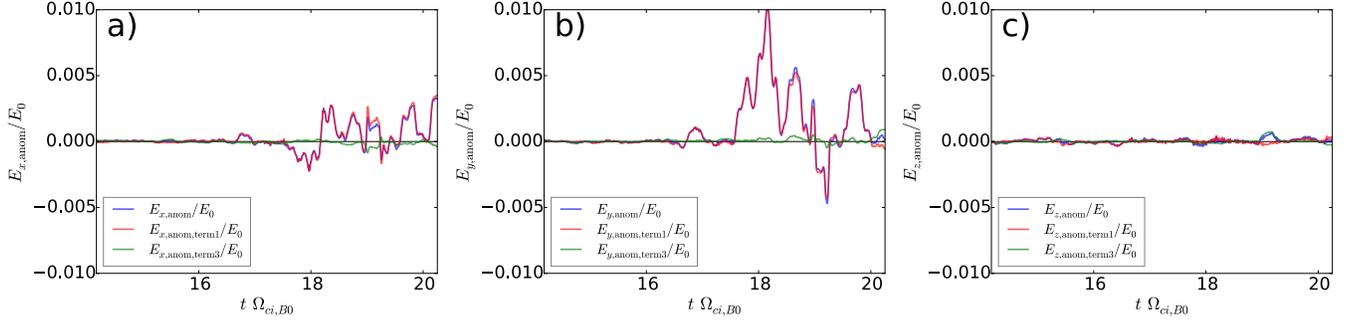

	\centering
	\includegraphics[width=1.0\linewidth]{{{./e_anom_total_timeseries_new}}}
	\caption{Time series of the three spatial components of the electric field due to the anomalous correlated fluctuations $\vec{E}_{{\rm anom}-\delta}$ (given in Eq.~\eqref{ohm_anomalous3}) for the case $b_g=3$ and for the components (a) ${E}_{x,{\rm anom}-\delta}$, (b) ${E}_{y,{\rm anom}-\delta}$, (c) ${E}_{z,{\rm anom}-\delta}$. The tracked point in the low density separatrix is located at $x=-4d_i$, $y=3d_i$, indicated in Fig.~\ref{fig:einertia_bg3_components}. Compare with $\vec{E}_{{\rm anom}-ddt}$ shown in Fig.~\ref{fig:edamping_timeseries_bg3}. Each component also shows the contribution from the first (electrostatic) and third (electromagnetic) term in Eq.~\eqref{ohm_anomalous3}. \label{fig:fluctuations_timeseries_bg3}}
\end{figure*}

Fig.~\ref{fig:fluctuations_timeseries_bg3} shows the time series of the electric field due to correlated fluctuations  $\vec{E}_{{\rm anom}-\delta}$ at the same point in the separatrices used for  $\vec{E}_{{\rm anom-ddt}}$ in Fig.~\ref{fig:edamping_timeseries_bg3}. Comparing with the latter, we can see that the electric field due to fluctuations shows peaks differing significantly among all the components. As expected, the component $E_{z,{\rm anom}-\delta}$ is very small compared to the in-plane components. Although the reason of such disagreement between the values of the anomalous resistivity given by $\vec{E}_{{\rm anom-ddt}}$ and $\vec{E}_{{\rm anom}-\delta}$ is not clear at the moment, we speculate that it might be due to the tensor nature of the effective resistivity since we consider it as an scalar quantity in all our calculations. In such a way, the electric field might have contributions from all components of the current density, modifying their final values, especially for $\vec{E}_{{\rm anom}-\delta}$.

Fig.~\ref{fig:fluctuations_timeseries_bg3} also shows that, in general,  $\vec{E}_{{\rm anom}-\delta}$ tends to have more positive than negative values for later times, providing a net transport due to anomalous resistivity, but mostly in the in-plane directions. The other critical difference by comparing  $\vec{E}_{{\rm anom}-\delta}$ and  $\vec{E}_{{\rm anom-ddt}}$ is that the latter (shown in Fig.~\ref{fig:edamping_timeseries_bg3}) has typical values larger by one order of magnitude. This fact illustrates that different methods to calculate anomalous resistivity based on (the same) time averages do not necessarily agree each other, a fact that has not been considered properly in previous works. This has to be taken into account when extrapolating conclusions based on this methods when applied to other scenarios with more significant anomalous resistivity.

In any case, these very small values due to correlated fluctuations are not enough to explain the deviations in the mean field generalized Ohm's law or to contribute significantly to the mean electric field. This is in agreement with Ref.~\onlinecite{Pritchett2013a}, which also found very small values of anomalous resistivity when performing a time average, in comparison with a calculation based on a spatial average.

We also show in Fig.~\ref{fig:fluctuations_timeseries_bg3} the relative contributions from the first and third term to the total electric field $\vec{E}_{{\rm anom}-\delta}$ in  Eq.~\eqref{ohm_anomalous3}. In general, the electromagnetic contributions from the fluctuations of the magnetic field $\delta B$ with the electron current $\delta (n_e\vec{V}_e)$  are one order of magnitude smaller than the electrostatic ones ($\delta n_e \delta \vec{E}$) for the in-plane components. This is to be expected since the electromagnetic correlated fluctuations were shown to be smaller in low-plasma-$\beta$ conditions,\cite{Davidson1975} like in our case with a strong guide field. In spite of this, we also find a significant contribution from the electromagnetic term for the $z-$ component of the anomalous resistivity calculated from the correlated fluctuations since $\langle\delta E_z\delta n_e\rangle$ is very small.

\subsection{Numerical convergence of correlated fluctuations and time averaging}\label{sec:convergence_fluctuations}
\begin{figure*}[!ht]
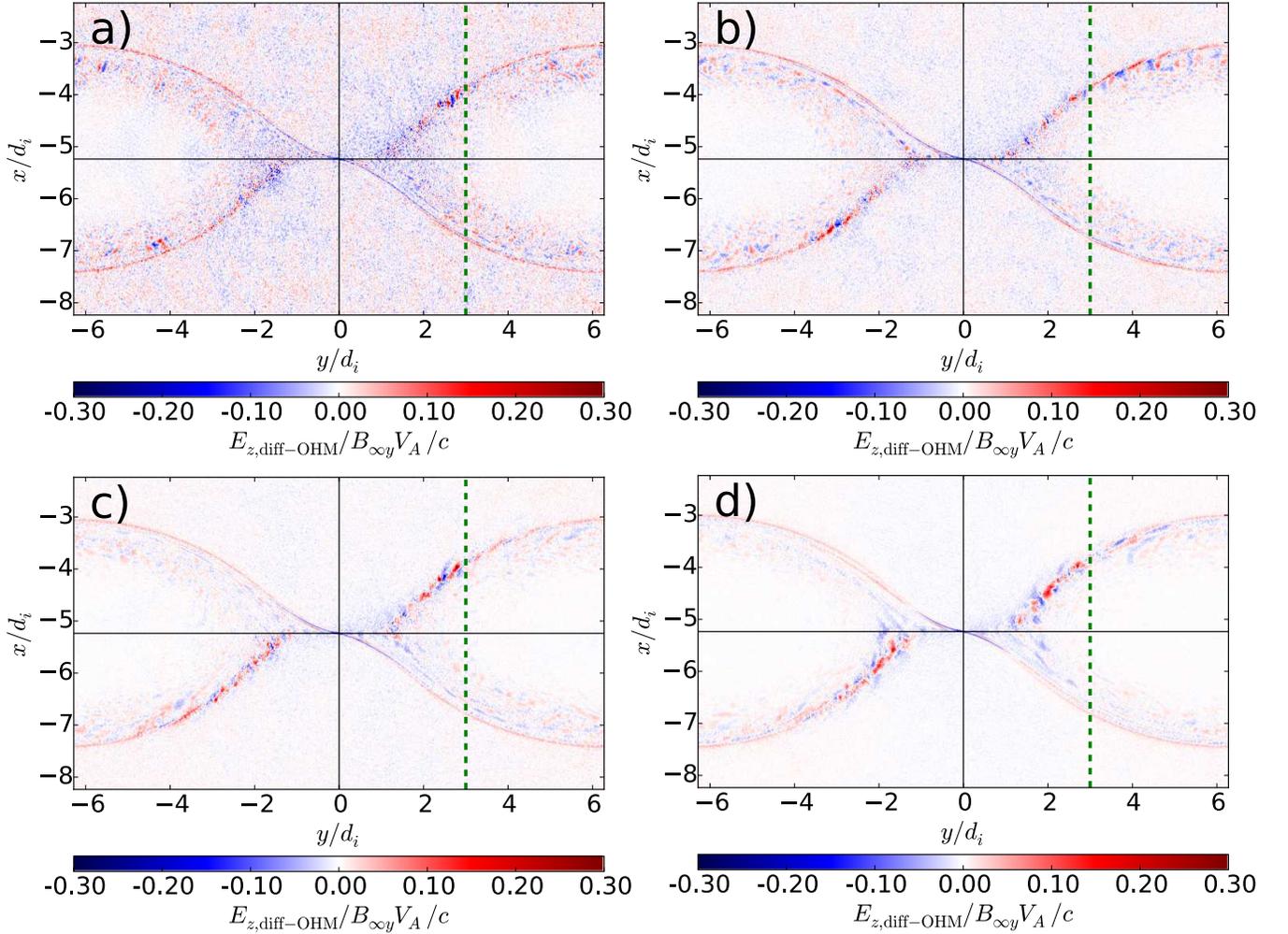

	\centering
	\includegraphics[width=1.0\linewidth]{{{./ohm_diff_bg3_convergence}}}
	\caption{Contour plot of the remaining term  (l.h.s. - r.h.s., excepting the anomalous term) of the $z-$component of the mean generalized Ohm's law Eq.~\eqref{eq:ohm_2fluid_meanfield} for guide field $b_g=3$ and different time average windows (a): $\Delta T=0.128\Omega_{ci}^{-1}=3\Omega_{LH}$, (b): $\Delta T=0.25\Omega_{ci}^{-1}=6\Omega_{LH}$ (same as the one used in all the previous plots) (c): $\Delta T=0.5\Omega_{ci}^{-1}=12\Omega_{LH}$. d) $\Delta T=1\Omega_{ci}^{-1}=25\Omega_{LH}$. The vertical dashed green lines are the x-cuts shown in Fig.~\ref{fig:ohm_diff_profile_convergence}.\label{fig:ohm_diff_convergence}}
\end{figure*}

All the calculations shown before rely on the choice of a given length for the time windows used for splitting the mean from fluctuating quantities, being somewhat arbitrary to define the separation between both. The values of all quantities shown before will vary by changing this time windows. This is because, in a kinetic approach, the separation of scales introduced by the mean and fluctuating quantities is not unique. This is specially critical when there is no clear scale separation between average and fluctuating quantities in Eq.~\eqref{eq:fluctuations}, making Eq.~\eqref{vlasov_anomalous} possibly invalid. As a consequence, the quasi-collision term in the Boltzmann equation do vary for different choices, affecting the macroscopic description. This fact was already considered since a long time ago\cite{Braginskii1965}  (see also Ref.~\onlinecite[p.~32]{Schindler2007}) and it continues being discussed nowadays when applied to in-situ measurements in the solar wind.\cite{Isaacs2015} For example, the same measurement process of electron distributions functions is by definition non-ergodic,\cite{Verscharen2011} therefore affecting any kind of calculations involving the macroscopic momenta of this quantity.

\begin{figure*}[!ht]
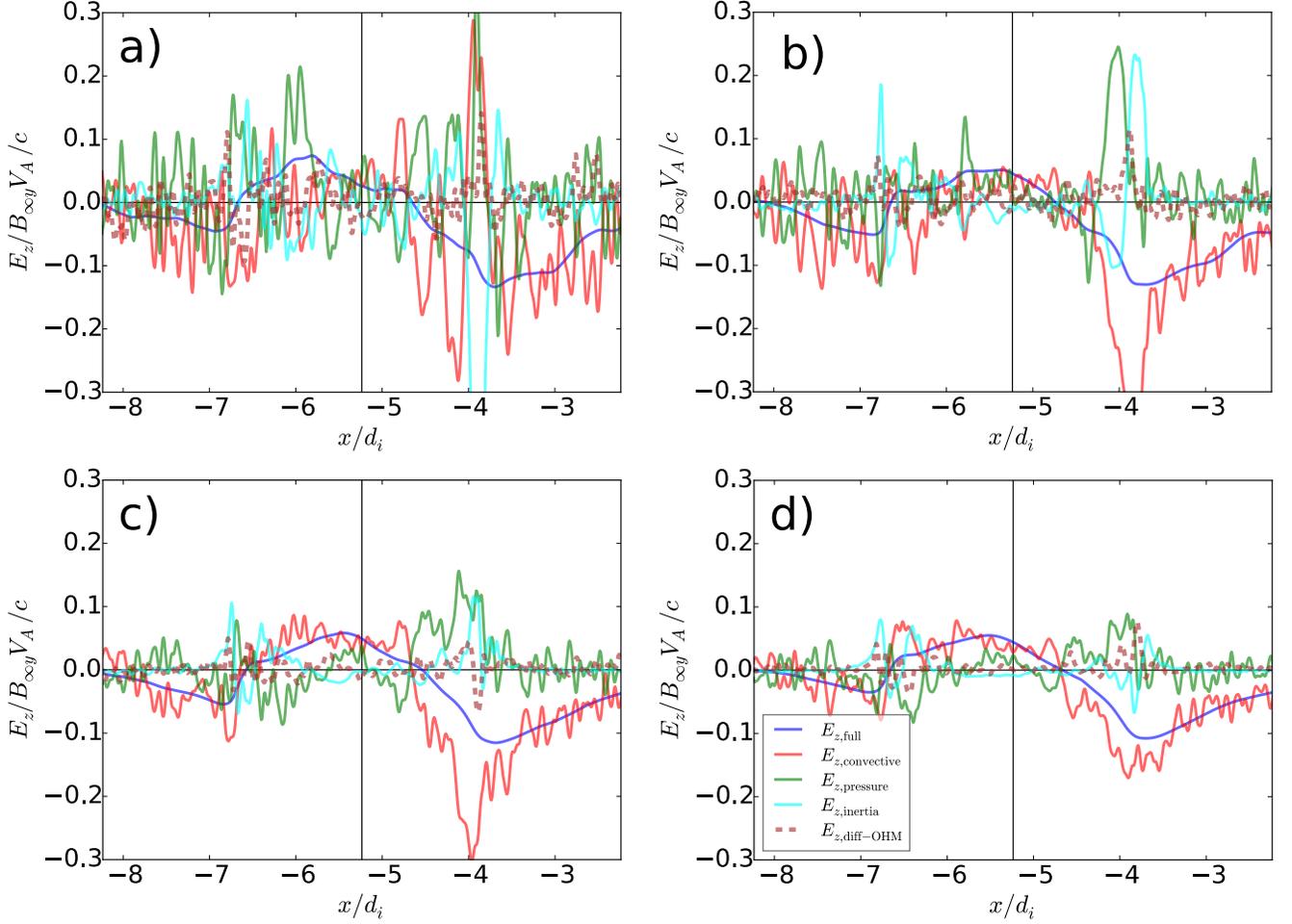

	\centering
	\includegraphics[width=1.0\linewidth]{{{./ohm_balance_bg3_convergence_new}}}
	\caption{$x$-cut of the $z-$component terms of the mean generalized Ohm's law for the guide field $b_g=3$ and different time average windows (a): $\Delta T=0.128\Omega_{ci}^{-1}$, (b): $\Delta T=0.25\Omega_{ci}^{-1}$ (same as the one used in all the previous plots) (c): $\Delta T=0.5\Omega_{ci}^{-1}$. d) $\Delta T=1\Omega_{ci}^{-1}$. The $x$-cuts were obtained at the location indicated in  Fig.~\ref{fig:ohm_diff_convergence}. \label{fig:ohm_diff_profile_convergence}}
\end{figure*}

In all our previous calculations, we chose $\Delta T=0.25\Omega_{ci}^{-1}$ as a representative time average windows, since it is one of the smallest values where the fluctuations due to the noise are significantly smaller than the fluctuations of interest (broadband up to the lower hybrid frequency). Also, the mean and fluctuating quantities are not too sensitive to small changes around this value. Fig.~\ref{fig:ohm_diff_convergence}  show the difference between the l.h.s. and r.h.s. of the $z-$component of the mean generalized Ohm's law (Eq.~\eqref{eq:ohm_2fluid_meanfield}) (same quantity shown in Fig.~\ref{fig:ohm_diff}(d)), for the case $b_g=3$ and varying time windows over one order of magnitude. Smaller time windows ($\Delta T=0.128\Omega_{ci}^{-1}$, Fig.~\ref{fig:ohm_diff_convergence}(a)) are unreliable because they produce quantities with mostly fast fluctuations, in regions away from the current sheet, due to numerical noise and not due to relevant physical processes. Larger time windows ($\Delta T=0.5-1.0\Omega_{ci}^{-1}$, Figs.~\ref{fig:ohm_diff_convergence}(c)-(d)) smooth out small scale structures in the separatrices. We do not show even larger values of time windows, close to reconnection time scales, since they would produce unreliable results because the current sheet and the location of separatrices change during the development of reconnection (see Fig.~\ref{fig:e_anom_ddt_stack}). A large time average would be meaningful under conditions of stationary reconnection, which would require a much larger simulation box and/or non-periodic boundary conditions.

The profiles shown in Fig.~\ref{fig:ohm_diff_profile_convergence} reveal complementary information to that displayed in Fig.~\ref{fig:ohm_diff_convergence}. Note that, similar to Fig.~\ref{fig:ohm_diff_profile}, we have also applied a spatial Savitzky-Golay filter. The peak values of the remaining term in the mean Ohm's law are always at the low density separatrix and decrease linearly for increasing time windows. On the other hand, and at the same time, the anomalous electric field becomes more comparable to the peak values of the pressure and inertia term contributions to the mean electric field when using larger time windows since its maximum value decreases faster than in a linear fashion. This can also be understood in the sense that larger time windows take into account a broader spectrum of fluctuating frequencies, and then the relative contribution to the mean electric field should be larger than smaller time windows,  which only consider high frequencies not caused by streaming instabilities in the lower hybrid frequency range.

\begin{figure*}[!ht]
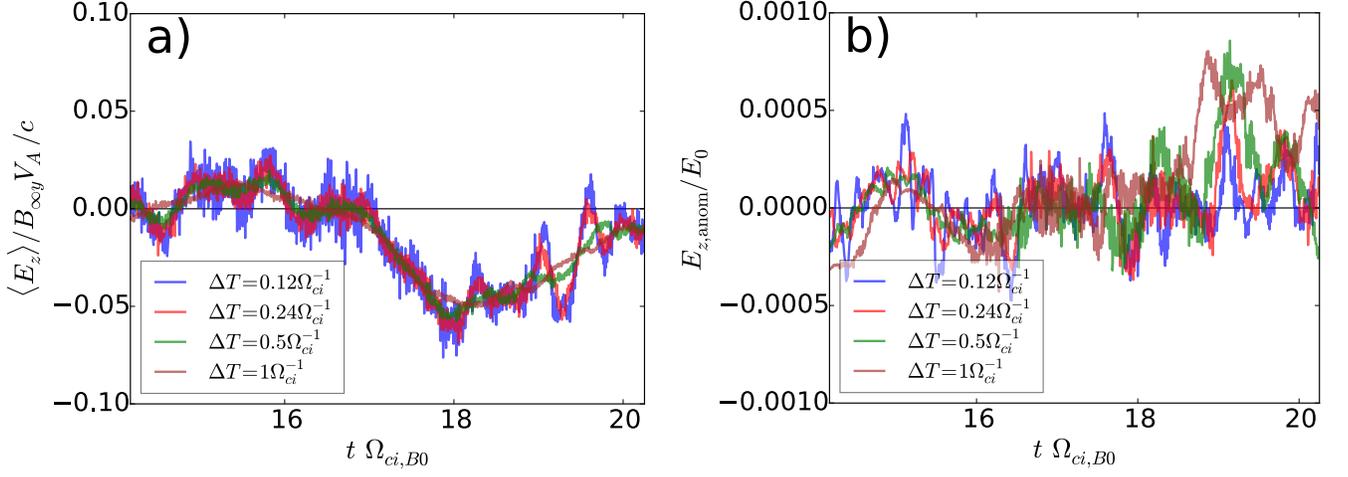

	\centering
	\includegraphics[width=1.0\linewidth]{{{./ez_mean_anom_timeseries_convergence_new}}}
	\caption{Time series of a) the mean value of the out-of-plane electric field component $\langle E_z\rangle$ and b) electric field anomalous correlated fluctuations $E_{z,{\rm anom-\delta}}$ according to Eq.~\eqref{ohm_anomalous3}. The calculations are for the case $b_g=3$, with the tracked point at $x=-4d_i$, $y=3d_i$ (indicated in Fig.~\ref{fig:einertia_bg3_components}) and different choices of the running average. Note that the $y$ range between both plots differ in two orders of magnitude. \label{fig:anom_ez_timeseries_bg3}}
\end{figure*}

\begin{figure*}[!ht]
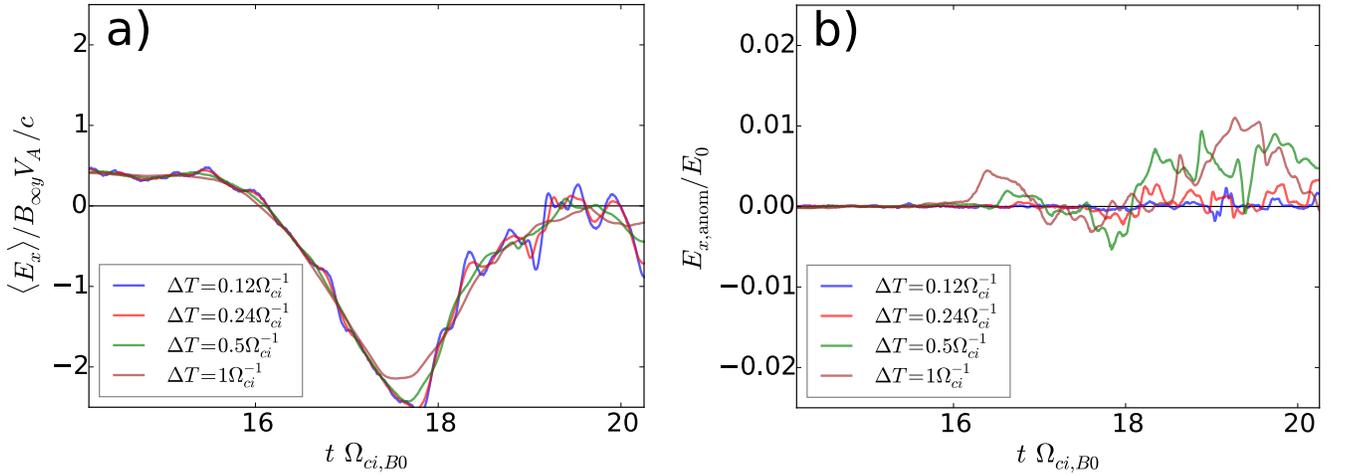

	\centering
	\includegraphics[width=1.0\linewidth]{{{./ex_mean_anom_timeseries_convergence_new}}}

	\caption{Time series of a) the mean value of the in-plane electric field component $\langle E_x\rangle$ and b) anomalous correlated fluctuations $E_{x,{\rm anom-\delta}}$ according to Eq.~\eqref{ohm_anomalous3}. The other details are similar to Fig.~\ref{fig:anom_ez_timeseries_bg3}, excepting the difference in (slightly more than) one order of magnitude in the $y$ range. \label{fig:anom_ex_timeseries_bg3}}
\end{figure*}

In order to see the time evolution of the correlated fluctuations,  in Fig.~\ref{fig:anom_ez_timeseries_bg3}(b) we show the electric field due to this contribution $E_{z,{\rm anom-\delta}}$. As expected, the peak values of this anomalous electric field do not contribute significantly to the mean electric field $\langle E_z\rangle$, shown in  Fig.~\ref{fig:anom_ez_timeseries_bg3}(a), or to the non-ideal electric field in the l.h.s. of the mean Ohm's law (Eq.~\eqref{eq:ohm_2fluid_meanfield}) (plot not shown here). The difference is about two orders of magnitude. Those plots also confirm that larger time windows exhibit larger values of the anomalous resistivity but in any case not varying significantly for later times, as well as more stable over large time periods and with a tendency toward positive values. This same behavior is also shown, but even in a more strongly manner, by the $E_{x,{\rm anom-\delta}}$ component of the anomalous resistivity due to correlated fluctuations in Fig.~\ref{fig:anom_ex_timeseries_bg3}(a). Note that all the values are enhanced in more than one order of magnitude compared to the $z-$component.

Finally, we have also verified that the time average of all fluctuating quantities ($\langle \delta A\rangle/A \sim 0$) in the chosen range of time windows is close to zero, i.e., the Reynolds rules (see, e.g., Appendix A2 in the textbook Ref.~\onlinecite{Sagaut2006}) are well satisfied (plots not shown here).

\section{Summary and conclusions}\label{sec:conclusion}

We investigated the consequences of the self-generated plasma-turbulence for the transport of the current-carrying charged particles in two-dimensional guide-field reconnection.
In order to derive an  ``anomalous'' effective resistivity due to the turbulence,
we utilized a mean-field approach.
For this sake, we split-off the fast, micro-turbulent fluctuations from the mean
slowly varying variables quantities describing the macroscopic evolution of the plasma.
We based our analysis on the results of fully kinetic 2.5D Particle-in-Cell (PiC)
code (ACRONYM) simulations.
Based on these data, we first calculated the contributions to the balanced slowly varying (mean) electric field in the framework of a generalized Ohm's law description.
We then compared our findings first with the slowing-down rate of the macroscopic (mean) current  due to a net momentum transfer from particles to waves and also by calculating the first order correlations of the micro-turbulent fluctuations.
The latter method provides a collision term due to the fluctuations in the right hand side of a Boltzmann equation for the mean quantities, derived from the Vlasov equation for the fully kinetic variables.

We found, by all the previous methods, that in two-dimensional collisionless guide-field reconnection, the self-generated kinetic-scale micro-turbulence does not significantly contribute to the balance of the mean electric fields, neither near the reconnection X-line nor near the separatrices and the exhaust region of reconnection.

For strong external guide magnetic fields ($1.5\lesssim b_g\lesssim 6$),
we found, however, that the self-generated turbulence near the ``low-density'' separatrices  provides non-negligible contributions to the mean electric field.
There, the energy exchange between the particles and the mean, averaged
over the turbulence, electric fields is ``patchy'' in space  and oscillating in time due to the unstable plasma waves caused by a counterstreaming electron beam instability.\cite{Munoz2016a}
Part of the energy exchange is, however, irreversible, and the corresponding slowly varying (mean) electric field can be related to be due to a ``anomalous'' resistivity produced by the turbulence

Nevertheless, for most of the reconnection regions in guide fields  in
the range $1.5\lesssim b_g\lesssim 6$, the electron inertia dominates the
slowly varying, macroscopic (mean) electric field.
In the Ohm's law, this contribution is dominant due to the large  Eulerian partial time-derivative of the mean electron bulk-drift velocity.
Note that only near the ``low density'' separatrices, the inertia term is comparable to the pressure and the convective electric field terms.
For smaller and vanishing guide fields $ b_g\lesssim 1.5$, however, the contribution of the electron inertia to the mean electric field is negligibly small, not only near the reconnection X-line but also at the separatrices and in the exhaust region of reconnection.

An independent estimate for the effective resistivity is the slowing-down rate of the mean velocity  of the main current carriers, the electrons, due to a net momentum transfer from particles to waves.
The resulting ``anomalous'' resistive (turbulent) contribution to the electric field is spatially spread.
It slightly differs from the results for the balanced mean electric field obtained by calculating the terms of the generalized Ohm's law. However, the slowing-down rate of the mean electron current  due to a net momentum transfer from particles to waves, associated with the electron-ion momentum transfer, is largest where the anomalous electric field via the calculation of the mean generalized Ohm's law is the strongest. The resulting slowing-down rate of the current due to a net momentum transfer from particles to waves, if attributed to an  effective ``collision frequency'', ranges from the ion-cyclotron to the lower-hybrid frequency.
Note that the electron inertia term is not  directly proportional to the electron-current slowing-down rate.

In addition to the out-of-plane electric field field $E_z$, we further investigated the contribution of the self-generated turbulence to the electric field balance and energy transfer via the in-plane electric field components $E_x$ and $E_y$.
The momentum-exchange between the electron and ions in those directions appeared to be comparable to the one in the $z$ direction near the ``low density separatrices'', i.e., there the momentum transfer
is isotropic.
A large portion of the energy exchange is reversible and quasi-periodic  as indicated by a changing sign of $\vec{j}\cdot\vec{E}$.
A smaller part of the energy exchange corresponds to a net irreversible energy loss, i.e., to dissipation.
Note that the latter should be detected with a time average window, which should be large enough to take into
account the broad turbulence spectrum below the lower-hybrid frequency, generated by the beam instability of electrons flowing along the separatrix.

We further verified the enhancement of the momentum transfer due to wave-particle interaction near the separatrices of strong-guide-field reconnection by another independent method, i.e., calculating the first
order correlations of field and plasma fluctuations.
We found that for strong guide fields, the electric field balanced due to
the quasi-collision term of the resulting Boltzmann equation is negligibly
small near the X-line of 2D reconnection.
As obtained by the other methods, the contributions to the correlated field-plasma fluctuations are  most important near the ``low density separatrices''.
The relevant fluctuations are mainly electrostatic with a characteristic frequency less than the lower-hybrid frequency.
Note that the contribution of the correlated fluctuations to the mean electric field is smaller and does not completely match the contributions obtained by the other two methods of determining the remaining term in the Ohm's law and of calculating the rate of the slowing down of the electron current due to a net momentum transfer from particles to waves.
This disagreement is, perhaps, due to the tensorial nature of the ``anomalous'' resistivity due to micro-turbulence, which involves all vector components of the current density.

Further, we found the criterion for an appropriate choice of the sliding window width for the averaging procedure of the mean field approach.
According to our results, one has to make sure that the window width has to include all the relevant frequency ranges of the turbulence for the anomalous transport; otherwise,  the result will be very sensitive on the chosen window-width.

Our findings provide a number of macroscopic signatures, allowing a quantification of the wave-particle interaction and the momentum transfer due to micro-turbulence in antiparallel and in collisionless guide-field magnetic reconnection.
In particular, we found that in guide-field magnetic reconnection, the self-generated turbulence is maximum near the separatrices rather than close to the X-line.
Our findings can be used, e.g., for the diagnostics of in-situ measurements in magnetic reconnection regions in space and in laboratory experiments.
Since all such measurements are averaging over space and time,
it is important to make sure which part of the turbulence is cut off,
which may essentially contributes to the ``anomalous'' transport.
This can be done by coordinated measurements of the mean quantities
readily available, such as the Ohm's law terms and the time  derivatives of the
mean electron current.
Such analysis became possible due to the availability of high resolution observations of fluctuations in space by the current MMS mission and it would become possible in an even better way with the envisioned THOR\cite{Vaivads2016a} spacecraft mission.

While we found that in the limit of two-dimensional collisionless
guide field reconnection, the self-generated micro-turbulence
does not significantly contribute to the mean slowly varying electric field near the X-line,
the situation will change if three-dimensional turbulence becomes relevant.
Our approach to obtain the turbulence-related contribution
to the mean macroscopically observable electric field and
the different terms of the generalized Ohm's law will, however, still apply.
This is particularly true if the kinetic turbulence is caused by
plasma micro-instabilities which can only arise in three dimensions including the direction of the main current flow. Then, it is expected that 3D turbulence will contribute more significantly
to the ``anomalous'' transport terms in the generalized
Ohm's law.

\begin{acknowledgments}
	We acknowledge the developers of the ACRONYM code (Verein zur F\"orderung kinetischer Plasmasimulationen e.V.). P.M. and J.B. acknowledge the financial support by the Max-Planck-Princeton Center for Plasma Physics. P.K. acknowledges support from the NRF and DST of South Africa through the following disclosure:\\
	This work is based upon research supported by the National Research Foundation and Department of Science and Technology. Any opinion, findings and conclusions or recommendations expressed in this material are those of the authors and therefore the NRF and DST do not accept any liability in regard thereto.

	All authors thank the referee for valuable and constructive comments which helped us to clarify and improve our explanations.
\end{acknowledgments}

\appendix
\section{Non ideal electric field in the exhaust for small guide fields}\label{sec:app}

In this appendix, we detail the processes balancing the non-ideal electric field in the exhaust of magnetic reconnection in cases of small finite guide field (see Fig.~\ref{fig:ohm_lhs}).
Some of these features and effects differ from those known for antiparallel reconnection, in particular, away from the X-line. In order to limit the scope of the discussion, we focus only on the case $b_g=0$.

\begin{figure*}[!ht]
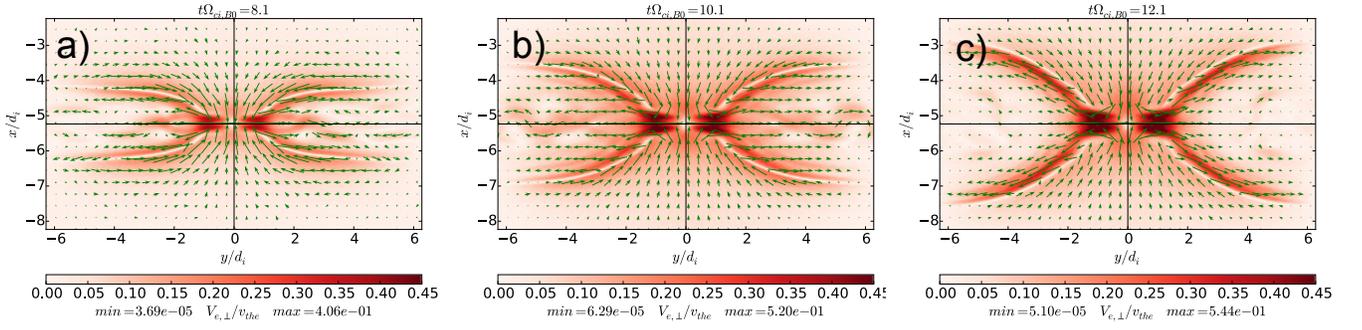

	\centering
	\includegraphics[width=1.0\linewidth]{{{./app_fig5_flows}}}
	\caption{Color-coded contour plot of the magnitude of the in-plane electron flow $|\vec{V}_{e,\perp}|$ for three characteristic times in the case $b_g=0$ (a) $t\Omega_{ci}=8$, b) $t\Omega_{ci}=10$, c) $t\Omega_{ci}=12$. $t\Omega_{ci}=12$ corresponds to the same time shown in Fig.~\ref{fig:ohm_lhs}. Arrows indicate the direction of the flow.\label{fig:elec_flows}}
\end{figure*}

\begin{figure*}[!ht]
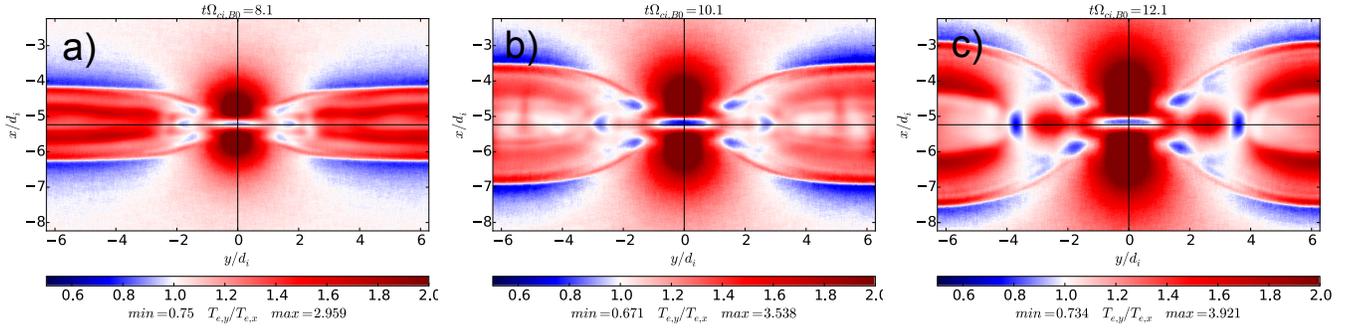

	\centering
	\includegraphics[width=1.0\linewidth]{{{./app_fig7_ani}}}
	\caption{Color-coded contour plot of the of the electron temperature anisotropy $T_{e,y}/T_{e,x}$ for three characteristic times (a) $t\Omega_{ci}=8$, b) $t\Omega_{ci}=10$, and c) $t\Omega_{ci}=12$.\label{fig:anisotropy}}
\end{figure*}

In the outflow region away from the X-line, the complex structure of the non-ideal electric field is mostly due to the convective electric field.
An oscillating electron flow is seen in the exhaust with alternating values of $V_{e,x}$, especially at $t\Omega_{ci}=10$ and with some remnants at $t\Omega_{ci}=12$ (Fig.~\ref{fig:elec_flows}).
Note that this is reflected in the alternating pattern of signs in the non-ideal electric field  $E_{z,nonideal-LHS}$ (Fig.~\ref{fig:ohm_lhs}).
This is due to a purely kinetic effect --- the Weibel instability, driven by electron temperature anisotropy.\cite{Weibel1959}
Note that this process  can only take place in a fully-kinetic consideration. Hybrid, two fluid, Hall-MHD or MHD approaches do not take it into account.
The Weibel instability is sometimes also called ``filamentation'' instability,\cite{Bret2005} especially when it is excited due to counterstreaming cold beams, which can be considered as a form of temperature anisotropy (see Sec.~9.10.2 in Ref.~\onlinecite{Krall1973}).
A Weibel instability is generated because inside the magnetic island, near the CS midplane, magnetic reconnection generates a temperature anisotropy by heating the electrons along the CS direction $y$, mostly due to the natural development of the tearing mode.
Fig.~\ref{fig:anisotropy} shows that the anisotropy $T_{e,y}/T_{e,x}$ for three different characteristic times for our antiparallel case ($b_g=0$).  The heating is along $y$, which will be considered the parallel direction in the following (and $x$ or $z$, the perpendicular, colder directions).
The Weibel instability has already been observed in antiparallel magnetic reconnection in previous investigations using similar parameters and geometry\cite{Lu2011,Schoeffler2013,Munoz2014a}.
It has also been more frequently detected in relativistic pair plasmas, because in them the space available for its growth inside the CS is larger (see Ref.~\onlinecite{Liu2009} and references therein). Note that the firehose instability, also driven by a temperature anisotropy,  fulfills the conditions to be excited in the exhaust of magnetic reconnection of our simulations with $b_g=0$. However, its growth rate is too small compared to the Weibel instability, in such a way that we have not found any characteristic signatures of this instability in our investigations.

\begin{figure*}[!ht]
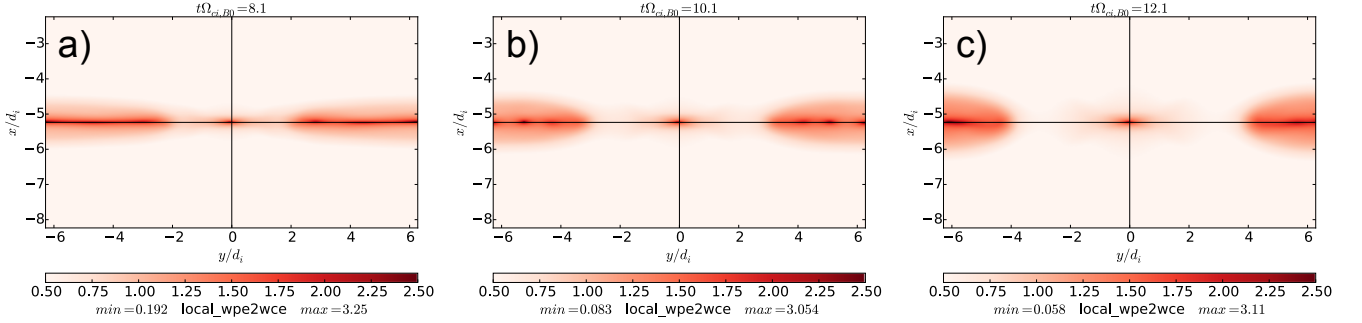

	\centering
	\includegraphics[width=1.0\linewidth]{{{./app_fig8_omegas}}}
	\caption{Color-coded contour plot of the logarithm of the local $\omega_{pe}/\Omega_{ce}$. Regions with high values (towards the dark red in the colortable scale) indicate unmagnetized regions. a) $t\Omega_{ci}=8$, b) $t\Omega_{ci}=10$, and c) $t\Omega_{ci}=12$. \label{fig:omegas}}
\end{figure*}

The electron temperature anisotropy drives the Weibel instability in the unmagnetized regions near the CS midplane inside magnetic islands. Although the anisotropy is even stronger in the two circular symmetric regions near the X-line, temperature  anisotropies driven instabilities like Weibel instability cannot grow there because the magnetization is too strong. The magnetic field (mostly along $y$ outside the CS) tends to suppress the mechanism that leads to this instability, due to the transfer of momentum between the parallel and perpendicular directions. This condition can be quantified as  $\omega_{pe}\gg\Omega_{ce}$ (see, e.g., Refs.~\onlinecite{Hededal2005,Stockem2006}), and is shown in Fig.~\ref{fig:omegas}.

\begin{figure*}[!ht]
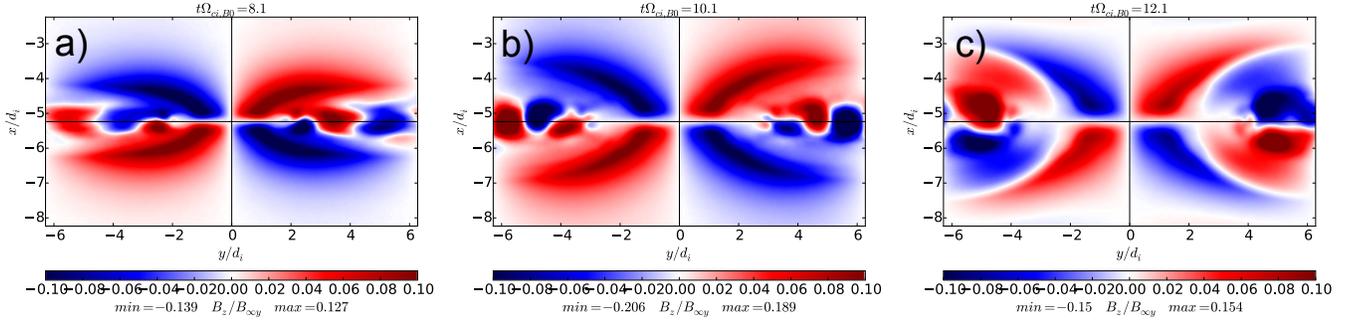

	\centering
	\includegraphics[width=1.0\linewidth]{{{./app_fig9_bz}}}
	\caption{Color-coded contour plot of $B_z$. a) $t\Omega_{ci}=8$, b) $t\Omega_{ci}=10$, c) $t\Omega_{ci}=12$. \label{fig:bz}}
\end{figure*}

The main effect of the Weibel instability is generating a magnetic field in the direction perpendicular to the higher temperature ($y$), i.e., mainly in the $z$ direction.\cite{Fried1959,Medvedev1999}
Although Weibel instability is a non-propagating instability ($\omega\sim 0$), these magnetic fields have a characteristic length scale also perpendicular to the higher temperature direction ($y$), which in our 2.5D geometry can only be the $x$-direction.
Fig.~\ref{fig:bz} shows this characteristic signature of alternating signs or ``checkerboard pattern'' of $B_z$. They are generated in regions of stronger anisotropy, to be then ejected outwards due to the outflows of reconnection and the compression of the magnetic island, being always located in regions with low-magnetization. Note that Fig.~\ref{fig:bz} also displays the quadrupolar Hall magnetic fields due to the decoupling of electron and ion motion near the X-line. However, from $t\Omega_{ci}=10$ to $t\Omega_{ci}=12$, the Weibel-generated $B_z$ expands to a larger area inside the magnetic island, forming an opposite quadrupolar structure to those of the Hall magnetic field.
Comparing  Fig.~\ref{fig:bz} with Fig.~\ref{fig:anisotropy}, one can see that as the magnetic field gets stronger, the electron anisotropy is reduced.

In order to quantify the identification of the Weibel instability responsible for the structure formation and dynamics in the outflow region, let us estimate some relevant quantities predicted by the linear theory (see, e.g., Sec.~9.10.2 of Ref.~\onlinecite{Krall1973}). The temperature anisotropy threshold, the wave number of the maximum unstable perturbation $k_{x,max}$ (in the  direction $x$ perpendicular to the hotter temperature along $y$), and maximum growth rate $\gamma_{max}$ are given by:
\begin{align}
\label{eq:weibel_threshold}	\left(\frac{T_{e,y}}{T_{e,x}}-1\right)  &> \left(\frac{k_xc}{\omega_{pe}}\right)^2,\\
\label{eq:weibel_kmax}	k_{x,max} &= \frac{\omega_{pe}}{c}\sqrt{\frac{1}{3}\left(\frac{T_{e,y}}{T_{e,x}}-1\right)},\\
\label{eq:weibel_kmax}	\gamma_{max}&=\sqrt{\frac{8}{27\pi}}\omega_{pe}\frac{v_{th,e}}{c}\frac{T_{e,x}}{T_{e,y}}\left(\frac{T_{e,y}}{T_{e,x}}-1\right)^{3/2}.
\end{align}
Note that we have considered only electron temperature anisotropy. Ions would require having a much stronger temperature anisotropy, not seen in our simulations, to drive instabilities.

\begin{figure*}[!ht]
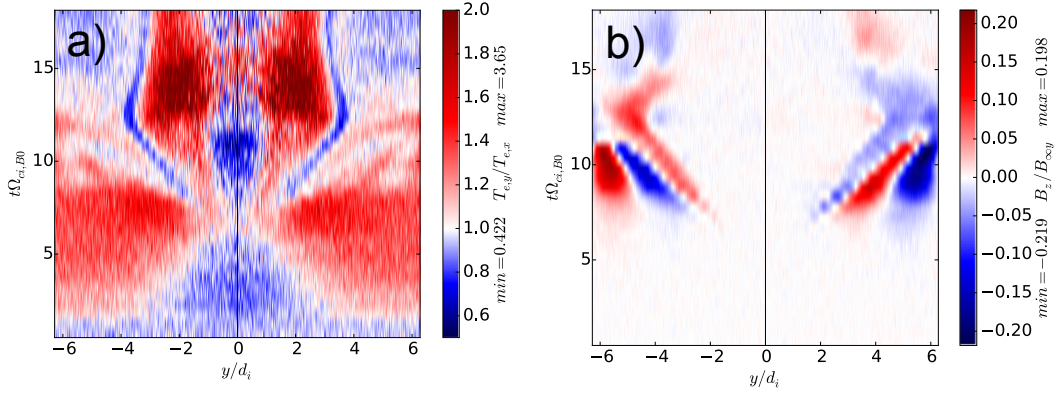

	\centering
	\includegraphics[width=0.8\linewidth]{{{./app_fig10_stacks}}}
	\caption{Stack plots  showing the time evolution in the CS midplane for a): $T_{e,y}/T_{e,x}$ and b) $B_z$. \label{fig:stacks}}
\end{figure*}

\begin{figure*}[!ht]
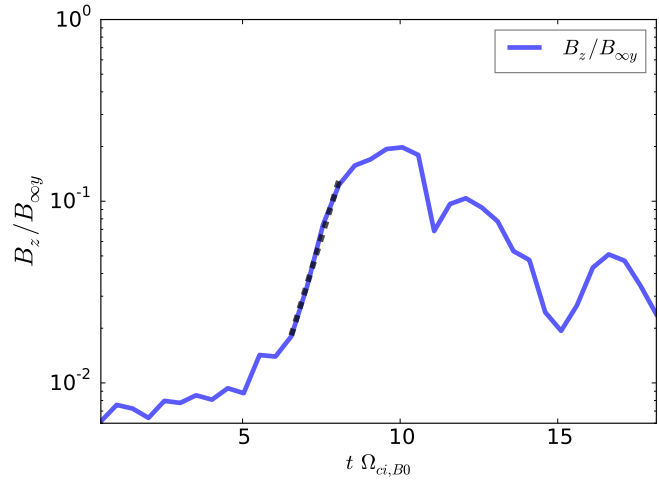

	\centering
		\includegraphics[width=0.5\linewidth]{{{./app_fig11_bz_evol}}}
	\caption{Time evolution of the maximum of the out-of-plane magnetic field along the CS center ${\rm max}(B_z(x=CS,y))$, with linear fit $\gamma/\Omega_{ci}=1.30$ during the exponential growth phase (dashed line).
	\label{fig:growth_rate_weibel}}
\end{figure*}

\begin{figure*}[!ht]
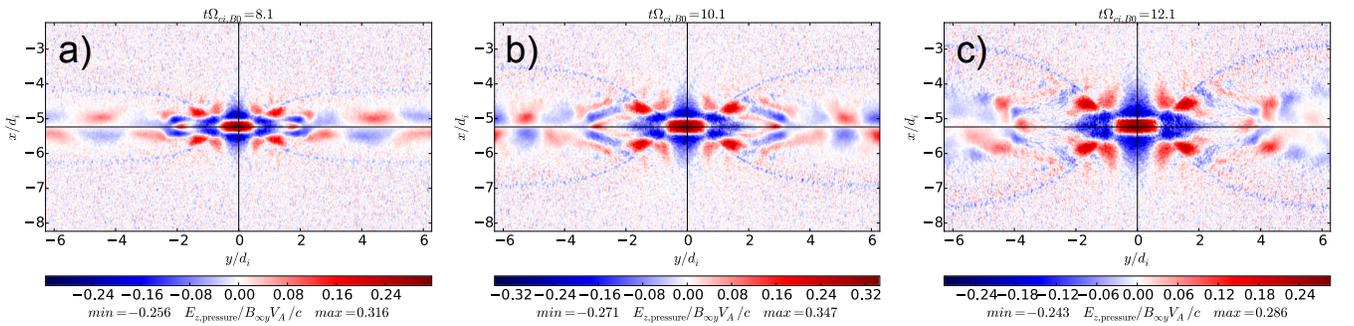

	\centering
	\includegraphics[width=1.0\linewidth]{{{./app_fig12_ez_pressure}}}
	\caption{Color-coded contour plots for the reconnection electric field ($E_z$) due to the electron pressure term in the Ohm's law. a) $t\Omega_{ci}=8$, b) $t\Omega_{ci}=10$, and c) $t\Omega_{ci}=12$. \label{fig:ez_pressure}}
\end{figure*}

\begin{figure*}[!ht]
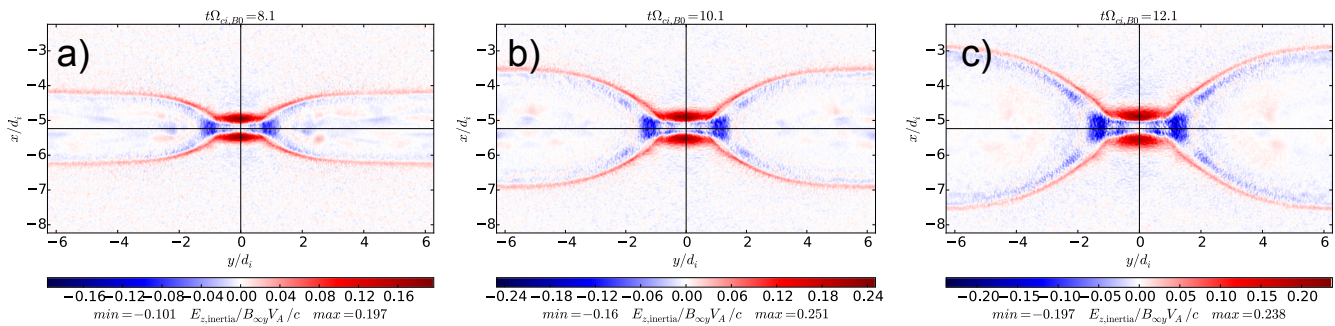

	\centering
	\includegraphics[width=1.0\linewidth]{{{./app_fig13_ez_inertia}}}
	\caption{Color-coded contour plots for the reconnection electric field ($E_z$) due to the electron inertia in the Ohm's law ($z-$ component). a) $t\Omega_{ci}=8$, b) $t\Omega_{ci}=10$, and c) $t\Omega_{ci}=12$. \label{fig:ez_inertia}}
\end{figure*}
The threshold in Eq.~\eqref{eq:weibel_threshold} implies that weaker anisotropies will produce larger Weibel unstable waves, but a minimum anisotropy is required to confine the waves inside the unmagnetized region of the CS. Taking $\lambda\sim d_i$ as the maximum distance across the CS that remains unmagnetized during the whole evolution (see Fig.~\ref{fig:omegas}), Eq.~\eqref{eq:weibel_threshold} implies that $T_{e,y}/T_{e,x}>1.4$ is required for the excitation of the Weibel instability inside the CS (for comparison, $\lambda\sim1.5d_i$ requires $T_{e,y}/T_{e,x}>1.18$).
This value is reached in the CS midplane at $t\Omega_{ci}\gtrsim 8$ as shown in the contour plot Fig.~\ref{fig:anisotropy}a), and in a more clear way in the stack plot along that line in Fig.~\ref{fig:stacks}a). That instant of time correlates well with the generation of the $B_z$ magnetic field component at those locations as seen in Fig.~\ref{fig:stacks}b), Fig.~\ref{fig:bz}a), and also with the associated reduction of anisotropy at later times.
Note how the $B_z$ magnetic field component generated by the Weibel instability propagates outwards due to the acceleration of the reconnection outflows and the compression of the magnetic island.
The wavelength associated with the wave number of maximum growth rate in Eq.~\eqref{eq:weibel_kmax} gives $\lambda_{x,max}\sim 1.7d_i$ (with $\lambda_{x,max}=2\pi/k_{x,max}$) for the same temperature anisotropy $T_{e,y}/T_{e,x}\sim 1.4$. Because no mode can grow larger than $\lambda\sim d_i$, the approximate width of the unmagnetized region of the CS, the observed wavelength is also $\lambda\sim d_i$ (see also Fig.~\ref{fig:bz}a)).
This corresponds to the predictions of the model of Ref.~\onlinecite{Liu2009} about the length scale of Weibel structures in a CS, depending on the electron gyroradius in the generated magnetic field.
Using the same value of anisotropy and the theoretical value Eq.~\eqref{eq:weibel_kmax}, we obtain the estimated growth rate $\gamma_{max}/\Omega_{ci}=2.77$ for our initial equilibrium parameters (and $\gamma_{max}/\Omega_{ci}=3.5$ considering the small corrections of $\omega_{pe}$ and $v_{th,e}$ due to compression and heating, respectively).
By means of a linear fitting during the exponential growth phase of the logarithm of the fastest growing $B_z$ at the CS midplane, we obtain $\gamma/\Omega_{ci}=1.30$ (see Fig.~\ref{fig:growth_rate_weibel}). This is just half of the theoretical estimated value, but a good approximation considering that the growth of Weibel instability should be slower due to the (small) magnetization inside the CS and the fact that the maximum wavelength of its unstable mode is constrained by the island size.\cite{Liu2009}
All these results evidence that we indeed observe a Weibel instability in the exhaust region of the CS in the case of antiparallel magnetic field $b_g=0$.
For finite but small guide fields, such as $b_g=0.26$ (shown in Fig.~\ref{fig:ohm_lhs}b)), we also see structures in the exhaust that agree with most of the aforementioned features of a Weibel instability. For even stronger guide fields (e.g., $b_g\gtrsim 1.0$), however, there are no signatures of Weibel structures. This should be expected because the magnetization in the exhaust of reconnection suppresses completely this unmagnetized plasma temperature anisotropy driven instability.

The balance of the non-ideal reconnecting electric field also agrees with most results of previous studies of antiparallel magnetic reconnection. Fig.~\ref{fig:ez_pressure} shows the spatial distribution of the contributions to the electric field (generalized Ohm's law) due to the non-gyrotropic electron pressure term, and Fig.~\ref{fig:ez_inertia} shows the electric field due the electron inertia (see Eq.~\eqref{eq:ohm_2fluid_meanfield}).
The non-gyrotropic pressure term (Fig.~\ref{fig:ez_pressure}) balances most of the convective electric field responsible for the Weibel structures in the exhaust (associated with the temperature anisotropy) and also, near the X-line, for the non-ideal electric field $E_{z,nonideal-LHS}$.
The pressure term displays a double peak structure near the X-line along $y$, with a dip at the very X-line.
The electron inertia term also reveals a double peak structure along $y$ near the X-line, although without the central dip. These features were shown and explained in a number of previous  studies (see, e.g., Refs.~\onlinecite{Kuznetsova1998,Pritchett2001,Ricci2004}).

The most relevant contribution to the non-ideal electric field comes from $E_{z,pressure}$ (Fig.~\ref{fig:ez_pressure}), both near the X-line and in the exhaust regions, balancing the structures formed by the Weibel instability.
The spatial structure of the two non-gyrotropic (off-diagonal) pressure tensor components of $E_{z,pressure}$  displays a double peak structure near the X-line  along $y$, together with a dip at the very X-line.
All these symmetries and features were predicted\cite{Vasyliunas1975,Dungey1989} and confirmed by  fully-kinetic\cite{Cai1994,Cai1997,Hesse1998a,Hesse1999,Pritchett2001} and hybrid code simulations with an evolution equation for the electron pressure tensor\cite{Hesse1994,Kuznetsova2001}.
They are due to the meandering motion of electrons near the neutral line, mediating momentum transport between regions close to the X-line to those outside, which corresponds to a quasi-viscosity.
They have a well known characteristic length scale which agrees with our results.\cite{Cai1997,Hesse1999}
Our results differ in the pressure term only by the presence of Weibel structures in the exhaust and  in the inflow region just next to the X-line.
The latter is, perhaps, due to the anisotropic heating of the background plasma population, different from the current carriers.
Note that low resolution PiC simulations, with low order shape functions, without current smoothing, too small number of particle per cells or not resolving well enough the smallest electron length scales might not display accurately the structure of the off-diagonal terms of the electron pressure or electron inertia, due to the enhanced effective collisionality or numerical heating effects\cite{Munoz2014a,Melzani2013}.

The change of behavior for finite guide fields, leading to increasing asymmetries of the generalized Ohm's law terms, has also been shown to be related to the non-gyrotropic pressure as well\cite{Ricci2004,Hesse2004,Hesse2006}.


\begin{thebibliography}{98}%
\makeatletter
\providecommand \@ifxundefined [1]{%
 \@ifx{#1\undefined}
}%
\providecommand \@ifnum [1]{%
 \ifnum #1\expandafter \@firstoftwo
 \else \expandafter \@secondoftwo
 \fi
}%
\providecommand \@ifx [1]{%
 \ifx #1\expandafter \@firstoftwo
 \else \expandafter \@secondoftwo
 \fi
}%
\providecommand \natexlab [1]{#1}%
\providecommand \enquote  [1]{``#1''}%
\providecommand \bibnamefont  [1]{#1}%
\providecommand \bibfnamefont [1]{#1}%
\providecommand \citenamefont [1]{#1}%
\providecommand \href@noop [0]{\@secondoftwo}%
\providecommand \href [0]{\begingroup \@sanitize@url \@href}%
\providecommand \@href[1]{\@@startlink{#1}\@@href}%
\providecommand \@@href[1]{\endgroup#1\@@endlink}%
\providecommand \@sanitize@url [0]{\catcode `\\12\catcode `\$12\catcode
  `\&12\catcode `\#12\catcode `\^12\catcode `\_12\catcode `\%12\relax}%
\providecommand \@@startlink[1]{}%
\providecommand \@@endlink[0]{}%
\providecommand \url  [0]{\begingroup\@sanitize@url \@url }%
\providecommand \@url [1]{\endgroup\@href {#1}{\urlprefix }}%
\providecommand \urlprefix  [0]{URL }%
\providecommand \Eprint [0]{\href }%
\providecommand \doibase [0]{http://dx.doi.org/}%
\providecommand \selectlanguage [0]{\@gobble}%
\providecommand \bibinfo  [0]{\@secondoftwo}%
\providecommand \bibfield  [0]{\@secondoftwo}%
\providecommand \translation [1]{[#1]}%
\providecommand \BibitemOpen [0]{}%
\providecommand \bibitemStop [0]{}%
\providecommand \bibitemNoStop [0]{.\EOS\space}%
\providecommand \EOS [0]{\spacefactor3000\relax}%
\providecommand \BibitemShut  [1]{\csname bibitem#1\endcsname}%
\let\auto@bib@innerbib\@empty
%
\bibitem [{\citenamefont {Burch}\ \emph {et~al.}(2016)\citenamefont {Burch},
  \citenamefont {Torbert}, \citenamefont {Phan},\ and\ \citenamefont {{et.
  al.}}}]{Burch2016}%
  \BibitemOpen
  \bibfield  {author} {\bibinfo {author} {\bibfnamefont {J.~L.}\ \bibnamefont
  {Burch}}, \bibinfo {author} {\bibfnamefont {R.~B.}\ \bibnamefont {Torbert}},
  \bibinfo {author} {\bibfnamefont {T.~D.}\ \bibnamefont {Phan}}, \ and\
  \bibinfo {author} {\bibnamefont {{et. al.}}},\ }\bibfield  {title} {\enquote
  {\bibinfo {title} {{Electron-scale measurements of magnetic reconnection in
  space}},}\ }\href {\doibase 10.1126/science.aaf2939} {\bibfield  {journal}
  {\bibinfo  {journal} {Science}\ }\textbf {\bibinfo {volume} {352}},\ \bibinfo
  {pages} {aaf2939} (\bibinfo {year} {2016})}\BibitemShut {NoStop}%
\bibitem [{\citenamefont {Yamada}\ \emph {et~al.}(1997)\citenamefont {Yamada},
  \citenamefont {Ji}, \citenamefont {Hsu}, \citenamefont {Carter},
  \citenamefont {Kulsrud}, \citenamefont {Bretz}, \citenamefont {Jobes},
  \citenamefont {Ono},\ and\ \citenamefont {Perkins}}]{Yamada1997}%
  \BibitemOpen
  \bibfield  {author} {\bibinfo {author} {\bibfnamefont {M.}~\bibnamefont
  {Yamada}}, \bibinfo {author} {\bibfnamefont {H.}~\bibnamefont {Ji}}, \bibinfo
  {author} {\bibfnamefont {S.}~\bibnamefont {Hsu}}, \bibinfo {author}
  {\bibfnamefont {T.}~\bibnamefont {Carter}}, \bibinfo {author} {\bibfnamefont
  {R.}~\bibnamefont {Kulsrud}}, \bibinfo {author} {\bibfnamefont
  {N.}~\bibnamefont {Bretz}}, \bibinfo {author} {\bibfnamefont
  {F.}~\bibnamefont {Jobes}}, \bibinfo {author} {\bibfnamefont
  {Y.}~\bibnamefont {Ono}}, \ and\ \bibinfo {author} {\bibfnamefont
  {F.}~\bibnamefont {Perkins}},\ }\bibfield  {title} {\enquote {\bibinfo
  {title} {{Study of driven magnetic reconnection in a laboratory plasma}},}\
  }\href {\doibase 10.1063/1.872336} {\bibfield  {journal} {\bibinfo  {journal}
  {Phys. Plasmas}\ }\textbf {\bibinfo {volume} {4}},\ \bibinfo {pages} {1936}
  (\bibinfo {year} {1997})}\BibitemShut {NoStop}%
\bibitem [{\citenamefont {Bohlin}\ \emph {et~al.}(2014)\citenamefont {Bohlin},
  \citenamefont {{Von Stechow}}, \citenamefont {Rahbarnia}, \citenamefont
  {Grulke},\ and\ \citenamefont {Klinger}}]{Bohlin2014}%
  \BibitemOpen
  \bibfield  {author} {\bibinfo {author} {\bibfnamefont {H.}~\bibnamefont
  {Bohlin}}, \bibinfo {author} {\bibfnamefont {A.}~\bibnamefont {{Von
  Stechow}}}, \bibinfo {author} {\bibfnamefont {K.}~\bibnamefont {Rahbarnia}},
  \bibinfo {author} {\bibfnamefont {O.}~\bibnamefont {Grulke}}, \ and\ \bibinfo
  {author} {\bibfnamefont {T.}~\bibnamefont {Klinger}},\ }\bibfield  {title}
  {\enquote {\bibinfo {title} {{VINETA II: A linear magnetic reconnection
  experiment}},}\ }\href {\doibase 10.1063/1.4861359} {\bibfield  {journal}
  {\bibinfo  {journal} {Rev. Sci. Instrum.}\ }\textbf {\bibinfo {volume}
  {85}},\ \bibinfo {pages} {023501} (\bibinfo {year} {2014})}\BibitemShut
  {NoStop}%
\bibitem [{\citenamefont {B{\"{u}}chner}\ and\ \citenamefont
  {Daughton}(2007)}]{Buchner2007a}%
  \BibitemOpen
  \bibfield  {author} {\bibinfo {author} {\bibfnamefont {J.}~\bibnamefont
  {B{\"{u}}chner}}\ and\ \bibinfo {author} {\bibfnamefont {W.}~\bibnamefont
  {Daughton}},\ }\bibfield  {title} {\enquote {\bibinfo {title} {{Basic theory
  of collisionless reconnection}},}\ }in\ \href
  {http://www.cambridge.org/us/academic/subjects/astronomy/astrophysics/reconnection-magnetic-fields-magnetohydrodynamics-and-collisionless-theory-and-observations}
  {\emph {\bibinfo {booktitle} {Reconnection of Magnetic Fields:
  Magnetohydrodynamics Collisionless Theory and Observations}}},\ \bibinfo
  {editor} {edited by\ \bibinfo {editor} {\bibfnamefont {J.}~\bibnamefont
  {Birn}}\ and\ \bibinfo {editor} {\bibfnamefont {E.~R.}\ \bibnamefont
  {Priest}}}\ (\bibinfo  {publisher} {Cambridge Univ. Press},\ \bibinfo {year}
  {2007})\ Chap.~\bibinfo {chapter} {3}, pp.\ \bibinfo {pages}
  {87--166}\BibitemShut {NoStop}%
\bibitem [{\citenamefont {Zweibel}\ and\ \citenamefont
  {Yamada}(2009)}]{Zweibel2009}%
  \BibitemOpen
  \bibfield  {author} {\bibinfo {author} {\bibfnamefont {E.~G.}\ \bibnamefont
  {Zweibel}}\ and\ \bibinfo {author} {\bibfnamefont {M.}~\bibnamefont
  {Yamada}},\ }\bibfield  {title} {\enquote {\bibinfo {title} {{Magnetic
  Reconnection in Astrophysical and Laboratory Plasmas}},}\ }\href {\doibase
  10.1146/annurev-astro-082708-101726} {\bibfield  {journal} {\bibinfo
  {journal} {Annu. Rev. Astron. Astrophys.}\ }\textbf {\bibinfo {volume}
  {47}},\ \bibinfo {pages} {291--332} (\bibinfo {year} {2009})}\BibitemShut
  {NoStop}%
\bibitem [{\citenamefont {Yamada}, \citenamefont {Kulsrud},\ and\ \citenamefont
  {Ji}(2010)}]{Yamada2010}%
  \BibitemOpen
  \bibfield  {author} {\bibinfo {author} {\bibfnamefont {M.}~\bibnamefont
  {Yamada}}, \bibinfo {author} {\bibfnamefont {R.~M.}\ \bibnamefont {Kulsrud}},
  \ and\ \bibinfo {author} {\bibfnamefont {H.}~\bibnamefont {Ji}},\ }\bibfield
  {title} {\enquote {\bibinfo {title} {{Magnetic reconnection}},}\ }\href
  {\doibase 10.1103/RevModPhys.82.603} {\bibfield  {journal} {\bibinfo
  {journal} {Rev. Mod. Phys.}\ }\textbf {\bibinfo {volume} {82}},\ \bibinfo
  {pages} {603--664} (\bibinfo {year} {2010})}\BibitemShut {NoStop}%
\bibitem [{\citenamefont {Treumann}\ and\ \citenamefont
  {Baumjohann}(2013)}]{Treumann2013b}%
  \BibitemOpen
  \bibfield  {author} {\bibinfo {author} {\bibfnamefont {R.~A.}\ \bibnamefont
  {Treumann}}\ and\ \bibinfo {author} {\bibfnamefont {W.}~\bibnamefont
  {Baumjohann}},\ }\bibfield  {title} {\enquote {\bibinfo {title}
  {{Collisionless magnetic reconnection in space plasmas}},}\ }\href {\doibase
  10.3389/fphy.2013.00031} {\bibfield  {journal} {\bibinfo  {journal} {Front.
  Phys.}\ }\textbf {\bibinfo {volume} {1}},\ \bibinfo {pages} {31} (\bibinfo
  {year} {2013})}\BibitemShut {NoStop}%
\bibitem [{\citenamefont {Gonzalez}\ and\ \citenamefont
  {Parker}(2016)}]{Gonzalez2016a}%
  \BibitemOpen
  \bibfield  {author} {\bibinfo {author} {\bibfnamefont {W.}~\bibnamefont
  {Gonzalez}}\ and\ \bibinfo {author} {\bibfnamefont {E.}~\bibnamefont
  {Parker}},\ }\href {\doibase 10.1007/978-3-319-26432-5} {\emph {\bibinfo
  {title} {{Magnetic Reconnection}}}},\ Astrophysics and Space Science Library\
  (\bibinfo  {publisher} {Springer International Publishing},\ \bibinfo {year}
  {2016})\BibitemShut {NoStop}%
\bibitem [{\citenamefont {Davidson}\ and\ \citenamefont
  {Gladd}(1975)}]{Davidson1975}%
  \BibitemOpen
  \bibfield  {author} {\bibinfo {author} {\bibfnamefont {R.~C.}\ \bibnamefont
  {Davidson}}\ and\ \bibinfo {author} {\bibfnamefont {N.~T.}\ \bibnamefont
  {Gladd}},\ }\bibfield  {title} {\enquote {\bibinfo {title} {{Anomalous
  transport properties associated with the lower-hybrid-drift instability}},}\
  }\href {\doibase 10.1063/1.861021} {\bibfield  {journal} {\bibinfo  {journal}
  {Phys. Fluids}\ }\textbf {\bibinfo {volume} {18}},\ \bibinfo {pages} {1327}
  (\bibinfo {year} {1975})}\BibitemShut {NoStop}%
\bibitem [{\citenamefont {Davidson}\ \emph {et~al.}(1977)\citenamefont
  {Davidson}, \citenamefont {Gladd}, \citenamefont {Wu},\ and\ \citenamefont
  {Huba}}]{Davidson1977}%
  \BibitemOpen
  \bibfield  {author} {\bibinfo {author} {\bibfnamefont {R.~C.}\ \bibnamefont
  {Davidson}}, \bibinfo {author} {\bibfnamefont {N.~T.}\ \bibnamefont {Gladd}},
  \bibinfo {author} {\bibfnamefont {C.~S.}\ \bibnamefont {Wu}}, \ and\ \bibinfo
  {author} {\bibfnamefont {J.~D.}\ \bibnamefont {Huba}},\ }\bibfield  {title}
  {\enquote {\bibinfo {title} {{Effects of finite plasma beta on the
  lower-hybrid-drift instability}},}\ }\href {\doibase 10.1063/1.861867}
  {\bibfield  {journal} {\bibinfo  {journal} {Phys. Fluids}\ }\textbf {\bibinfo
  {volume} {20}},\ \bibinfo {pages} {301} (\bibinfo {year} {1977})}\BibitemShut
  {NoStop}%
\bibitem [{\citenamefont {Davidson}\ and\ \citenamefont
  {Krall}(1977)}]{Davidson1977a}%
  \BibitemOpen
  \bibfield  {author} {\bibinfo {author} {\bibfnamefont {R.}~\bibnamefont
  {Davidson}}\ and\ \bibinfo {author} {\bibfnamefont {N.}~\bibnamefont
  {Krall}},\ }\bibfield  {title} {\enquote {\bibinfo {title} {{Anomalous
  transport in high-temperature plasmas with applications to solenoidal fusion
  systems}},}\ }\href {\doibase 10.1088/0029-5515/17/6/017} {\bibfield
  {journal} {\bibinfo  {journal} {Nucl. Fusion}\ }\textbf {\bibinfo {volume}
  {17}},\ \bibinfo {pages} {1313--1372} (\bibinfo {year} {1977})}\BibitemShut
  {NoStop}%
\bibitem [{\citenamefont {Papadopoulos}(1977)}]{Papadopoulos1977}%
  \BibitemOpen
  \bibfield  {author} {\bibinfo {author} {\bibfnamefont {K.}~\bibnamefont
  {Papadopoulos}},\ }\bibfield  {title} {\enquote {\bibinfo {title} {{A review
  of anomalous resistivity for the ionosphere}},}\ }\href {\doibase
  10.1029/RG015i001p00113} {\bibfield  {journal} {\bibinfo  {journal} {Rev.
  Geophys.}\ }\textbf {\bibinfo {volume} {15}},\ \bibinfo {pages} {113}
  (\bibinfo {year} {1977})}\BibitemShut {NoStop}%
\bibitem [{\citenamefont {Galeev}(1979)}]{Galeev1979}%
  \BibitemOpen
  \bibfield  {author} {\bibinfo {author} {\bibfnamefont {A.}~\bibnamefont
  {Galeev}},\ }\bibfield  {title} {\enquote {\bibinfo {title} {{Reconnection in
  the magnetotail}},}\ }\href {\doibase 10.1007/BF00172248} {\bibfield
  {journal} {\bibinfo  {journal} {Space Sci. Rev.}\ }\textbf {\bibinfo {volume}
  {23}},\ \bibinfo {pages} {411--425} (\bibinfo {year} {1979})}\BibitemShut
  {NoStop}%
\bibitem [{\citenamefont {Sagdeev}(1979)}]{Sagdeev1979}%
  \BibitemOpen
  \bibfield  {author} {\bibinfo {author} {\bibfnamefont {R.~Z.}\ \bibnamefont
  {Sagdeev}},\ }\bibfield  {title} {\enquote {\bibinfo {title} {{The 1976
  Oppenheimer lectures: Critical problems in plasma astrophysics. I. Turbulence
  and nonlinear waves}},}\ }\href {\doibase 10.1103/RevModPhys.51.1} {\bibfield
   {journal} {\bibinfo  {journal} {Rev. Mod. Phys.}\ }\textbf {\bibinfo
  {volume} {51}},\ \bibinfo {pages} {1--9} (\bibinfo {year}
  {1979})}\BibitemShut {NoStop}%
\bibitem [{\citenamefont {Spitzer}(1965)}]{Spitzer1965}%
  \BibitemOpen
  \bibfield  {author} {\bibinfo {author} {\bibfnamefont {L.}~\bibnamefont
  {Spitzer}},\ }\href@noop {} {\emph {\bibinfo {title} {{Physics of fully
  ionized gases}}}},\ \bibinfo {edition} {2nd}\ ed.\ (\bibinfo  {publisher}
  {Interscience Publication},\ \bibinfo {address} {New York},\ \bibinfo {year}
  {1965})\BibitemShut {NoStop}%
\bibitem [{\citenamefont {Braginskii}(1965)}]{Braginskii1965}%
  \BibitemOpen
  \bibfield  {author} {\bibinfo {author} {\bibfnamefont {S.~I.}\ \bibnamefont
  {Braginskii}},\ }\bibfield  {title} {\enquote {\bibinfo {title} {{Transport
  Processes in a Plasma}},}\ }in\ \href@noop {} {\emph {\bibinfo {booktitle}
  {Reviews of Plasma Physics 1}}}\ (\bibinfo  {publisher} {Consultants
  Bureau},\ \bibinfo {year} {1965})\ pp.\ \bibinfo {pages}
  {205--311}\BibitemShut {NoStop}%
\bibitem [{\citenamefont {Boris}\ \emph {et~al.}(1970)\citenamefont {Boris},
  \citenamefont {Dawson}, \citenamefont {Orens},\ and\ \citenamefont
  {Roberts}}]{Boris1970a}%
  \BibitemOpen
  \bibfield  {author} {\bibinfo {author} {\bibfnamefont {J.}~\bibnamefont
  {Boris}}, \bibinfo {author} {\bibfnamefont {J.~M.}\ \bibnamefont {Dawson}},
  \bibinfo {author} {\bibfnamefont {J.}~\bibnamefont {Orens}}, \ and\ \bibinfo
  {author} {\bibfnamefont {K.}~\bibnamefont {Roberts}},\ }\bibfield  {title}
  {\enquote {\bibinfo {title} {{Computations on Anomalous Resistance}},}\
  }\href {\doibase 10.1103/PhysRevLett.25.706} {\bibfield  {journal} {\bibinfo
  {journal} {Phys. Rev. Lett.}\ }\textbf {\bibinfo {volume} {25}},\ \bibinfo
  {pages} {706--710} (\bibinfo {year} {1970})}\BibitemShut {NoStop}%
\bibitem [{\citenamefont {Biskamp}\ and\ \citenamefont
  {Chodura}(1971)}]{Biskamp1971a}%
  \BibitemOpen
  \bibfield  {author} {\bibinfo {author} {\bibfnamefont {D.}~\bibnamefont
  {Biskamp}}\ and\ \bibinfo {author} {\bibfnamefont {R.}~\bibnamefont
  {Chodura}},\ }\bibfield  {title} {\enquote {\bibinfo {title} {{Computer
  Simulation of Anomalous dc Resistivity}},}\ }\href {\doibase
  10.1103/PhysRevLett.27.1553} {\bibfield  {journal} {\bibinfo  {journal}
  {Phys. Rev. Lett.}\ }\textbf {\bibinfo {volume} {27}},\ \bibinfo {pages}
  {1553--1556} (\bibinfo {year} {1971})}\BibitemShut {NoStop}%
\bibitem [{\citenamefont {Winske}\ and\ \citenamefont
  {Liewer}(1978)}]{Winske1978}%
  \BibitemOpen
  \bibfield  {author} {\bibinfo {author} {\bibfnamefont {D.}~\bibnamefont
  {Winske}}\ and\ \bibinfo {author} {\bibfnamefont {P.~C.}\ \bibnamefont
  {Liewer}},\ }\bibfield  {title} {\enquote {\bibinfo {title} {{Particle
  simulation studies of the lower hybrid drift instability}},}\ }\href
  {\doibase 10.1063/1.862321} {\bibfield  {journal} {\bibinfo  {journal} {Phys.
  Fluids}\ }\textbf {\bibinfo {volume} {21}},\ \bibinfo {pages} {1017}
  (\bibinfo {year} {1978})}\BibitemShut {NoStop}%
\bibitem [{\citenamefont {Sato}\ and\ \citenamefont {Okuda}(1981)}]{Sato1981}%
  \BibitemOpen
  \bibfield  {author} {\bibinfo {author} {\bibfnamefont {T.}~\bibnamefont
  {Sato}}\ and\ \bibinfo {author} {\bibfnamefont {H.}~\bibnamefont {Okuda}},\
  }\bibfield  {title} {\enquote {\bibinfo {title} {{Numerical simulations on
  ion acoustic double layers}},}\ }\href {\doibase 10.1029/JA086iA05p03357}
  {\bibfield  {journal} {\bibinfo  {journal} {J. Geophys. Res.}\ }\textbf
  {\bibinfo {volume} {86}},\ \bibinfo {pages} {3357} (\bibinfo {year}
  {1981})}\BibitemShut {NoStop}%
\bibitem [{\citenamefont {Panov}\ \emph
  {et~al.}(2006{\natexlab{a}})\citenamefont {Panov}, \citenamefont
  {B{\"{u}}chner}, \citenamefont {Fr{\"{a}}nz}, \citenamefont {Korth},
  \citenamefont {Khotyaintsev}, \citenamefont {Nikutowski}, \citenamefont
  {Savin}, \citenamefont {Forna{\c{c}}on}, \citenamefont {Dandouras},\ and\
  \citenamefont {R{\`{e}}me}}]{Panov2006}%
  \BibitemOpen
  \bibfield  {author} {\bibinfo {author} {\bibfnamefont {E.}~\bibnamefont
  {Panov}}, \bibinfo {author} {\bibfnamefont {J.}~\bibnamefont
  {B{\"{u}}chner}}, \bibinfo {author} {\bibfnamefont {M.}~\bibnamefont
  {Fr{\"{a}}nz}}, \bibinfo {author} {\bibfnamefont {A.}~\bibnamefont {Korth}},
  \bibinfo {author} {\bibfnamefont {Y.}~\bibnamefont {Khotyaintsev}}, \bibinfo
  {author} {\bibfnamefont {B.}~\bibnamefont {Nikutowski}}, \bibinfo {author}
  {\bibfnamefont {S.}~\bibnamefont {Savin}}, \bibinfo {author} {\bibfnamefont
  {K.-H.}\ \bibnamefont {Forna{\c{c}}on}}, \bibinfo {author} {\bibfnamefont
  {I.}~\bibnamefont {Dandouras}}, \ and\ \bibinfo {author} {\bibfnamefont
  {H.}~\bibnamefont {R{\`{e}}me}},\ }\bibfield  {title} {\enquote {\bibinfo
  {title} {{CLUSTER spacecraft observation of a thin current sheet at the
  Earth's magnetopause}},}\ }\href {\doibase 10.1016/j.asr.2005.08.024}
  {\bibfield  {journal} {\bibinfo  {journal} {Adv. Space Res.}\ }\textbf
  {\bibinfo {volume} {37}},\ \bibinfo {pages} {1363--1372} (\bibinfo {year}
  {2006}{\natexlab{a}})}\BibitemShut {NoStop}%
\bibitem [{\citenamefont {Panov}\ \emph
  {et~al.}(2006{\natexlab{b}})\citenamefont {Panov}, \citenamefont
  {B{\"{u}}chner}, \citenamefont {Fr{\"{a}}nz}, \citenamefont {Korth},
  \citenamefont {Savin}, \citenamefont {Forna{\c{c}}on}, \citenamefont
  {Dandouras},\ and\ \citenamefont {R{\`{e}}me}}]{Panov2006a}%
  \BibitemOpen
  \bibfield  {author} {\bibinfo {author} {\bibfnamefont {E.~V.}\ \bibnamefont
  {Panov}}, \bibinfo {author} {\bibfnamefont {J.}~\bibnamefont
  {B{\"{u}}chner}}, \bibinfo {author} {\bibfnamefont {M.}~\bibnamefont
  {Fr{\"{a}}nz}}, \bibinfo {author} {\bibfnamefont {A.}~\bibnamefont {Korth}},
  \bibinfo {author} {\bibfnamefont {S.~P.}\ \bibnamefont {Savin}}, \bibinfo
  {author} {\bibfnamefont {K.-H.}\ \bibnamefont {Forna{\c{c}}on}}, \bibinfo
  {author} {\bibfnamefont {I.}~\bibnamefont {Dandouras}}, \ and\ \bibinfo
  {author} {\bibfnamefont {H.}~\bibnamefont {R{\`{e}}me}},\ }\bibfield  {title}
  {\enquote {\bibinfo {title} {{CLUSTER observation of collisionless transport
  at the magnetopause}},}\ }\href {\doibase 10.1029/2006GL026556} {\bibfield
  {journal} {\bibinfo  {journal} {Geophys. Res. Lett.}\ }\textbf {\bibinfo
  {volume} {33}},\ \bibinfo {pages} {L15109} (\bibinfo {year}
  {2006}{\natexlab{b}})}\BibitemShut {NoStop}%
\bibitem [{\citenamefont {Lui}\ \emph {et~al.}(2007)\citenamefont {Lui},
  \citenamefont {Zheng}, \citenamefont {R{\`{e}}me}, \citenamefont {Dunlop},
  \citenamefont {Gustafsson},\ and\ \citenamefont {Owen}}]{Lui2007}%
  \BibitemOpen
  \bibfield  {author} {\bibinfo {author} {\bibfnamefont {A.~T.~Y.}\
  \bibnamefont {Lui}}, \bibinfo {author} {\bibfnamefont {Y.}~\bibnamefont
  {Zheng}}, \bibinfo {author} {\bibfnamefont {H.}~\bibnamefont {R{\`{e}}me}},
  \bibinfo {author} {\bibfnamefont {M.~W.}\ \bibnamefont {Dunlop}}, \bibinfo
  {author} {\bibfnamefont {G.}~\bibnamefont {Gustafsson}}, \ and\ \bibinfo
  {author} {\bibfnamefont {C.~J.}\ \bibnamefont {Owen}},\ }\bibfield  {title}
  {\enquote {\bibinfo {title} {{Breakdown of the frozen-in condition in the
  Earth's magnetotail}},}\ }\href {\doibase 10.1029/2006JA012000} {\bibfield
  {journal} {\bibinfo  {journal} {J. Geophys. Res. Space Phys.}\ }\textbf
  {\bibinfo {volume} {112}},\ \bibinfo {pages} {A04215} (\bibinfo {year}
  {2007})}\BibitemShut {NoStop}%
\bibitem [{\citenamefont {Torbert}\ \emph {et~al.}(2016)\citenamefont
  {Torbert}, \citenamefont {Burch}, \citenamefont {Giles}, \citenamefont
  {Gershman}, \citenamefont {Pollock}, \citenamefont {Dorelli}, \citenamefont
  {Avanov}, \citenamefont {Argall}, \citenamefont {Shuster}, \citenamefont
  {Strangeway}, \citenamefont {Russell}, \citenamefont {Ergun}, \citenamefont
  {Wilder}, \citenamefont {Goodrich}, \citenamefont {Faith}, \citenamefont
  {Farrugia}, \citenamefont {Lindqvist}, \citenamefont {Phan}, \citenamefont
  {Khotyaintsev}, \citenamefont {Moore}, \citenamefont {Marklund},
  \citenamefont {Daughton}, \citenamefont {Magnes}, \citenamefont {Kletzing},\
  and\ \citenamefont {Bounds}}]{Torbert2016}%
  \BibitemOpen
  \bibfield  {author} {\bibinfo {author} {\bibfnamefont {R.~B.}\ \bibnamefont
  {Torbert}}, \bibinfo {author} {\bibfnamefont {J.}~\bibnamefont {Burch}},
  \bibinfo {author} {\bibfnamefont {B.}~\bibnamefont {Giles}}, \bibinfo
  {author} {\bibfnamefont {D.}~\bibnamefont {Gershman}}, \bibinfo {author}
  {\bibfnamefont {C.}~\bibnamefont {Pollock}}, \bibinfo {author} {\bibfnamefont
  {J.}~\bibnamefont {Dorelli}}, \bibinfo {author} {\bibfnamefont
  {L.}~\bibnamefont {Avanov}}, \bibinfo {author} {\bibfnamefont
  {M.}~\bibnamefont {Argall}}, \bibinfo {author} {\bibfnamefont
  {J.}~\bibnamefont {Shuster}}, \bibinfo {author} {\bibfnamefont {R.~J.}\
  \bibnamefont {Strangeway}}, \bibinfo {author} {\bibfnamefont {C.~T.}\
  \bibnamefont {Russell}}, \bibinfo {author} {\bibfnamefont {R.}~\bibnamefont
  {Ergun}}, \bibinfo {author} {\bibfnamefont {F.}~\bibnamefont {Wilder}},
  \bibinfo {author} {\bibfnamefont {K.}~\bibnamefont {Goodrich}}, \bibinfo
  {author} {\bibfnamefont {H.}~\bibnamefont {Faith}}, \bibinfo {author}
  {\bibfnamefont {C.}~\bibnamefont {Farrugia}}, \bibinfo {author}
  {\bibfnamefont {P.-A.}\ \bibnamefont {Lindqvist}}, \bibinfo {author}
  {\bibfnamefont {T.}~\bibnamefont {Phan}}, \bibinfo {author} {\bibfnamefont
  {Y.}~\bibnamefont {Khotyaintsev}}, \bibinfo {author} {\bibfnamefont
  {T.}~\bibnamefont {Moore}}, \bibinfo {author} {\bibfnamefont
  {G.}~\bibnamefont {Marklund}}, \bibinfo {author} {\bibfnamefont
  {W.}~\bibnamefont {Daughton}}, \bibinfo {author} {\bibfnamefont
  {W.}~\bibnamefont {Magnes}}, \bibinfo {author} {\bibfnamefont
  {C.}~\bibnamefont {Kletzing}}, \ and\ \bibinfo {author} {\bibfnamefont
  {S.}~\bibnamefont {Bounds}},\ }\bibfield  {title} {\enquote {\bibinfo {title}
  {{Estimates of Terms in Ohm's Law during an Encounter with an Electron
  Diffusion Region}},}\ }\href {\doibase 10.1002/2016GL069553} {\bibfield
  {journal} {\bibinfo  {journal} {Geophys. Res. Lett.}\ }\textbf {\bibinfo
  {volume} {43}},\ \bibinfo {pages} {5918} (\bibinfo {year}
  {2016})}\BibitemShut {NoStop}%
\bibitem [{\citenamefont {Bale}, \citenamefont {Mozer},\ and\ \citenamefont
  {Phan}(2002)}]{Bale2002}%
  \BibitemOpen
  \bibfield  {author} {\bibinfo {author} {\bibfnamefont {S.~D.}\ \bibnamefont
  {Bale}}, \bibinfo {author} {\bibfnamefont {F.~S.}\ \bibnamefont {Mozer}}, \
  and\ \bibinfo {author} {\bibfnamefont {T.~D.}\ \bibnamefont {Phan}},\
  }\bibfield  {title} {\enquote {\bibinfo {title} {{Observation of lower hybrid
  drift instability in the diffusion region at a reconnecting magnetopause}},}\
  }\href {\doibase 10.1029/2002GL016113} {\bibfield  {journal} {\bibinfo
  {journal} {Geophys. Res. Lett.}\ }\textbf {\bibinfo {volume} {29}},\ \bibinfo
  {pages} {2180} (\bibinfo {year} {2002})}\BibitemShut {NoStop}%
\bibitem [{\citenamefont {Eastwood}\ \emph {et~al.}(2009)\citenamefont
  {Eastwood}, \citenamefont {Phan}, \citenamefont {Bale},\ and\ \citenamefont
  {Tjulin}}]{Eastwood2009}%
  \BibitemOpen
  \bibfield  {author} {\bibinfo {author} {\bibfnamefont {J.~P.}\ \bibnamefont
  {Eastwood}}, \bibinfo {author} {\bibfnamefont {T.~D.}\ \bibnamefont {Phan}},
  \bibinfo {author} {\bibfnamefont {S.~D.}\ \bibnamefont {Bale}}, \ and\
  \bibinfo {author} {\bibfnamefont {A.}~\bibnamefont {Tjulin}},\ }\bibfield
  {title} {\enquote {\bibinfo {title} {{Observations of Turbulence Generated by
  Magnetic Reconnection}},}\ }\href {\doibase 10.1103/PhysRevLett.102.035001}
  {\bibfield  {journal} {\bibinfo  {journal} {Phys. Rev. Lett.}\ }\textbf
  {\bibinfo {volume} {102}},\ \bibinfo {pages} {035001} (\bibinfo {year}
  {2009})}\BibitemShut {NoStop}%
\bibitem [{\citenamefont {Mozer}, \citenamefont {Wilber},\ and\ \citenamefont
  {Drake}(2011)}]{Mozer2011}%
  \BibitemOpen
  \bibfield  {author} {\bibinfo {author} {\bibfnamefont {F.~S.}\ \bibnamefont
  {Mozer}}, \bibinfo {author} {\bibfnamefont {M.}~\bibnamefont {Wilber}}, \
  and\ \bibinfo {author} {\bibfnamefont {J.~F.}\ \bibnamefont {Drake}},\
  }\bibfield  {title} {\enquote {\bibinfo {title} {{Wave associated anomalous
  drag during magnetic field reconnection}},}\ }\href {\doibase
  10.1063/1.3647508} {\bibfield  {journal} {\bibinfo  {journal} {Phys.
  Plasmas}\ }\textbf {\bibinfo {volume} {18}},\ \bibinfo {pages} {102902}
  (\bibinfo {year} {2011})}\BibitemShut {NoStop}%
\bibitem [{\citenamefont {Mozer}\ \emph {et~al.}(2011)\citenamefont {Mozer},
  \citenamefont {Sundkvist}, \citenamefont {McFadden}, \citenamefont
  {Pritchett},\ and\ \citenamefont {Roth}}]{Mozer2011a}%
  \BibitemOpen
  \bibfield  {author} {\bibinfo {author} {\bibfnamefont {F.~S.}\ \bibnamefont
  {Mozer}}, \bibinfo {author} {\bibfnamefont {D.}~\bibnamefont {Sundkvist}},
  \bibinfo {author} {\bibfnamefont {J.~P.}\ \bibnamefont {McFadden}}, \bibinfo
  {author} {\bibfnamefont {P.~L.}\ \bibnamefont {Pritchett}}, \ and\ \bibinfo
  {author} {\bibfnamefont {I.}~\bibnamefont {Roth}},\ }\bibfield  {title}
  {\enquote {\bibinfo {title} {{Satellite observations of plasma physics near
  the magnetic field reconnection X line}},}\ }\href {\doibase
  10.1029/2011JA017109} {\bibfield  {journal} {\bibinfo  {journal} {J. Geophys.
  Res. Space Phys.}\ }\textbf {\bibinfo {volume} {116}},\ \bibinfo {pages}
  {A12224} (\bibinfo {year} {2011})}\BibitemShut {NoStop}%
\bibitem [{\citenamefont {Dorfman}\ \emph {et~al.}(2014)\citenamefont
  {Dorfman}, \citenamefont {Ji}, \citenamefont {Yamada}, \citenamefont {Yoo},
  \citenamefont {Lawrence}, \citenamefont {Myers},\ and\ \citenamefont
  {Tharp}}]{Dorfman2014}%
  \BibitemOpen
  \bibfield  {author} {\bibinfo {author} {\bibfnamefont {S.}~\bibnamefont
  {Dorfman}}, \bibinfo {author} {\bibfnamefont {H.}~\bibnamefont {Ji}},
  \bibinfo {author} {\bibfnamefont {M.}~\bibnamefont {Yamada}}, \bibinfo
  {author} {\bibfnamefont {J.}~\bibnamefont {Yoo}}, \bibinfo {author}
  {\bibfnamefont {E.}~\bibnamefont {Lawrence}}, \bibinfo {author}
  {\bibfnamefont {C.}~\bibnamefont {Myers}}, \ and\ \bibinfo {author}
  {\bibfnamefont {T.~D.}\ \bibnamefont {Tharp}},\ }\bibfield  {title} {\enquote
  {\bibinfo {title} {{Experimental observation of 3-D, impulsive reconnection
  events in a laboratory plasma}},}\ }\href {\doibase 10.1063/1.4862039}
  {\bibfield  {journal} {\bibinfo  {journal} {Phys. Plasmas}\ }\textbf
  {\bibinfo {volume} {21}},\ \bibinfo {pages} {012109} (\bibinfo {year}
  {2014})}\BibitemShut {NoStop}%
\bibitem [{\citenamefont {Silin}, \citenamefont {B{\"{u}}chner},\ and\
  \citenamefont {Vaivads}(2005)}]{Silin2005}%
  \BibitemOpen
  \bibfield  {author} {\bibinfo {author} {\bibfnamefont {I.}~\bibnamefont
  {Silin}}, \bibinfo {author} {\bibfnamefont {J.}~\bibnamefont
  {B{\"{u}}chner}}, \ and\ \bibinfo {author} {\bibfnamefont {A.}~\bibnamefont
  {Vaivads}},\ }\bibfield  {title} {\enquote {\bibinfo {title} {{Anomalous
  resistivity due to nonlinear lower-hybrid drift waves}},}\ }\href {\doibase
  10.1063/1.1927096} {\bibfield  {journal} {\bibinfo  {journal} {Phys.
  Plasmas}\ }\textbf {\bibinfo {volume} {12}},\ \bibinfo {pages} {062902}
  (\bibinfo {year} {2005})}\BibitemShut {NoStop}%
\bibitem [{\citenamefont {Yoon}\ and\ \citenamefont {Lui}(2006)}]{Yoon2006}%
  \BibitemOpen
  \bibfield  {author} {\bibinfo {author} {\bibfnamefont {P.~H.}\ \bibnamefont
  {Yoon}}\ and\ \bibinfo {author} {\bibfnamefont {A.~T.~Y.}\ \bibnamefont
  {Lui}},\ }\bibfield  {title} {\enquote {\bibinfo {title} {{Quasi-linear
  theory of anomalous resistivity}},}\ }\href {\doibase 10.1029/2005JA011482}
  {\bibfield  {journal} {\bibinfo  {journal} {J. Geophys. Res.}\ }\textbf
  {\bibinfo {volume} {111}},\ \bibinfo {pages} {A02203} (\bibinfo {year}
  {2006})}\BibitemShut {NoStop}%
\bibitem [{\citenamefont {B{\"{u}}chner}(2006)}]{Buchner2006b}%
  \BibitemOpen
  \bibfield  {author} {\bibinfo {author} {\bibfnamefont {J.}~\bibnamefont
  {B{\"{u}}chner}},\ }\bibfield  {title} {\enquote {\bibinfo {title} {{Theory
  and Simulation of Reconnection}},}\ }\href {\doibase
  10.1007/s11214-006-9094-x} {\bibfield  {journal} {\bibinfo  {journal} {Space
  Sci. Rev.}\ }\textbf {\bibinfo {volume} {124}},\ \bibinfo {pages} {345--360}
  (\bibinfo {year} {2006})}\BibitemShut {NoStop}%
\bibitem [{\citenamefont {B{\"{u}}chner}(2007)}]{Buchner2007}%
  \BibitemOpen
  \bibfield  {author} {\bibinfo {author} {\bibfnamefont {J.}~\bibnamefont
  {B{\"{u}}chner}},\ }\bibfield  {title} {\enquote {\bibinfo {title}
  {{Astrophysical reconnection and collisionless dissipation}},}\ }\href
  {\doibase 10.1088/0741-3335/49/12B/S30} {\bibfield  {journal} {\bibinfo
  {journal} {Plasma Phys. Control. Fusion}\ }\textbf {\bibinfo {volume} {49}},\
  \bibinfo {pages} {B325--B339} (\bibinfo {year} {2007})}\BibitemShut {NoStop}%
\bibitem [{\citenamefont {Treumann}\ and\ \citenamefont
  {Baumjohann}(2001)}]{Treumann2001a}%
  \BibitemOpen
  \bibfield  {author} {\bibinfo {author} {\bibfnamefont {R.~A.}\ \bibnamefont
  {Treumann}}\ and\ \bibinfo {author} {\bibfnamefont {W.}~\bibnamefont
  {Baumjohann}},\ }\href {\doibase 10.1142/p020} {\emph {\bibinfo {title}
  {{Advanced Space Plasma Physics}}}}\ (\bibinfo  {publisher} {Published by
  Imperial College Press and distributed by World Scientific Publishing Co.},\
  \bibinfo {year} {2001})\BibitemShut {NoStop}%
\bibitem [{\citenamefont {Galeev}\ and\ \citenamefont
  {Sudan}(1984)}]{Galeev1984}%
  \BibitemOpen
  \bibfield  {author} {\bibinfo {author} {\bibfnamefont {A.}~\bibnamefont
  {Galeev}}\ and\ \bibinfo {author} {\bibfnamefont {R.}~\bibnamefont {Sudan}},\
  }\href@noop {} {\emph {\bibinfo {title} {{Basic Plasma Physics II}}}}\
  (\bibinfo  {publisher} {North-Holland Pub.},\ \bibinfo {address} {Amsterdam ;
  New York : New York, N.Y},\ \bibinfo {year} {1984})\BibitemShut {NoStop}%
\bibitem [{\citenamefont {Treumann}(2001)}]{Treumann2001}%
  \BibitemOpen
  \bibfield  {author} {\bibinfo {author} {\bibfnamefont {R.~A.}\ \bibnamefont
  {Treumann}},\ }\bibfield  {title} {\enquote {\bibinfo {title} {{Origin of
  resistivity in reconnection}},}\ }\href {\doibase 10.1186/BF03353256}
  {\bibfield  {journal} {\bibinfo  {journal} {Earth Planets Space}\ }\textbf
  {\bibinfo {volume} {53}},\ \bibinfo {pages} {453--462} (\bibinfo {year}
  {2001})}\BibitemShut {NoStop}%
\bibitem [{\citenamefont {Biskamp}(2000)}]{Biskamp2000}%
  \BibitemOpen
  \bibfield  {author} {\bibinfo {author} {\bibfnamefont {D.}~\bibnamefont
  {Biskamp}},\ }\href
  {https://www.worldcat.org/title/magnetic-reconnection-in-plasmas/oclc/43076847}
  {\emph {\bibinfo {title} {{Magnetic Reconnection in Plasmas}}}}\ (\bibinfo
  {publisher} {Cambridge University Press},\ \bibinfo {address} {Cambridge ;
  New York},\ \bibinfo {year} {2000})\BibitemShut {NoStop}%
\bibitem [{\citenamefont {Mu{\~{n}}oz}\ and\ \citenamefont
  {B{\"{u}}chner}(2016)}]{Munoz2016a}%
  \BibitemOpen
  \bibfield  {author} {\bibinfo {author} {\bibfnamefont {P.~A.}\ \bibnamefont
  {Mu{\~{n}}oz}}\ and\ \bibinfo {author} {\bibfnamefont {J.}~\bibnamefont
  {B{\"{u}}chner}},\ }\bibfield  {title} {\enquote {\bibinfo {title}
  {{Non-Maxwellian electron distribution functions due to self-generated
  turbulence in collisionless guide-field reconnection}},}\ }\href {\doibase
  10.1063/1.4963773} {\bibfield  {journal} {\bibinfo  {journal} {Phys.
  Plasmas}\ }\textbf {\bibinfo {volume} {23}},\ \bibinfo {pages} {102103}
  (\bibinfo {year} {2016})},\ \Eprint {http://arxiv.org/abs/1608.03110}
  {arXiv:1608.03110} \BibitemShut {NoStop}%
\bibitem [{\citenamefont {Kilian}, \citenamefont {Burkart},\ and\ \citenamefont
  {Spanier}(2012)}]{Kilian2012}%
  \BibitemOpen
  \bibfield  {author} {\bibinfo {author} {\bibfnamefont {P.}~\bibnamefont
  {Kilian}}, \bibinfo {author} {\bibfnamefont {T.}~\bibnamefont {Burkart}}, \
  and\ \bibinfo {author} {\bibfnamefont {F.}~\bibnamefont {Spanier}},\
  }\bibfield  {title} {\enquote {\bibinfo {title} {{The Influence of the Mass
  Ratio on Particle Acceleration by the Filamentation Instability}},}\ }in\
  \href {\doibase 10.1007/978-3-642-23869-7} {\emph {\bibinfo {booktitle} {High
  Performance Computing in Science and Engineering '11}}},\ \bibinfo {editor}
  {edited by\ \bibinfo {editor} {\bibfnamefont {W.~E.}\ \bibnamefont {Nagel}},
  \bibinfo {editor} {\bibfnamefont {D.~B.}\ \bibnamefont {Kr{\"{o}}ner}}, \
  and\ \bibinfo {editor} {\bibfnamefont {M.~M.}\ \bibnamefont {Resch}}}\
  (\bibinfo  {publisher} {Springer Berlin Heidelberg},\ \bibinfo {address}
  {Berlin, Heidelberg},\ \bibinfo {year} {2012})\ pp.\ \bibinfo {pages}
  {5--13}\BibitemShut {NoStop}%
\bibitem [{\citenamefont {Harris}(1962)}]{Harris1962}%
  \BibitemOpen
  \bibfield  {author} {\bibinfo {author} {\bibfnamefont {E.~G.}\ \bibnamefont
  {Harris}},\ }\bibfield  {title} {\enquote {\bibinfo {title} {{On a plasma
  sheath separating regions of oppositely directed magnetic field}},}\ }\href
  {\doibase 10.1007/BF02733547} {\bibfield  {journal} {\bibinfo  {journal}
  {Nuovo Cim.}\ }\textbf {\bibinfo {volume} {23}},\ \bibinfo {pages} {115--121}
  (\bibinfo {year} {1962})}\BibitemShut {NoStop}%
\bibitem [{\citenamefont {Horiuchi}\ and\ \citenamefont
  {Sato}(1997)}]{Horiuchi1997}%
  \BibitemOpen
  \bibfield  {author} {\bibinfo {author} {\bibfnamefont {R.}~\bibnamefont
  {Horiuchi}}\ and\ \bibinfo {author} {\bibfnamefont {T.}~\bibnamefont
  {Sato}},\ }\bibfield  {title} {\enquote {\bibinfo {title} {{Particle
  simulation study of collisionless driven reconnection in a sheared magnetic
  field}},}\ }\href {\doibase 10.1063/1.872088} {\bibfield  {journal} {\bibinfo
   {journal} {Phys. Plasmas}\ }\textbf {\bibinfo {volume} {4}},\ \bibinfo
  {pages} {277} (\bibinfo {year} {1997})}\BibitemShut {NoStop}%
\bibitem [{\citenamefont {Pritchett}(2005)}]{Pritchett2005}%
  \BibitemOpen
  \bibfield  {author} {\bibinfo {author} {\bibfnamefont {P.~L.}\ \bibnamefont
  {Pritchett}},\ }\bibfield  {title} {\enquote {\bibinfo {title} {{Onset and
  saturation of guide-field magnetic reconnection}},}\ }\href {\doibase
  10.1063/1.1914309} {\bibfield  {journal} {\bibinfo  {journal} {Phys.
  Plasmas}\ }\textbf {\bibinfo {volume} {12}},\ \bibinfo {pages} {062301}
  (\bibinfo {year} {2005})}\BibitemShut {NoStop}%
\bibitem [{\citenamefont {Ricci}\ \emph {et~al.}(2004)\citenamefont {Ricci},
  \citenamefont {Brackbill}, \citenamefont {Daughton},\ and\ \citenamefont
  {Lapenta}}]{Ricci2004}%
  \BibitemOpen
  \bibfield  {author} {\bibinfo {author} {\bibfnamefont {P.}~\bibnamefont
  {Ricci}}, \bibinfo {author} {\bibfnamefont {J.~U.}\ \bibnamefont
  {Brackbill}}, \bibinfo {author} {\bibfnamefont {W.}~\bibnamefont {Daughton}},
  \ and\ \bibinfo {author} {\bibfnamefont {G.}~\bibnamefont {Lapenta}},\
  }\bibfield  {title} {\enquote {\bibinfo {title} {{Collisionless magnetic
  reconnection in the presence of a guide field}},}\ }\href {\doibase
  10.1063/1.1768552} {\bibfield  {journal} {\bibinfo  {journal} {Phys.
  Plasmas}\ }\textbf {\bibinfo {volume} {11}},\ \bibinfo {pages} {4102}
  (\bibinfo {year} {2004})}\BibitemShut {NoStop}%
\bibitem [{\citenamefont {Huba}(2005)}]{Huba2005}%
  \BibitemOpen
  \bibfield  {author} {\bibinfo {author} {\bibfnamefont {J.~D.}\ \bibnamefont
  {Huba}},\ }\bibfield  {title} {\enquote {\bibinfo {title} {{Hall magnetic
  reconnection: Guide field dependence}},}\ }\href {\doibase 10.1063/1.1834592}
  {\bibfield  {journal} {\bibinfo  {journal} {Phys. Plasmas}\ }\textbf
  {\bibinfo {volume} {12}},\ \bibinfo {pages} {012322} (\bibinfo {year}
  {2005})}\BibitemShut {NoStop}%
\bibitem [{\citenamefont {Le}\ \emph {et~al.}(2013)\citenamefont {Le},
  \citenamefont {Egedal}, \citenamefont {Ohia}, \citenamefont {Daughton},
  \citenamefont {Karimabadi},\ and\ \citenamefont {Lukin}}]{Le2013}%
  \BibitemOpen
  \bibfield  {author} {\bibinfo {author} {\bibfnamefont {A.}~\bibnamefont
  {Le}}, \bibinfo {author} {\bibfnamefont {J.}~\bibnamefont {Egedal}}, \bibinfo
  {author} {\bibfnamefont {O.}~\bibnamefont {Ohia}}, \bibinfo {author}
  {\bibfnamefont {W.}~\bibnamefont {Daughton}}, \bibinfo {author}
  {\bibfnamefont {H.}~\bibnamefont {Karimabadi}}, \ and\ \bibinfo {author}
  {\bibfnamefont {V.~S.}\ \bibnamefont {Lukin}},\ }\bibfield  {title} {\enquote
  {\bibinfo {title} {{Regimes of the Electron Diffusion Region in Magnetic
  Reconnection}},}\ }\href {\doibase 10.1103/PhysRevLett.110.135004} {\bibfield
   {journal} {\bibinfo  {journal} {Phys. Rev. Lett.}\ }\textbf {\bibinfo
  {volume} {110}},\ \bibinfo {pages} {135004} (\bibinfo {year}
  {2013})}\BibitemShut {NoStop}%
\bibitem [{\citenamefont {Kleva}, \citenamefont {Drake},\ and\ \citenamefont
  {Waelbroeck}(1995)}]{Kleva1995}%
  \BibitemOpen
  \bibfield  {author} {\bibinfo {author} {\bibfnamefont {R.~G.}\ \bibnamefont
  {Kleva}}, \bibinfo {author} {\bibfnamefont {J.~F.}\ \bibnamefont {Drake}}, \
  and\ \bibinfo {author} {\bibfnamefont {F.~L.}\ \bibnamefont {Waelbroeck}},\
  }\bibfield  {title} {\enquote {\bibinfo {title} {{Fast reconnection in high
  temperature plasmas}},}\ }\href {\doibase 10.1063/1.871095} {\bibfield
  {journal} {\bibinfo  {journal} {Phys. Plasmas}\ }\textbf {\bibinfo {volume}
  {2}},\ \bibinfo {pages} {23--34} (\bibinfo {year} {1995})}\BibitemShut
  {NoStop}%
\bibitem [{\citenamefont {Rogers}, \citenamefont {Denton},\ and\ \citenamefont
  {Drake}(2003)}]{Rogers2003}%
  \BibitemOpen
  \bibfield  {author} {\bibinfo {author} {\bibfnamefont {B.~N.}\ \bibnamefont
  {Rogers}}, \bibinfo {author} {\bibfnamefont {R.~E.}\ \bibnamefont {Denton}},
  \ and\ \bibinfo {author} {\bibfnamefont {J.~F.}\ \bibnamefont {Drake}},\
  }\bibfield  {title} {\enquote {\bibinfo {title} {{Signatures of collisionless
  magnetic reconnection}},}\ }\href {\doibase 10.1029/2002JA009699} {\bibfield
  {journal} {\bibinfo  {journal} {J. Geophys. Res.}\ }\textbf {\bibinfo
  {volume} {108}},\ \bibinfo {pages} {1111} (\bibinfo {year}
  {2003})}\BibitemShut {NoStop}%
\bibitem [{\citenamefont {Pritchett}(2013)}]{Pritchett2013a}%
  \BibitemOpen
  \bibfield  {author} {\bibinfo {author} {\bibfnamefont {P.~L.}\ \bibnamefont
  {Pritchett}},\ }\bibfield  {title} {\enquote {\bibinfo {title} {{The
  influence of intense electric fields on three-dimensional asymmetric magnetic
  reconnection}},}\ }\href {\doibase 10.1063/1.4811123} {\bibfield  {journal}
  {\bibinfo  {journal} {Phys. Plasmas}\ }\textbf {\bibinfo {volume} {20}},\
  \bibinfo {pages} {061204} (\bibinfo {year} {2013})}\BibitemShut {NoStop}%
\bibitem [{\citenamefont {Wendel}\ \emph {et~al.}(2016)\citenamefont {Wendel},
  \citenamefont {Hesse}, \citenamefont {Bessho}, \citenamefont {Adrian},\ and\
  \citenamefont {Kuznetsova}}]{Wendel2016}%
  \BibitemOpen
  \bibfield  {author} {\bibinfo {author} {\bibfnamefont {D.~E.}\ \bibnamefont
  {Wendel}}, \bibinfo {author} {\bibfnamefont {M.}~\bibnamefont {Hesse}},
  \bibinfo {author} {\bibfnamefont {N.}~\bibnamefont {Bessho}}, \bibinfo
  {author} {\bibfnamefont {M.~L.}\ \bibnamefont {Adrian}}, \ and\ \bibinfo
  {author} {\bibfnamefont {M.}~\bibnamefont {Kuznetsova}},\ }\bibfield  {title}
  {\enquote {\bibinfo {title} {{Nongyrotropic electrons in guide field
  reconnection}},}\ }\href {\doibase 10.1063/1.4942031} {\bibfield  {journal}
  {\bibinfo  {journal} {Phys. Plasmas}\ }\textbf {\bibinfo {volume} {23}},\
  \bibinfo {pages} {022114} (\bibinfo {year} {2016})}\BibitemShut {NoStop}%
\bibitem [{\citenamefont {Pritchett}\ and\ \citenamefont
  {Mozer}(2009)}]{Pritchett2009}%
  \BibitemOpen
  \bibfield  {author} {\bibinfo {author} {\bibfnamefont {P.~L.}\ \bibnamefont
  {Pritchett}}\ and\ \bibinfo {author} {\bibfnamefont {F.~S.}\ \bibnamefont
  {Mozer}},\ }\bibfield  {title} {\enquote {\bibinfo {title} {{Asymmetric
  magnetic reconnection in the presence of a guide field}},}\ }\href {\doibase
  10.1029/2009JA014343} {\bibfield  {journal} {\bibinfo  {journal} {J. Geophys.
  Res.}\ }\textbf {\bibinfo {volume} {114}},\ \bibinfo {pages} {A11210}
  (\bibinfo {year} {2009})}\BibitemShut {NoStop}%
\bibitem [{\citenamefont {Che}, \citenamefont {Drake},\ and\ \citenamefont
  {Swisdak}(2011)}]{Che2011}%
  \BibitemOpen
  \bibfield  {author} {\bibinfo {author} {\bibfnamefont {H.}~\bibnamefont
  {Che}}, \bibinfo {author} {\bibfnamefont {J.~F.}\ \bibnamefont {Drake}}, \
  and\ \bibinfo {author} {\bibfnamefont {M.}~\bibnamefont {Swisdak}},\
  }\bibfield  {title} {\enquote {\bibinfo {title} {{A current filamentation
  mechanism for breaking magnetic field lines during reconnection}},}\ }\href
  {\doibase 10.1038/nature10091} {\bibfield  {journal} {\bibinfo  {journal}
  {Nature}\ }\textbf {\bibinfo {volume} {474}},\ \bibinfo {pages} {184--187}
  (\bibinfo {year} {2011})}\BibitemShut {NoStop}%
\bibitem [{\citenamefont {Moritaka}, \citenamefont {Horiuchi},\ and\
  \citenamefont {Ohtani}(2007)}]{Moritaka2007}%
  \BibitemOpen
  \bibfield  {author} {\bibinfo {author} {\bibfnamefont {T.}~\bibnamefont
  {Moritaka}}, \bibinfo {author} {\bibfnamefont {R.}~\bibnamefont {Horiuchi}},
  \ and\ \bibinfo {author} {\bibfnamefont {H.}~\bibnamefont {Ohtani}},\
  }\bibfield  {title} {\enquote {\bibinfo {title} {{Anomalous resistivity due
  to kink modes in a thin current sheet}},}\ }\href {\doibase
  10.1063/1.2767623} {\bibfield  {journal} {\bibinfo  {journal} {Phys.
  Plasmas}\ }\textbf {\bibinfo {volume} {14}},\ \bibinfo {pages} {102109}
  (\bibinfo {year} {2007})}\BibitemShut {NoStop}%
\bibitem [{\citenamefont {Bellan}(2006)}]{Bellan2006}%
  \BibitemOpen
  \bibfield  {author} {\bibinfo {author} {\bibfnamefont {P.~M.}\ \bibnamefont
  {Bellan}},\ }\href@noop {} {\emph {\bibinfo {title} {{Fundamentals of Plasma
  Physics}}}}\ (\bibinfo  {publisher} {Cambridge University Press},\ \bibinfo
  {address} {Cambridge, UK ; New York},\ \bibinfo {year} {2006})\BibitemShut
  {NoStop}%
\bibitem [{\citenamefont {Watt}, \citenamefont {Horne},\ and\ \citenamefont
  {Freeman}(2002)}]{Watt2002}%
  \BibitemOpen
  \bibfield  {author} {\bibinfo {author} {\bibfnamefont {C.~E.~J.}\
  \bibnamefont {Watt}}, \bibinfo {author} {\bibfnamefont {R.~B.}\ \bibnamefont
  {Horne}}, \ and\ \bibinfo {author} {\bibfnamefont {M.~P.}\ \bibnamefont
  {Freeman}},\ }\bibfield  {title} {\enquote {\bibinfo {title} {{Ion-acoustic
  resistivity in plasmas with similar ion and electron temperatures}},}\ }\href
  {\doibase 10.1029/2001GL013451} {\bibfield  {journal} {\bibinfo  {journal}
  {Geophys. Res. Lett.}\ }\textbf {\bibinfo {volume} {29}},\ \bibinfo {pages}
  {1004} (\bibinfo {year} {2002})}\BibitemShut {NoStop}%
\bibitem [{\citenamefont {Petkaki}(2003)}]{Petkaki2003}%
  \BibitemOpen
  \bibfield  {author} {\bibinfo {author} {\bibfnamefont {P.}~\bibnamefont
  {Petkaki}},\ }\bibfield  {title} {\enquote {\bibinfo {title} {{Anomalous
  resistivity in non-Maxwellian plasmas}},}\ }\href {\doibase
  10.1029/2003JA010092} {\bibfield  {journal} {\bibinfo  {journal} {J. Geophys.
  Res.}\ }\textbf {\bibinfo {volume} {108}},\ \bibinfo {pages} {1442} (\bibinfo
  {year} {2003})}\BibitemShut {NoStop}%
\bibitem [{\citenamefont {Hellinger}, \citenamefont
  {Tr{\'{a}}vn{\'{i}}{\v{c}}ek},\ and\ \citenamefont
  {Menietti}(2004)}]{Hellinger2004}%
  \BibitemOpen
  \bibfield  {author} {\bibinfo {author} {\bibfnamefont {P.}~\bibnamefont
  {Hellinger}}, \bibinfo {author} {\bibfnamefont {P.}~\bibnamefont
  {Tr{\'{a}}vn{\'{i}}{\v{c}}ek}}, \ and\ \bibinfo {author} {\bibfnamefont
  {J.~D.}\ \bibnamefont {Menietti}},\ }\bibfield  {title} {\enquote {\bibinfo
  {title} {{Effective collision frequency due to ion-acoustic instability:
  Theory and simulations}},}\ }\href {\doibase 10.1029/2004GL020028} {\bibfield
   {journal} {\bibinfo  {journal} {Geophys. Res. Lett.}\ }\textbf {\bibinfo
  {volume} {31}},\ \bibinfo {pages} {L10806} (\bibinfo {year}
  {2004})}\BibitemShut {NoStop}%
\bibitem [{\citenamefont {Petkaki}\ \emph {et~al.}(2006)\citenamefont
  {Petkaki}, \citenamefont {Freeman}, \citenamefont {Kirk}, \citenamefont
  {Watt},\ and\ \citenamefont {Horne}}]{Petkaki2006}%
  \BibitemOpen
  \bibfield  {author} {\bibinfo {author} {\bibfnamefont {P.}~\bibnamefont
  {Petkaki}}, \bibinfo {author} {\bibfnamefont {M.~P.}\ \bibnamefont
  {Freeman}}, \bibinfo {author} {\bibfnamefont {T.}~\bibnamefont {Kirk}},
  \bibinfo {author} {\bibfnamefont {C.~E.~J.}\ \bibnamefont {Watt}}, \ and\
  \bibinfo {author} {\bibfnamefont {R.~B.}\ \bibnamefont {Horne}},\ }\bibfield
  {title} {\enquote {\bibinfo {title} {{Anomalous resistivity and the nonlinear
  evolution of the ion-acoustic instability}},}\ }\href {\doibase
  10.1029/2004JA010793} {\bibfield  {journal} {\bibinfo  {journal} {J. Geophys.
  Res.}\ }\textbf {\bibinfo {volume} {111}},\ \bibinfo {pages} {A01205}
  (\bibinfo {year} {2006})}\BibitemShut {NoStop}%
\bibitem [{\citenamefont {B{\"{u}}chner}\ and\ \citenamefont
  {Elkina}(2005)}]{Buchner2005a}%
  \BibitemOpen
  \bibfield  {author} {\bibinfo {author} {\bibfnamefont {J.}~\bibnamefont
  {B{\"{u}}chner}}\ and\ \bibinfo {author} {\bibfnamefont {N.}~\bibnamefont
  {Elkina}},\ }\bibfield  {title} {\enquote {\bibinfo {title} {{Vlasov Code
  Simulation of Anomalous Resistivity}},}\ }\href {\doibase
  10.1007/s11214-006-6542-6} {\bibfield  {journal} {\bibinfo  {journal} {Space
  Sci. Rev.}\ }\textbf {\bibinfo {volume} {121}},\ \bibinfo {pages} {237--252}
  (\bibinfo {year} {2005})}\BibitemShut {NoStop}%
\bibitem [{\citenamefont {B{\"{u}}chner}\ and\ \citenamefont
  {Elkina}(2006)}]{Buchner2006}%
  \BibitemOpen
  \bibfield  {author} {\bibinfo {author} {\bibfnamefont {J.}~\bibnamefont
  {B{\"{u}}chner}}\ and\ \bibinfo {author} {\bibfnamefont {N.}~\bibnamefont
  {Elkina}},\ }\bibfield  {title} {\enquote {\bibinfo {title} {{Anomalous
  resistivity of current-driven isothermal plasmas due to phase space
  structuring}},}\ }\href {\doibase 10.1063/1.2209611} {\bibfield  {journal}
  {\bibinfo  {journal} {Phys. Plasmas}\ }\textbf {\bibinfo {volume} {13}},\
  \bibinfo {pages} {082304} (\bibinfo {year} {2006})}\BibitemShut {NoStop}%
\bibitem [{\citenamefont {Petkaki}\ and\ \citenamefont
  {Freeman}(2008)}]{Petkaki2008}%
  \BibitemOpen
  \bibfield  {author} {\bibinfo {author} {\bibfnamefont {P.}~\bibnamefont
  {Petkaki}}\ and\ \bibinfo {author} {\bibfnamefont {M.}~\bibnamefont
  {Freeman}},\ }\bibfield  {title} {\enquote {\bibinfo {title} {{Nonlinear
  dependence of ion-acoustic anomalous resistivity on electron drift
  velocity}},}\ }\href {\doibase 10.1086/590654} {\bibfield  {journal}
  {\bibinfo  {journal} {Astrophys. J.}\ }\textbf {\bibinfo {volume} {686}},\
  \bibinfo {pages} {686} (\bibinfo {year} {2008})}\BibitemShut {NoStop}%
\bibitem [{\citenamefont {Wu}, \citenamefont {Huang},\ and\ \citenamefont
  {Ji}(2010)}]{Wu2010a}%
  \BibitemOpen
  \bibfield  {author} {\bibinfo {author} {\bibfnamefont {G.}~\bibnamefont
  {Wu}}, \bibinfo {author} {\bibfnamefont {G.}~\bibnamefont {Huang}}, \ and\
  \bibinfo {author} {\bibfnamefont {H.}~\bibnamefont {Ji}},\ }\bibfield
  {title} {\enquote {\bibinfo {title} {{Dependence of the Anomalous Resistivity
  on the Induced Electric Field in Solar Flares}},}\ }\href {\doibase
  10.1088/0004-637X/720/1/771} {\bibfield  {journal} {\bibinfo  {journal}
  {Astrophys. J.}\ }\textbf {\bibinfo {volume} {720}},\ \bibinfo {pages}
  {771--775} (\bibinfo {year} {2010})}\BibitemShut {NoStop}%
\bibitem [{\citenamefont {Zenitani}\ \emph {et~al.}(2011)\citenamefont
  {Zenitani}, \citenamefont {Hesse}, \citenamefont {Klimas},\ and\
  \citenamefont {Kuznetsova}}]{Zenitani2011}%
  \BibitemOpen
  \bibfield  {author} {\bibinfo {author} {\bibfnamefont {S.}~\bibnamefont
  {Zenitani}}, \bibinfo {author} {\bibfnamefont {M.}~\bibnamefont {Hesse}},
  \bibinfo {author} {\bibfnamefont {A.}~\bibnamefont {Klimas}}, \ and\ \bibinfo
  {author} {\bibfnamefont {M.}~\bibnamefont {Kuznetsova}},\ }\bibfield  {title}
  {\enquote {\bibinfo {title} {{New Measure of the Dissipation Region in
  Collisionless Magnetic Reconnection}},}\ }\href {\doibase
  10.1103/PhysRevLett.106.195003} {\bibfield  {journal} {\bibinfo  {journal}
  {Phys. Rev. Lett.}\ }\textbf {\bibinfo {volume} {106}},\ \bibinfo {pages}
  {195003} (\bibinfo {year} {2011})},\ \Eprint {http://arxiv.org/abs/1104.3846}
  {arXiv:1104.3846} \BibitemShut {NoStop}%
\bibitem [{\citenamefont {Tanaka}\ and\ \citenamefont
  {Sato}(1981)}]{Tanaka1981}%
  \BibitemOpen
  \bibfield  {author} {\bibinfo {author} {\bibfnamefont {M.}~\bibnamefont
  {Tanaka}}\ and\ \bibinfo {author} {\bibfnamefont {T.}~\bibnamefont {Sato}},\
  }\bibfield  {title} {\enquote {\bibinfo {title} {{Simulations on lower hybrid
  drift instability and anomalous resistivity in the magnetic neutral
  sheet}},}\ }\href {\doibase 10.1029/JA086iA07p05541} {\bibfield  {journal}
  {\bibinfo  {journal} {J. Geophys. Res.}\ }\textbf {\bibinfo {volume} {86}},\
  \bibinfo {pages} {5541} (\bibinfo {year} {1981})}\BibitemShut {NoStop}%
\bibitem [{\citenamefont {Drake}\ \emph {et~al.}(2003)\citenamefont {Drake},
  \citenamefont {Swisdak}, \citenamefont {Cattell}, \citenamefont {Shay},
  \citenamefont {Rogers},\ and\ \citenamefont {Zeiler}}]{Drake2003}%
  \BibitemOpen
  \bibfield  {author} {\bibinfo {author} {\bibfnamefont {J.~F.}\ \bibnamefont
  {Drake}}, \bibinfo {author} {\bibfnamefont {M.}~\bibnamefont {Swisdak}},
  \bibinfo {author} {\bibfnamefont {C.}~\bibnamefont {Cattell}}, \bibinfo
  {author} {\bibfnamefont {M.~A.}\ \bibnamefont {Shay}}, \bibinfo {author}
  {\bibfnamefont {B.~N.}\ \bibnamefont {Rogers}}, \ and\ \bibinfo {author}
  {\bibfnamefont {A.}~\bibnamefont {Zeiler}},\ }\bibfield  {title} {\enquote
  {\bibinfo {title} {{Formation of electron holes and particle energization
  during magnetic reconnection.}}}\ }\href {\doibase 10.1126/science.1080333}
  {\bibfield  {journal} {\bibinfo  {journal} {Science}\ }\textbf {\bibinfo
  {volume} {299}},\ \bibinfo {pages} {873--7} (\bibinfo {year}
  {2003})}\BibitemShut {NoStop}%
\bibitem [{\citenamefont {Roytershteyn}\ \emph {et~al.}(2012)\citenamefont
  {Roytershteyn}, \citenamefont {Daughton}, \citenamefont {Karimabadi},\ and\
  \citenamefont {Mozer}}]{Roytershteyn2012}%
  \BibitemOpen
  \bibfield  {author} {\bibinfo {author} {\bibfnamefont {V.}~\bibnamefont
  {Roytershteyn}}, \bibinfo {author} {\bibfnamefont {W.}~\bibnamefont
  {Daughton}}, \bibinfo {author} {\bibfnamefont {H.}~\bibnamefont
  {Karimabadi}}, \ and\ \bibinfo {author} {\bibfnamefont {F.~S.}\ \bibnamefont
  {Mozer}},\ }\bibfield  {title} {\enquote {\bibinfo {title} {{Influence of the
  Lower-Hybrid Drift Instability on Magnetic Reconnection in Asymmetric
  Configurations}},}\ }\href {\doibase 10.1103/PhysRevLett.108.185001}
  {\bibfield  {journal} {\bibinfo  {journal} {Phys. Rev. Lett.}\ }\textbf
  {\bibinfo {volume} {108}},\ \bibinfo {pages} {185001} (\bibinfo {year}
  {2012})}\BibitemShut {NoStop}%
\bibitem [{\citenamefont {Fujimoto}\ and\ \citenamefont
  {Sydora}(2012)}]{Fujimoto2012}%
  \BibitemOpen
  \bibfield  {author} {\bibinfo {author} {\bibfnamefont {K.}~\bibnamefont
  {Fujimoto}}\ and\ \bibinfo {author} {\bibfnamefont {R.~D.}\ \bibnamefont
  {Sydora}},\ }\bibfield  {title} {\enquote {\bibinfo {title}
  {{Plasmoid-Induced Turbulence in Collisionless Magnetic Reconnection}},}\
  }\href {\doibase 10.1103/PhysRevLett.109.265004} {\bibfield  {journal}
  {\bibinfo  {journal} {Phys. Rev. Lett.}\ }\textbf {\bibinfo {volume} {109}},\
  \bibinfo {pages} {265004} (\bibinfo {year} {2012})}\BibitemShut {NoStop}%
\bibitem [{\citenamefont {Liu}\ \emph {et~al.}(2013)\citenamefont {Liu},
  \citenamefont {Daughton}, \citenamefont {Karimabadi}, \citenamefont {Li},\
  and\ \citenamefont {Roytershteyn}}]{Liu2013}%
  \BibitemOpen
  \bibfield  {author} {\bibinfo {author} {\bibfnamefont {Y.-H.}\ \bibnamefont
  {Liu}}, \bibinfo {author} {\bibfnamefont {W.}~\bibnamefont {Daughton}},
  \bibinfo {author} {\bibfnamefont {H.}~\bibnamefont {Karimabadi}}, \bibinfo
  {author} {\bibfnamefont {H.}~\bibnamefont {Li}}, \ and\ \bibinfo {author}
  {\bibfnamefont {V.}~\bibnamefont {Roytershteyn}},\ }\bibfield  {title}
  {\enquote {\bibinfo {title} {{Bifurcated Structure of the Electron Diffusion
  Region in Three-Dimensional Magnetic Reconnection}},}\ }\href {\doibase
  10.1103/PhysRevLett.110.265004} {\bibfield  {journal} {\bibinfo  {journal}
  {Phys. Rev. Lett.}\ }\textbf {\bibinfo {volume} {110}},\ \bibinfo {pages}
  {265004} (\bibinfo {year} {2013})}\BibitemShut {NoStop}%
\bibitem [{\citenamefont {Che}(2014)}]{Che2014}%
  \BibitemOpen
  \bibfield  {author} {\bibinfo {author} {\bibfnamefont {H.}~\bibnamefont
  {Che}},\ }\bibfield  {title} {\enquote {\bibinfo {title} {{Two-fluid
  description of wave-particle interactions in strong Buneman turbulence}},}\
  }\href {\doibase 10.1063/1.4882677} {\bibfield  {journal} {\bibinfo
  {journal} {Phys. Plasmas}\ }\textbf {\bibinfo {volume} {21}},\ \bibinfo
  {pages} {062305} (\bibinfo {year} {2014})},\ \Eprint
  {http://arxiv.org/abs/1403.4552} {arXiv:1403.4552} \BibitemShut {NoStop}%
\bibitem [{\citenamefont {Tsinober}(2014)}]{Tsinober2014}%
  \BibitemOpen
  \bibfield  {author} {\bibinfo {author} {\bibfnamefont {A.}~\bibnamefont
  {Tsinober}},\ }\href {\doibase 10.1007/978-94-007-7180-2} {\emph {\bibinfo
  {title} {{The Essence of Turbulence as a Physical Phenomenon}}}}\ (\bibinfo
  {publisher} {Springer Netherlands},\ \bibinfo {address} {Dordrecht},\
  \bibinfo {year} {2014})\BibitemShut {NoStop}%
\bibitem [{\citenamefont {Schindler}(2007)}]{Schindler2007}%
  \BibitemOpen
  \bibfield  {author} {\bibinfo {author} {\bibfnamefont {K.}~\bibnamefont
  {Schindler}},\ }\href
  {http://www.cambridge.org/catalogue/catalogue.asp?isbn=9780521858977} {\emph
  {\bibinfo {title} {{Physics of Space Plasma Activity}}}}\ (\bibinfo
  {publisher} {Cambridge University Press},\ \bibinfo {address} {Cambridge ;
  New York},\ \bibinfo {year} {2007})\BibitemShut {NoStop}%
\bibitem [{\citenamefont {Isaacs}, \citenamefont {Tessein},\ and\ \citenamefont
  {Matthaeus}(2015)}]{Isaacs2015}%
  \BibitemOpen
  \bibfield  {author} {\bibinfo {author} {\bibfnamefont {J.~J.}\ \bibnamefont
  {Isaacs}}, \bibinfo {author} {\bibfnamefont {J.~A.}\ \bibnamefont {Tessein}},
  \ and\ \bibinfo {author} {\bibfnamefont {W.~H.}\ \bibnamefont {Matthaeus}},\
  }\bibfield  {title} {\enquote {\bibinfo {title} {{Systematic averaging
  interval effects on solar wind statistics}},}\ }\href {\doibase
  10.1002/2014JA020661} {\bibfield  {journal} {\bibinfo  {journal} {J. Geophys.
  Res. Space Phys.}\ }\textbf {\bibinfo {volume} {120}},\ \bibinfo {pages}
  {868--879} (\bibinfo {year} {2015})}\BibitemShut {NoStop}%
\bibitem [{\citenamefont {Verscharen}\ and\ \citenamefont
  {Marsch}(2011)}]{Verscharen2011}%
  \BibitemOpen
  \bibfield  {author} {\bibinfo {author} {\bibfnamefont {D.}~\bibnamefont
  {Verscharen}}\ and\ \bibinfo {author} {\bibfnamefont {E.}~\bibnamefont
  {Marsch}},\ }\bibfield  {title} {\enquote {\bibinfo {title} {{Apparent
  temperature anisotropies due to wave activity in the solar wind}},}\ }\href
  {\doibase 10.5194/angeo-29-909-2011} {\bibfield  {journal} {\bibinfo
  {journal} {Ann. Geophys.}\ }\textbf {\bibinfo {volume} {29}},\ \bibinfo
  {pages} {909--917} (\bibinfo {year} {2011})}\BibitemShut {NoStop}%
\bibitem [{\citenamefont {Sagaut}(2006)}]{Sagaut2006}%
  \BibitemOpen
  \bibfield  {author} {\bibinfo {author} {\bibfnamefont {P.}~\bibnamefont
  {Sagaut}},\ }\href {\doibase 10.1007/b137536} {\emph {\bibinfo {title}
  {{Large Eddy Simulation for Incompressible Flows}}}},\ Scientific
  Computation\ (\bibinfo  {publisher} {Springer-Verlag},\ \bibinfo {address}
  {Berlin/Heidelberg},\ \bibinfo {year} {2006})\BibitemShut {NoStop}%
\bibitem [{\citenamefont {Vaivads}\ \emph {et~al.}(2016)\citenamefont
  {Vaivads}, \citenamefont {Retin{\`{o}}}, \citenamefont {Soucek},\ and\
  \citenamefont {{et. al.}}}]{Vaivads2016a}%
  \BibitemOpen
  \bibfield  {author} {\bibinfo {author} {\bibfnamefont {A.}~\bibnamefont
  {Vaivads}}, \bibinfo {author} {\bibfnamefont {A.}~\bibnamefont
  {Retin{\`{o}}}}, \bibinfo {author} {\bibfnamefont {J.}~\bibnamefont
  {Soucek}}, \ and\ \bibinfo {author} {\bibnamefont {{et. al.}}},\ }\bibfield
  {title} {\enquote {\bibinfo {title} {{Turbulence Heating ObserveR ---
  satellite mission proposal}},}\ }\href {\doibase 10.1017/S0022377816000775}
  {\bibfield  {journal} {\bibinfo  {journal} {J. Plasma Phys.}\ }\textbf
  {\bibinfo {volume} {82}},\ \bibinfo {pages} {905820501} (\bibinfo {year}
  {2016})}\BibitemShut {NoStop}%
\bibitem [{\citenamefont {Weibel}(1959)}]{Weibel1959}%
  \BibitemOpen
  \bibfield  {author} {\bibinfo {author} {\bibfnamefont {E.}~\bibnamefont
  {Weibel}},\ }\bibfield  {title} {\enquote {\bibinfo {title} {{Spontaneously
  Growing Transverse Waves in a Plasma Due to an Anisotropic Velocity
  Distribution}},}\ }\href {\doibase 10.1103/PhysRevLett.2.83} {\bibfield
  {journal} {\bibinfo  {journal} {Phys. Rev. Lett.}\ }\textbf {\bibinfo
  {volume} {2}},\ \bibinfo {pages} {83--84} (\bibinfo {year}
  {1959})}\BibitemShut {NoStop}%
\bibitem [{\citenamefont {Bret}, \citenamefont {Firpo},\ and\ \citenamefont
  {Deutsch}(2005)}]{Bret2005}%
  \BibitemOpen
  \bibfield  {author} {\bibinfo {author} {\bibfnamefont {A.}~\bibnamefont
  {Bret}}, \bibinfo {author} {\bibfnamefont {M.-C.}\ \bibnamefont {Firpo}}, \
  and\ \bibinfo {author} {\bibfnamefont {C.}~\bibnamefont {Deutsch}},\
  }\bibfield  {title} {\enquote {\bibinfo {title} {{Electromagnetic
  instabilities for relativistic beam-plasma interaction in whole k space:
  Nonrelativistic beam and plasma temperature effects}},}\ }\href {\doibase
  10.1103/PhysRevE.72.016403} {\bibfield  {journal} {\bibinfo  {journal} {Phys.
  Rev. E}\ }\textbf {\bibinfo {volume} {72}},\ \bibinfo {pages} {016403}
  (\bibinfo {year} {2005})}\BibitemShut {NoStop}%
\bibitem [{\citenamefont {Krall}\ and\ \citenamefont
  {Trivelpiece}(1973)}]{Krall1973}%
  \BibitemOpen
  \bibfield  {author} {\bibinfo {author} {\bibfnamefont {N.~A.}\ \bibnamefont
  {Krall}}\ and\ \bibinfo {author} {\bibfnamefont {A.~W.}\ \bibnamefont
  {Trivelpiece}},\ }\href@noop {} {\emph {\bibinfo {title} {{Principles of
  Plasma Physics}}}}\ (\bibinfo  {publisher} {McGraw-Hill},\ \bibinfo {year}
  {1973})\BibitemShut {NoStop}%
\bibitem [{\citenamefont {Lu}\ \emph {et~al.}(2011)\citenamefont {Lu},
  \citenamefont {Lu}, \citenamefont {Shao}, \citenamefont {Yoon},\ and\
  \citenamefont {Wang}}]{Lu2011}%
  \BibitemOpen
  \bibfield  {author} {\bibinfo {author} {\bibfnamefont {S.}~\bibnamefont
  {Lu}}, \bibinfo {author} {\bibfnamefont {Q.}~\bibnamefont {Lu}}, \bibinfo
  {author} {\bibfnamefont {X.}~\bibnamefont {Shao}}, \bibinfo {author}
  {\bibfnamefont {P.~H.}\ \bibnamefont {Yoon}}, \ and\ \bibinfo {author}
  {\bibfnamefont {S.}~\bibnamefont {Wang}},\ }\bibfield  {title} {\enquote
  {\bibinfo {title} {{Weibel instability and structures of magnetic island in
  anti-parallel collisionless magnetic reconnection}},}\ }\href {\doibase
  10.1063/1.3605029} {\bibfield  {journal} {\bibinfo  {journal} {Phys.
  Plasmas}\ }\textbf {\bibinfo {volume} {18}},\ \bibinfo {pages} {072105}
  (\bibinfo {year} {2011})}\BibitemShut {NoStop}%
\bibitem [{\citenamefont {Schoeffler}\ \emph {et~al.}(2013)\citenamefont
  {Schoeffler}, \citenamefont {Drake}, \citenamefont {Swisdak},\ and\
  \citenamefont {Knizhnik}}]{Schoeffler2013}%
  \BibitemOpen
  \bibfield  {author} {\bibinfo {author} {\bibfnamefont {K.~M.}\ \bibnamefont
  {Schoeffler}}, \bibinfo {author} {\bibfnamefont {J.~F.}\ \bibnamefont
  {Drake}}, \bibinfo {author} {\bibfnamefont {M.}~\bibnamefont {Swisdak}}, \
  and\ \bibinfo {author} {\bibfnamefont {K.}~\bibnamefont {Knizhnik}},\
  }\bibfield  {title} {\enquote {\bibinfo {title} {{The Role of Pressure
  Anisotropy on Particle Acceleration during Magnetic Reconnection}},}\ }\href
  {\doibase 10.1088/0004-637X/764/2/126} {\bibfield  {journal} {\bibinfo
  {journal} {Astrophys. J.}\ }\textbf {\bibinfo {volume} {764}},\ \bibinfo
  {pages} {126} (\bibinfo {year} {2013})}\BibitemShut {NoStop}%
\bibitem [{\citenamefont {Mu{\~{n}}oz}, \citenamefont {Kilian},\ and\
  \citenamefont {B{\"{u}}chner}(2014)}]{Munoz2014a}%
  \BibitemOpen
  \bibfield  {author} {\bibinfo {author} {\bibfnamefont {P.~A.}\ \bibnamefont
  {Mu{\~{n}}oz}}, \bibinfo {author} {\bibfnamefont {P.}~\bibnamefont {Kilian}},
  \ and\ \bibinfo {author} {\bibfnamefont {J.}~\bibnamefont {B{\"{u}}chner}},\
  }\bibfield  {title} {\enquote {\bibinfo {title} {{Instabilities of
  collisionless current sheets revisited: the role of anisotropic heating}},}\
  }\href {\doibase 10.1063/1.4901033} {\bibfield  {journal} {\bibinfo
  {journal} {Phys. Plasmas}\ }\textbf {\bibinfo {volume} {21}},\ \bibinfo
  {pages} {112106} (\bibinfo {year} {2014})},\ \Eprint
  {http://arxiv.org/abs/1501.06022v1} {arXiv:1501.06022v1} \BibitemShut
  {NoStop}%
\bibitem [{\citenamefont {Liu}, \citenamefont {Swisdak},\ and\ \citenamefont
  {Drake}(2009)}]{Liu2009}%
  \BibitemOpen
  \bibfield  {author} {\bibinfo {author} {\bibfnamefont {Y.-H.}\ \bibnamefont
  {Liu}}, \bibinfo {author} {\bibfnamefont {M.}~\bibnamefont {Swisdak}}, \ and\
  \bibinfo {author} {\bibfnamefont {J.~F.}\ \bibnamefont {Drake}},\ }\bibfield
  {title} {\enquote {\bibinfo {title} {{The Weibel instability inside the
  electron-positron Harris sheet}},}\ }\href {\doibase 10.1063/1.3097474}
  {\bibfield  {journal} {\bibinfo  {journal} {Phys. Plasmas}\ }\textbf
  {\bibinfo {volume} {16}},\ \bibinfo {pages} {042101} (\bibinfo {year}
  {2009})}\BibitemShut {NoStop}%
\bibitem [{\citenamefont {Hededal}\ and\ \citenamefont
  {Nishikawa}(2005)}]{Hededal2005}%
  \BibitemOpen
  \bibfield  {author} {\bibinfo {author} {\bibfnamefont {C.~B.}\ \bibnamefont
  {Hededal}}\ and\ \bibinfo {author} {\bibfnamefont {K.-I.}\ \bibnamefont
  {Nishikawa}},\ }\bibfield  {title} {\enquote {\bibinfo {title} {{The
  Influence of an Ambient Magnetic Field on Relativistic collisionless Plasma
  Shocks}},}\ }\href {\doibase 10.1086/430253} {\bibfield  {journal} {\bibinfo
  {journal} {Astrophys. J.}\ }\textbf {\bibinfo {volume} {623}},\ \bibinfo
  {pages} {L89--L92} (\bibinfo {year} {2005})}\BibitemShut {NoStop}%
\bibitem [{\citenamefont {Stockem}, \citenamefont {Lerche},\ and\ \citenamefont
  {Schlickeiser}(2006)}]{Stockem2006}%
  \BibitemOpen
  \bibfield  {author} {\bibinfo {author} {\bibfnamefont {A.}~\bibnamefont
  {Stockem}}, \bibinfo {author} {\bibfnamefont {I.}~\bibnamefont {Lerche}}, \
  and\ \bibinfo {author} {\bibfnamefont {R.}~\bibnamefont {Schlickeiser}},\
  }\bibfield  {title} {\enquote {\bibinfo {title} {{On the Physical Realization
  of Two-dimensional Turbulence Fields in Magnetized Interplanetary
  Plasmas}},}\ }\href {\doibase 10.1086/507461} {\bibfield  {journal} {\bibinfo
   {journal} {Astrophys. J.}\ }\textbf {\bibinfo {volume} {651}},\ \bibinfo
  {pages} {584--589} (\bibinfo {year} {2006})}\BibitemShut {NoStop}%
\bibitem [{\citenamefont {Fried}(1959)}]{Fried1959}%
  \BibitemOpen
  \bibfield  {author} {\bibinfo {author} {\bibfnamefont {B.~D.}\ \bibnamefont
  {Fried}},\ }\bibfield  {title} {\enquote {\bibinfo {title} {{Mechanism for
  Instability of Transverse Plasma Waves}},}\ }\href {\doibase
  10.1063/1.1705933} {\bibfield  {journal} {\bibinfo  {journal} {Phys. Fluids}\
  }\textbf {\bibinfo {volume} {2}},\ \bibinfo {pages} {337} (\bibinfo {year}
  {1959})}\BibitemShut {NoStop}%
\bibitem [{\citenamefont {Medvedev}\ and\ \citenamefont
  {Loeb}(1999)}]{Medvedev1999}%
  \BibitemOpen
  \bibfield  {author} {\bibinfo {author} {\bibfnamefont {M.~V.}\ \bibnamefont
  {Medvedev}}\ and\ \bibinfo {author} {\bibfnamefont {A.}~\bibnamefont
  {Loeb}},\ }\bibfield  {title} {\enquote {\bibinfo {title} {{Generation of
  Magnetic Fields in the Relativistic Shock of Gamma-Ray Burst Sources}},}\
  }\href {\doibase 10.1086/308038} {\bibfield  {journal} {\bibinfo  {journal}
  {Astrophys. J.}\ }\textbf {\bibinfo {volume} {526}},\ \bibinfo {pages}
  {697--706} (\bibinfo {year} {1999})}\BibitemShut {NoStop}%
\bibitem [{\citenamefont {Kuznetsova}, \citenamefont {Hesse},\ and\
  \citenamefont {Winske}(1998)}]{Kuznetsova1998}%
  \BibitemOpen
  \bibfield  {author} {\bibinfo {author} {\bibfnamefont {M.~M.}\ \bibnamefont
  {Kuznetsova}}, \bibinfo {author} {\bibfnamefont {M.}~\bibnamefont {Hesse}}, \
  and\ \bibinfo {author} {\bibfnamefont {D.}~\bibnamefont {Winske}},\
  }\bibfield  {title} {\enquote {\bibinfo {title} {{Kinetic quasi-viscous and
  bulk flow inertia effects in collisionless magnetotail reconnection}},}\
  }\href {\doibase 10.1029/97JA02699} {\bibfield  {journal} {\bibinfo
  {journal} {J. Geophys. Res.}\ }\textbf {\bibinfo {volume} {103}},\ \bibinfo
  {pages} {199} (\bibinfo {year} {1998})}\BibitemShut {NoStop}%
\bibitem [{\citenamefont {Pritchett}(2001)}]{Pritchett2001}%
  \BibitemOpen
  \bibfield  {author} {\bibinfo {author} {\bibfnamefont {P.~L.}\ \bibnamefont
  {Pritchett}},\ }\bibfield  {title} {\enquote {\bibinfo {title} {{Geospace
  Environment Modeling magnetic reconnection challenge: Simulations with a full
  particle electromagnetic code}},}\ }\href {\doibase 10.1029/1999JA001006}
  {\bibfield  {journal} {\bibinfo  {journal} {J. Geophys. Res.}\ }\textbf
  {\bibinfo {volume} {106}},\ \bibinfo {pages} {3783--3798} (\bibinfo {year}
  {2001})}\BibitemShut {NoStop}%
\bibitem [{\citenamefont {Vasyliunas}(1975)}]{Vasyliunas1975}%
  \BibitemOpen
  \bibfield  {author} {\bibinfo {author} {\bibfnamefont {V.~M.}\ \bibnamefont
  {Vasyliunas}},\ }\bibfield  {title} {\enquote {\bibinfo {title} {{Theoretical
  models of magnetic field line merging}},}\ }\href {\doibase
  10.1029/RG013i001p00303} {\bibfield  {journal} {\bibinfo  {journal} {Rev.
  Geophys.}\ }\textbf {\bibinfo {volume} {13}},\ \bibinfo {pages} {303}
  (\bibinfo {year} {1975})}\BibitemShut {NoStop}%
\bibitem [{\citenamefont {Dungey}(1989)}]{Dungey1989}%
  \BibitemOpen
  \bibfield  {author} {\bibinfo {author} {\bibfnamefont {J.}~\bibnamefont
  {Dungey}},\ }\bibfield  {title} {\enquote {\bibinfo {title} {{Noise-free
  neutral sheets}},}\ }in\ \href@noop {} {\emph {\bibinfo {booktitle}
  {Proceedings of an International School and Workshop on Reconnection in Space
  Plasma, Vol II}}}\ (\bibinfo  {publisher} {ESA},\ \bibinfo {address}
  {Potsdam},\ \bibinfo {year} {1989})\ pp.\ \bibinfo {pages}
  {15--19}\BibitemShut {NoStop}%
\bibitem [{\citenamefont {Cai}, \citenamefont {Ding},\ and\ \citenamefont
  {Lee}(1994)}]{Cai1994}%
  \BibitemOpen
  \bibfield  {author} {\bibinfo {author} {\bibfnamefont {H.~J.}\ \bibnamefont
  {Cai}}, \bibinfo {author} {\bibfnamefont {D.~Q.}\ \bibnamefont {Ding}}, \
  and\ \bibinfo {author} {\bibfnamefont {L.~C.}\ \bibnamefont {Lee}},\
  }\bibfield  {title} {\enquote {\bibinfo {title} {{Momentum transport near a
  magnetic X line in collisionless reconnection}},}\ }\href {\doibase
  10.1029/93JA02519} {\bibfield  {journal} {\bibinfo  {journal} {J. Geophys.
  Res.}\ }\textbf {\bibinfo {volume} {99}},\ \bibinfo {pages} {35} (\bibinfo
  {year} {1994})}\BibitemShut {NoStop}%
\bibitem [{\citenamefont {Cai}\ and\ \citenamefont {Lee}(1997)}]{Cai1997}%
  \BibitemOpen
  \bibfield  {author} {\bibinfo {author} {\bibfnamefont {H.~J.}\ \bibnamefont
  {Cai}}\ and\ \bibinfo {author} {\bibfnamefont {L.~C.}\ \bibnamefont {Lee}},\
  }\bibfield  {title} {\enquote {\bibinfo {title} {{The generalized Ohm's law
  in collisionless magnetic reconnection}},}\ }\href {\doibase
  10.1063/1.872178} {\bibfield  {journal} {\bibinfo  {journal} {Phys. Plasmas}\
  }\textbf {\bibinfo {volume} {4}},\ \bibinfo {pages} {509} (\bibinfo {year}
  {1997})}\BibitemShut {NoStop}%
\bibitem [{\citenamefont {Hesse}\ and\ \citenamefont
  {Winske}(1998)}]{Hesse1998a}%
  \BibitemOpen
  \bibfield  {author} {\bibinfo {author} {\bibfnamefont {M.}~\bibnamefont
  {Hesse}}\ and\ \bibinfo {author} {\bibfnamefont {D.}~\bibnamefont {Winske}},\
  }\bibfield  {title} {\enquote {\bibinfo {title} {{Electron dissipation in
  collisionless magnetic reconnection}},}\ }\href {\doibase 10.1029/98JA01570}
  {\bibfield  {journal} {\bibinfo  {journal} {J. Geophys. Res.}\ }\textbf
  {\bibinfo {volume} {103}},\ \bibinfo {pages} {26479} (\bibinfo {year}
  {1998})}\BibitemShut {NoStop}%
\bibitem [{\citenamefont {Hesse}\ \emph {et~al.}(1999)\citenamefont {Hesse},
  \citenamefont {Schindler}, \citenamefont {Birn},\ and\ \citenamefont
  {Kuznetsova}}]{Hesse1999}%
  \BibitemOpen
  \bibfield  {author} {\bibinfo {author} {\bibfnamefont {M.}~\bibnamefont
  {Hesse}}, \bibinfo {author} {\bibfnamefont {K.}~\bibnamefont {Schindler}},
  \bibinfo {author} {\bibfnamefont {J.}~\bibnamefont {Birn}}, \ and\ \bibinfo
  {author} {\bibfnamefont {M.}~\bibnamefont {Kuznetsova}},\ }\bibfield  {title}
  {\enquote {\bibinfo {title} {{The diffusion region in collisionless magnetic
  reconnection}},}\ }\href {\doibase 10.1063/1.873436} {\bibfield  {journal}
  {\bibinfo  {journal} {Phys. Plasmas}\ }\textbf {\bibinfo {volume} {6}},\
  \bibinfo {pages} {1781} (\bibinfo {year} {1999})}\BibitemShut {NoStop}%
\bibitem [{\citenamefont {Hesse}\ and\ \citenamefont
  {Winske}(1994)}]{Hesse1994}%
  \BibitemOpen
  \bibfield  {author} {\bibinfo {author} {\bibfnamefont {M.}~\bibnamefont
  {Hesse}}\ and\ \bibinfo {author} {\bibfnamefont {D.}~\bibnamefont {Winske}},\
  }\bibfield  {title} {\enquote {\bibinfo {title} {{Hybrid simulations of
  collisionless reconnection in current sheets}},}\ }\href {\doibase
  10.1029/94JA00676} {\bibfield  {journal} {\bibinfo  {journal} {J. Geophys.
  Res.}\ }\textbf {\bibinfo {volume} {99}},\ \bibinfo {pages} {11177} (\bibinfo
  {year} {1994})}\BibitemShut {NoStop}%
\bibitem [{\citenamefont {Kuznetsova}, \citenamefont {Hesse},\ and\
  \citenamefont {Winske}(2001)}]{Kuznetsova2001}%
  \BibitemOpen
  \bibfield  {author} {\bibinfo {author} {\bibfnamefont {M.~M.}\ \bibnamefont
  {Kuznetsova}}, \bibinfo {author} {\bibfnamefont {M.}~\bibnamefont {Hesse}}, \
  and\ \bibinfo {author} {\bibfnamefont {D.}~\bibnamefont {Winske}},\
  }\bibfield  {title} {\enquote {\bibinfo {title} {{Collisionless reconnection
  supported by nongyrotropic pressure effects in hybrid and particle
  simulations}},}\ }\href {\doibase 10.1029/1999JA001003} {\bibfield  {journal}
  {\bibinfo  {journal} {J. Geophys. Res.}\ }\textbf {\bibinfo {volume} {106}},\
  \bibinfo {pages} {3799--3810} (\bibinfo {year} {2001})}\BibitemShut {NoStop}%
\bibitem [{\citenamefont {Melzani}\ \emph {et~al.}(2013)\citenamefont
  {Melzani}, \citenamefont {Winisdoerffer}, \citenamefont {Walder},
  \citenamefont {Folini}, \citenamefont {Favre}, \citenamefont {Krastanov},\
  and\ \citenamefont {Messmer}}]{Melzani2013}%
  \BibitemOpen
  \bibfield  {author} {\bibinfo {author} {\bibfnamefont {M.}~\bibnamefont
  {Melzani}}, \bibinfo {author} {\bibfnamefont {C.}~\bibnamefont
  {Winisdoerffer}}, \bibinfo {author} {\bibfnamefont {R.}~\bibnamefont
  {Walder}}, \bibinfo {author} {\bibfnamefont {D.}~\bibnamefont {Folini}},
  \bibinfo {author} {\bibfnamefont {J.~M.}\ \bibnamefont {Favre}}, \bibinfo
  {author} {\bibfnamefont {S.}~\bibnamefont {Krastanov}}, \ and\ \bibinfo
  {author} {\bibfnamefont {P.}~\bibnamefont {Messmer}},\ }\bibfield  {title}
  {\enquote {\bibinfo {title} {{Apar-T: code, validation, and physical
  interpretation of particle-in-cell results}},}\ }\href {\doibase
  10.1051/0004-6361/201321557} {\bibfield  {journal} {\bibinfo  {journal}
  {Astron. Astrophys.}\ }\textbf {\bibinfo {volume} {558}},\ \bibinfo {pages}
  {A133} (\bibinfo {year} {2013})}\BibitemShut {NoStop}%
\bibitem [{\citenamefont {Hesse}, \citenamefont {Kuznetsova},\ and\
  \citenamefont {Birn}(2004)}]{Hesse2004}%
  \BibitemOpen
  \bibfield  {author} {\bibinfo {author} {\bibfnamefont {M.}~\bibnamefont
  {Hesse}}, \bibinfo {author} {\bibfnamefont {M.}~\bibnamefont {Kuznetsova}}, \
  and\ \bibinfo {author} {\bibfnamefont {J.}~\bibnamefont {Birn}},\ }\bibfield
  {title} {\enquote {\bibinfo {title} {{The role of electron heat flux in
  guide-field magnetic reconnection}},}\ }\href {\doibase 10.1063/1.1795991}
  {\bibfield  {journal} {\bibinfo  {journal} {Phys. Plasmas}\ }\textbf
  {\bibinfo {volume} {11}},\ \bibinfo {pages} {5387} (\bibinfo {year}
  {2004})}\BibitemShut {NoStop}%
\bibitem [{\citenamefont {Hesse}(2006)}]{Hesse2006}%
  \BibitemOpen
  \bibfield  {author} {\bibinfo {author} {\bibfnamefont {M.}~\bibnamefont
  {Hesse}},\ }\bibfield  {title} {\enquote {\bibinfo {title} {{Dissipation in
  magnetic reconnection with a guide magnetic field}},}\ }\href {\doibase
  10.1063/1.2403784} {\bibfield  {journal} {\bibinfo  {journal} {Phys.
  Plasmas}\ }\textbf {\bibinfo {volume} {13}},\ \bibinfo {pages} {122107}
  (\bibinfo {year} {2006})}\BibitemShut {NoStop}%
\end{thebibliography}
\end{document}